# Geometry of Interaction for MALL via Hughes-Van Glabbeek Proof-Nets

MASAHIRO HAMANO, Institute of Information Science, Academia Sinica, Taiwan

This paper presents, for the first time, a Geometry of Interaction (GoI) interpretation inspired from Hughes-Van Glabbeek (HvG) proof-nets for multiplicative additive linear logic (MALL). Our GoI dynamically captures HvG's geometric correctness criterion–the toggling cycle condition–in terms of algebraic operators. Our new ingredient is a scalar extension of the $*$-algebra in Girard's $*$-ring of partial isometries over a boolean polynomial ring with literals of eigenweights as indeterminates. In order to capture feedback arising from cuts, we construct a finer grained execution formula. The expansion of this execution formula is longer than that for collections of slices for multiplicative GoI, hence it is harder to prove termination. Our GoI gives a dynamical, semantical account of boolean valuations (in particular, pruning sub-proofs), conversion of weights (in particular, $\alpha$-conversion), and additive (co)contraction, peculiar to additive proof-theory. Termination of our execution formula is shown to correspond to HvG's toggling criterion. The slice-wise restriction of our execution formula (by collapsing the boolean structure) yields the well known correspondence, explicit or implicit in previous works on multiplicative GoI, between the convergence of execution formulas and acyclicity of proof-nets. Feedback arising from the execution formula by restricting to the boolean polynomial structure yields autonomous definability of eigenweights among cuts from the rest of the eigenweights.

## Introduction

Girard's Geometry of Interaction (GoI) [8] is a dynamical semantics which models the Gentzen cut-elimination procedure for proofs of Linear Logic (LL). While GoI is one of various kinds of semantics for proofs themselves (rather than the weaker notion of provability), the singular feature of GoI is to provide modelling of the dynamical aspect of the cut-elimination procedure itself. This aspect is either collapsed intrinsically in denotational models or is less compact in graph theoretical models (e.g., globally chasing paths). GoI is an operator-theoretical interpretation of proofs, in which certain calculations (e.g., solving feed-back equations algebraically or taking traces category-theoretically) extract the dynamics of proofs beyond their static aspect. The operator is modelled in terms of the execution formula, which stipulates how to solve feedback arising

Author's address: Masahiro HAMANO, Institute of Information Science, Academia Sinica, 128 Academia Road, Section 2, Nangang, Taipei, 115, Taiwan, hamano@jaist.ac.jp.

while resolving cuts. The execution formula was formulated originally by Girard [8] in terms of partial isometries of Hilbert spaces, and category theoretically by Haghverdi-Scott [13] in terms of traced monoidal categories [19]. More recently, Girard [11] analyzed feedback and GoI using von Neumann algebras.

The heart of GoI is Multiplicative Linear Logic (MLL) because of its success in syntactical free constructions for the fragment, in which tight connections are observed, implicitly or explicitly, to the other kinds of (categorical) models of denotational semantics and the graph theory of proof-nets. The former connection is that GoI subsumes denotational semantics, which is known to be an invariant of cut-elimination [8, 12], and the latter one is that expanding the Execution formula corresponds to chasing certain paths in proof-nets,; this yields the correspondence between nilpotency of the operator and Danos-Regnier's (DR) correctness criterion of acyclicity [4] for graphs representing proofs.

With a motivation to generalizing the connection to proof-nets for the additives, this paper presents a MALL GoI, for the first time, for Hughes-Van Glabbeek's (HvG) system of proof-nets [17, 18] so as to faithfully capture, in terms of GoI semantics, the HvG criteria, in particular their toggling condition. The HvG system of proof-nets stems from their breakthrough in answering the long-standing open problem of how to relax Girard's monomially weighted MALL proof-nets [10], generated by boolean eigenweights, to allow polynomially weighted ones. The polynomial weights provide a canonical representation of the proofs. For the sake of the relaxation, Hughes-Van Glabbeek have discovered stronger criteria than the canonical combination of the acyclicity and jumps, the latter of which was invented in Girard [10] to draw additional edges over those determined by argument edges of sub-formulas and by axioms. Let us quote HvG's two crucial criteria (P2) and (P3) (with our naming for each) for a set $\theta$ of linkings to represent a MALL proof:

(P2) Every switching of every linking of $\theta$ is acyclic and connected. (Slice-wise DR)
(P3) Every set $\Lambda$ of $\geq 2$ linkings of $\theta$ toggles a & that is not in any switching cycle of $\Lambda$. (Toggling)

While (P2) is point-wise MLL correctness (as shown in our naming), (P3) is the heart of HvG's stronger condition of toggling, in which under the non-singleton set $\Lambda$ of linkings, one has to chase paths in superposition of different slices, yielding many more cycles. Hence one needs to distinguish the *legal cycles* among them: here legal cycles mean those occurring in proofs.

The aim of the GoI presented in this paper is to faithfully accommodate (P3) to GoI semantics, together with (P2), by capturing the dynamics of feedback involving cuts in MALL proofs, inherent in the stronger HvG criteria. This is accomplished in terms of our execution formula, whose algebraic expansion, especially its nilpotency, corresponds not only to acyclicity of (P2) but also to legal cycles for (P3).

Our technical ingredient is a *scalar extension* (à la Bourbaki [2]) of Girard's $*$-ring of partial isometries over the polynomial ring $\mathbb{Z}_2[\mathcal{L}]$ with literals $\mathcal{L}$ of eigenweights as the indeterminates. The polynomial ring enables us to investigate a finer grained algebraic structure, in order to capture (P3). When we uniformly impose $a\bar{a} = 0$ for eigenweights $a$'s in $\mathcal{L}$, this is sufficient to capture the acyclicity of (P2). The previous MALL GoI works (e.g., [7, 16, 20, 24]), starting from Girard's GoI III [9], are not sound enough to reflect (P3), simply because none of them directly employs HvG as their representation of MALL proofs. Also some of them restrict their interest only to slice-wise GoI, which is merely point-wise collections of multiplicative GoI, hence they collapse (P3) in an adhoc manner. Some of the literature employs different flavored additive syntax (e.g., token machines and interaction graphs), whose relationship to HvG remains unknown. Since the HvG system of proof-nets is known to provide a canonical representation of MALL proofs, we believe that our GoI directly stemming from HvG clearly extracts the semantic aspect of the dynamics of MALL proofs.

This paper starts, in Section 1, with observing how a legal cycle on (P3) arises in representing a proof, while superposing its different slices.
Section 2 concerns the algebraic ingredients of the paper, in which we construct a $*$-algebra $\mathcal{A}_{(\mathbb{Z}_2[\mathcal{L}])}$ on Girard's $*$-ring $\mathcal{A}$ of partial isometries over a boolean ring. The $*$-ring $\mathcal{A}$ is the origin of Girard's discovery of GoI I [8]; the boolean ring is a ring-theoretic representation of a boolean algebra, which uses *exclusive or* instead of the usual *or* for the sum. The boolean structure is compatible with HvG's polynomial boolean weights and our ingredient is the polynomial ring $\mathbb{Z}_2[\mathcal{L}]$ over the literals $\mathcal{L}$ of eigenweights as indeterminates. Rather than the *idempotency* $x^2 = x$ required to be a boolean ring, we start with a weaker condition of *nilpotency* ( representing the exclusive-or) so that the boolean one is reached by quotient by ideals.

Section 3 interprets each MALL proof as an operator $[\![ \pi ]\!]$ using $\mathcal{A}_{(\mathbb{Z}_2[\mathcal{L}])}$, which naturally leads to a kind of universal construction of Girard's GoI III [9], generalising his implicit use of monomial weights into our explicit algebraic structure of polynomials. First, a *quasi-execution formula* $\mathsf{qEx}\,(\sigma_\Delta, [\![ \pi ]\!])$ is formulated, capturing point-wise DR of (P2) in terms of nilpotency of the operator under the coarsest algebraic reduction of the scalars. Second, a legitimate *execution formula* is obtained together with a ring homomorphism among the scalars $\mathbb{Z}_2[\mathcal{L}]$, which makes eigenweights of cut-formulas definable from the rest. Invariance of the execution formula (under cut-elimination) is shown up to *replacement*, *valuation* and *superposition* of indeterminates, realized by the ring homomorphism on the scalars, whose definition in this section is confined to external one.

Section 4 goes into constructing a *finer grained execution formula* for the HvG criterion (P3). The construction provides an interpretation of superposition of slices in order to accommodate the legal cycles. We augment a measure $\mathfrak{m}_\pi$ (on top of $[\![ \pi ]\!]$) as an operator using the *semiring* of formal languages over $\mathcal{L}$. The constructed $\mathfrak{m}_\pi$ controls which quotient to be taken in the scalars $\mathbb{Z}_2[\mathcal{L}]$ for each element of $[\![ \pi ]\!]$. To the pair $[\![ \pi ]\!] : \mathfrak{m}_\pi$, a quasi-execution formula $\mathfrak{qex}\,(\sigma_\Delta, \mathfrak{m}_\pi)$ is run simultaneously so that the nilpotency of the pair is shown to correspond to (P3).

Section 5 finally concerns a refinement of Section 3 on the status of ring homomorphisms among the scalars. The ring homomorphisms are improved to arise internally as solutions of autonomous equational systems $\mathsf{eq}(\sigma_\Delta, [\![ \pi ]\!])$, which we formulate parallel to the operators of the quasi-execution formulas, only dependent on $\pi$, but free from the cut elimination for $\pi$.

(GoI only has soundness:)
In spite of the advantages of GoI semantics to other kinds of semantics, in general GoI lacks any completeness theorem, only providing a sound interpretation of proofs. An analog of completeness at the level of proofs, the converse of soundness, has been established in some other frameworks, as *full completeness* of certain denotational models (e.g. [1, 5]) and as *correctness* in the geometry of proof-nets (e.g., [10, 18]). The former (resp. the latter) distinguishes those morphisms (resp. proof-nets) which are the denotations of proofs between general semantic objects in a category (resp. in a system of proof-structures). The MALL GoI presented in this paper, similarly to other GoI, does not touch the completeness problem. In particular, this means that the present paper does not model Girard's proof-net notion of jumps [10] (also employed by HvG [17]). This is a notion invented for the sake of obtaining completeness (i.e. correctness for proof-structures) by augmenting additional edges in each *single* slice. We think that the notion of jump, as far as GoI itself is concerned (at the level of soundness) is not crucial, though we leave it open how to interpret the jump in GoI.

# 1 Motivation: Legal Cycles Arising in HvG MALL Proof-Nets with Cuts

## 1.1 Hughes-Van Glabbeek MALL proof-net on a cut sequent ([17, 18])

### MALL syntax

MALL is the multiplicative additive fragment of linear logic. Formulas are built from propositional variables and their negations by the binary connectives $\otimes$ (tensor), $⅋$ (par), & (with) and $\oplus$ (plus). Negation $(-)^\perp$ extends to arbitrary formulas by de Morgan duality. A MALL sequent $\vdash [\Delta], \Gamma$ with a cut-list inside $[-]$ is a set $\Gamma$ of formula occurrences together with pair-wise dual formula occurrences $\Delta$. Each dual pair of $\Delta$ is written explicitly by the binary connective $*$ (cut). Sequents are proved using the following rules:

$$\frac{}{\vdash A, A^\perp}\ ax \qquad \frac{\vdash [\Delta_1],\ \Gamma_1, A \quad \vdash [\Delta_2],\ \Gamma_2, B}{\vdash [\Delta_1, \Delta_2],\ \Gamma_1, \Gamma_2, A \otimes B}\ \otimes$$

$$\frac{\vdash [\Delta],\ \Gamma, A, B}{\vdash [\Delta],\ \Gamma, A ⅋ B}\ ⅋ \qquad \frac{\vdash [\Delta_1],\ \Gamma_1, A \quad \vdash [\Delta_2],\ \Gamma_2, A^\perp}{\vdash [\Delta_1, \Delta_2, A * A^\perp],\ \Gamma_1, \Gamma_2}\ \text{cut}$$

$$\frac{\vdash [\Delta],\ \Gamma, A}{\vdash [\Delta],\ \Gamma, A \oplus B}\ \oplus_1 \qquad \frac{\vdash [\Delta],\ \Gamma, B}{\vdash [\Delta],\ \Gamma, A \oplus B}\ \oplus_2$$

$$\frac{\vdash [\Delta_1, \Sigma],\ \Gamma, A \quad \vdash [\Delta_2, \Sigma],\ \Gamma, B}{\vdash [\Delta_1, \Delta_2, \Sigma],\ \Gamma, A \,\&\, B}\ \&$$

**Note:** In &-rule, not only $\Gamma$ in the conclusion is superposed but also so is $\Sigma$ in the cut-list. $\Sigma$ is not deterministically chosen in the premises so that $\Sigma$ in general is neither empty (i.e., never superimpose cuts) nor maximal (i.e., superimpose as many cuts as possible). The deterministic choices are observed in [18] to be less harmonious with the HvG proof-nets in terms either of the canonicity of the graph transformation of syntactic sequents or of the simplicity of the graphical correctness criteria. The superposition among cut-formulas (as well as context), made explicit by the syntax with cut-formulas, causes the well-known *additive (co)contraction* arising in MALL cut-elimination. The exchange-rule is eliminated under the assumption that formula occurrences are implicitly tracked between premises and conclusion of a rule.

As a motivation for the paper, we recall the definition of the HvG system of proof-nets, into which every MALL proof is translated canonically. The image of the translation is characterized non-syntactically by geometric criteria (P0) ~(P3) (Theorem 1.3 below). For the cut-free proofs with $\Delta$ being empty, (P0) is omitted:

**Definition 1.1 ( Hughes-Van Glabbeek MALL proof-net on a cut sequent** $\vdash [\Delta], \Gamma$ **([17, 18]) ).**

We start with the terminology necessary for the definition, in which $\Xi$ denotes $[\Delta], \Gamma$.

-**Cut:** is a pair $\{A, A^\perp\}$ of complementary MALL formulas.

-&-**resolution:** Deletion of one argument subtree of each &.

-**Cut-Additive resolution:** Deletion of some cuts and one argument subtree of each & and $\oplus$.

-**Axiom link:** Pair of complementary leaves (complementary literals of propositional variables)

-**Linking** $\lambda$**:** Partitioning of the leaves of a cut-additive resolution $\Xi \upharpoonright \lambda$ into axiom links.

-A set $\Lambda$ of linkings **toggles** a & occurrence $a$ if both arguments of $a$ are present in $\bigcup_{\lambda \in \Lambda} \Xi \upharpoonright \lambda$.

- An axiom link $l$ **depends on** $a$ of & occurrence within $\Lambda$ if $\exists \lambda, \lambda' \in \Lambda$ such that $l \in \lambda \setminus \lambda'$ and $a$ is the only & toggled by $\{\lambda, \lambda'\}$.

-**Graph** $\mathcal{G}_\Lambda$**:** $\bigcup_{\lambda \in \Lambda} \Xi \upharpoonright \lambda + \bigcup \Lambda +$ jump edges between each axiom link in $\Lambda$ and any & occurrences, on which it depends within $\Lambda$.

-**Jump:** Edge from a &-vertex $a$ to an axiom link $l$ depending on $a$.

-**Switching cycle:** Cycle with $\leq 1$ **switch edge** (=jump or argument edge) of each $⅋$/&

Then we define:
-A set $\theta$ of linkings on $[\Delta], \Gamma$ is a **proof-net** if it satisfies:
**(P0)** At least one literal occurrence of every cut is in $\theta$. (Cut)
**(P1)** Every one linking of $\theta$ is on any given &-resolution. (Resolution)
**(P2)** Every switching of every linking of $\theta$ is acyclic and connected. (Slice-wise DR)
**(P3)** Every set $\Lambda$ of $\geq 2$ linkings of $\theta$ toggles a & that is not in any switching cycle of $\Lambda$. (Toggling)

The MALL derivation rules are abstracted graph theoretically in terms of the proof nets:

**Definition 1.2 (Translation of MALL proofs into the proof-nets).** Every MALL proof is inductively translated into the set of linkings on $\vdash [\Delta], \Gamma$ accordingly to the MALL rules. In the following $\theta_1$ (resp,. $\theta_2$) denotes the translation of the left (resp. right) premise of the respective (binary) rule.
(ax) The axiom is translated into the set $\{\overset{\frown}{A \quad A^{\perp}}\}$ consisting of the single axiom link.
($\parr$ and $\oplus_i$) The transformation remains the same as that of the premise.
($\otimes$ and cut) The proof is translated into the set $\{\lambda_1 \cup \lambda_2 \mid \lambda_i \in \theta_i\}$ of the linkings.
(&) The proof is translated into the set $\theta_1 \cup \theta_2$ of the linkings.

The main theorem of Hughes-Van Glabbeek states that the image of the transformation of Definition 1.2 is characterised geometrically without referring syntactic rules:

**Theorem 1.3 (Hughes-Van Glabbeek's sequentialisation [17, 18]).** *A set of linkings is a transformation of a* MALL *proof iff it is a proof-net.*

**Eigenweight associated with** &**:**
In what follows in this paper, each & occurrence is identified with an associated eigenweight, denoted $a, b, \ldots$. Different (occurrences of) &s are associated (often subscripted) with distinguished eigenweights, denoted $\&_a, \&_b, \ldots$. Each $\&_a$ is assigned left or right by choosing the argument subtree of the respective premise. The eigenweight $a$ and its negation $\bar{a}$ are read respectively as follows: $\&_a$ is assigned left and right. This directly implies the following conventions in interpreting proofs in Def 1.2.
($\otimes$-rule and cut) The eigenweights in the two premises are distinct.
(&-rule) The eigenweights in the superposed contexts $\Sigma$ and $\Gamma$ are superposed (i.e., contracted via the superposition), while the other eigenweights are distinct. In particular, a superposed formula may contain a &, whereby the respective eigenweights for a same & (e.g., $a$ and $b$ for the &) in the respective premises are contracted into one (e.g., a fresh $c$ for the suiperposed &) in the conclusion.

## 1.2 Motivation

After the above subsection having recalled the HvG proof-nets, this paper starts with observing a phenomenon peculiar to the HvG system of proof-nets for the additives. This motivates our GoI to be presented in this paper.

**A legal cycle in a MALL proof**

The following Fig 1 is a proof-net interpreting a MALL proof of

$$\vdash [X * Y^{\perp}, X^{\perp} * Y] \quad X^{\perp}, X \otimes Y^{\perp}, Y \&_a Y$$

in which $X$ and $Y$ denote the same formula:

The proof-net is obtained from the following two (first and second) proofs by applying a &-rule introducing $Y \&_a Y$. The cut lists of the two proofs are the same, and in applying the &-rule, all the cut-pairs (in the same lists) are superposed (i.e., taking the $\Delta_1$ and the $\Delta_2$ empty of the &-rule).

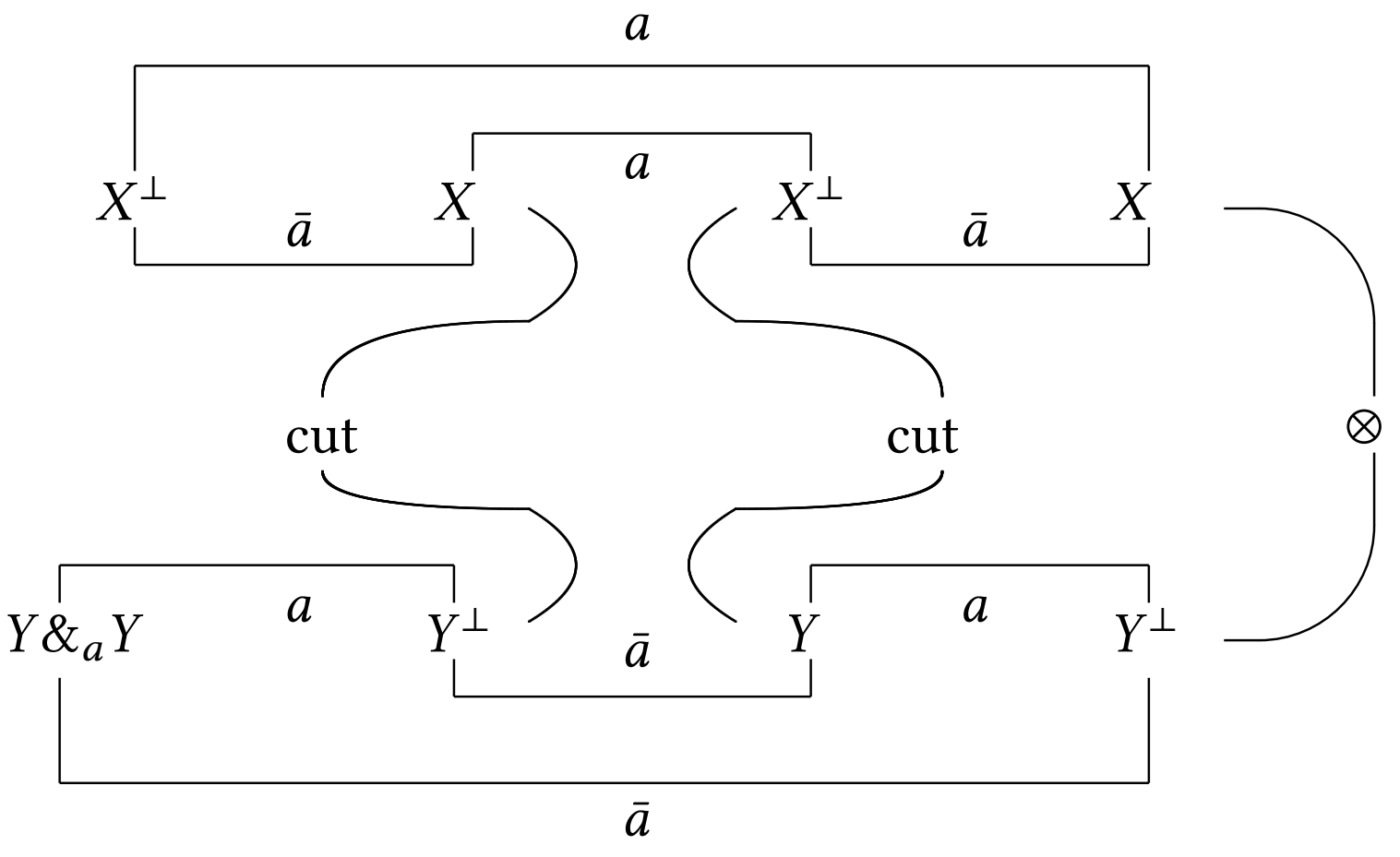

Fig. 1. A Legal Cycle in MALL Proof-Net

In the following left (res. right) $Y$ of $Y \,\&\, Y$ is written by $Y_a$ (res. by $Y_{\bar{a}}$). The linkings in the first (res. second) proof, written above (res. below) the sequent, become weighted with $a$ (res. $\bar{a}$) in Fig 1. (The unique proofs are omitted to derive the premises of the cuts).

- $$\dfrac{\vdash Y_a \quad [Y^\perp * X] \quad X^\perp \qquad \vdash X^\perp \quad X \otimes Y^\perp \quad Y}{\vdash Y_a \quad [Y^\perp * X, \;\; X^\perp * Y] \quad X^\perp \quad X \otimes Y^\perp}\; cut$$

- $$\dfrac{\vdash X^\perp \quad X \otimes Y^\perp \quad Y_{\bar{a}} \qquad \vdash X^\perp \quad [X * Y^\perp] \quad Y}{\vdash Y_{\bar{a}} \quad X^\perp \quad [X * Y^\perp, \;\; Y * X^\perp] \quad X \otimes Y^\perp}\; cut$$

The so-obtained proof-net of Fig 1 yields a cycle passing through the two-cuts with the upper-middle and lower-middle axioms weighted respectively by $a$ and by $\bar{a}$. The cycle is legal since it toggles the &, which lies outside the cycle, and on which no switching cycle passes. We leave it to the reader to check that whatever be the choice of superposition for cut formulas in applying the last &-rule (i.e., whatever be the division of the $\Delta_i$ and the $\Sigma$ in the &-rule), a variant of the same cycle appears outside the $\&_a$.

The observation reveals that the legal cycle becomes visible only when superposing the two slices for $a = 1$ and $a = 0$; hence, slice-wise imposing $a\bar{a} = 0$, without the communication between the two slices, fails to interpret the cycle.

This paper concerns a construction of GoI to faithfully interpret such legal cycles in execution formulas for interpretations of proofs.

## 2 The $*$-algebra $\mathcal{A}_{(\mathbb{Z}_2[\mathcal{L}])}$ on $\mathcal{A}$ of Partial Isometries over Polynomials $\mathbb{Z}_2[\mathcal{L}]$

**Definition 2.1 ((pseudo) boolean ring).** A *boolean ring* is a ring for which the following holds for every element $x$;

$$x^2 = x \quad \text{(idempotency)}$$

Idempotency implies that for every $x$, we have;

$$x + x = 0 \quad (\textit{nilpotency})$$

The nilpotency comes from a property of *exclusive-or*, which is the + in the boolean ring representation of boolean algebras. A *pseudo-boolean ring* is a ring only with the nilpotency condition.

**Definition 2.2 ($*$-ring).** A *$*$-ring* (also called *involutive ring*) is a ring with an involution $* : A \longrightarrow A$, for which the following holds for all $x, y \in A$:

$$\begin{array}{ll} (x+y)^* = x^* + y^* & (xy)^* = y^* x^* \\ 1^* = 1 & (x^*)^* = x \end{array}$$

**Definition 2.3 ($*$-algebra $A$ over commutative ring $R$).** An *(associative) algebra* over a commutative ring $R$ is a ring $A$ which is also a module over $R$ such that the ring and module multiplications are compatible as follows for all $x, y \in A$ and $v \in R$:

$$v(xy) = (vx)y = x(vy)$$

A *$*$-algebra* (also called *involutive algebra*) is an algebra over a commutative $*$-ring $R$ with involution $(\ )'$ such that

$$(rx)^* = r' x^*$$

**Definition 2.4 (The set $\mathcal{L}$ of the literals of the eigenweights).** *Eigenweights*, denoted by $a, b, \ldots$, are formal boolean variables. *Literals of eigenweights* are either eigenweights or their formal negations, denoted by $\bar{a}, \bar{b}, \ldots$. The set of literals is denoted by $\mathcal{L}$. In the following Sections from 3, each (atomic) eigenweight $a$ is associated with each occurrence of the logical connective & in MALL formulas. $a$ and $\bar{a}$ mean respectively the left and right of the $\&_a$.

**Definition 2.5 (The polynomial ring of $\mathcal{L}$ over $\mathbb{Z}_2$).** $\mathbb{Z}_2$ denotes the ring $\mathbb{Z}/2\mathbb{Z}$ of integers $\mathbb{Z}$ under addition and multiplication mod 2. For a set $\mathcal{L}$ of literals of eigenweights, $\mathbb{Z}_2[\mathcal{L}]$ denotes the polynomial ring over $\mathbb{Z}_2$ in the indeterminates $\mathcal{L}$. $\mathbb{Z}_2[\mathcal{L}]$ is automatically pseudo-boolean because its characteristic 2 represents the nilpotency of Definition 2.1.

In order to obtain a boolean ring as a certain quotient of $\mathbb{Z}_2[\mathcal{L}]$, we impose two conditions respectively on addition and on multiplication for complementary literals $a$ and $\bar{a}$:

**Definition 2.6 (quotient $\mathbb{Z}_2[\mathcal{L}]_{\mathtt{C}_+}$ by complementary addition $a + \bar{a} = 1$).** $\mathbb{Z}_2[\mathcal{L}]_{\mathtt{C}_+}$ denotes the quotient ring;

$$\mathbb{Z}_2[\mathcal{L}]_{\mathtt{C}_+} := \mathbb{Z}_2[\mathcal{L}]/\mathcal{L}_{\mathtt{C}_+}$$

by the ideal $\mathcal{L}_{\mathtt{C}_+} := \bigoplus_{a \in \mathcal{L}} \langle a + \bar{a} - 1 \rangle$, where $\langle x \rangle$ is the *principal ideal* generated by $x \in \mathbb{Z}_2[\mathcal{L}]$.

In $\mathbb{Z}_2[\mathcal{L}]_{\mathtt{C}_+}$, for every eigenweight $a \in \mathcal{L}$, $a^2 = a$ is equivalent to $a\bar{a} = 0 = \bar{a}a$. Hence the two principal ideals $\langle a^2 - a \rangle$ and $\langle a\bar{a} \rangle$ coincide there.

**Definition 2.7 (quotient $\mathbb{Z}_2[\mathcal{L}]_{\mathtt{C}_{+\times}}$ by complementary multiplication $a\bar{a} = 0$).** $\mathbb{Z}_2[\mathcal{L}]_{\mathtt{C}_{+\times}}$ denotes the quotient ring;

$$\mathbb{Z}_2[\mathcal{L}]_{\mathtt{C}_{+\times}} = \mathbb{Z}_2[\mathcal{L}]_{\mathtt{C}_+}/\mathcal{L}_{\mathtt{C}_\times} = \mathbb{Z}_2[\mathcal{L}]/(\mathcal{L}_{\mathtt{C}_+} + \mathcal{L}_{\mathtt{C}_\times}),$$

where the ideal $\mathcal{L}_{\mathtt{C}_\times} := \bigoplus_{a \in \mathcal{L}} \langle a\bar{a} \rangle$.
Note that $\mathcal{L}_{\mathtt{C}_+} + \mathcal{L}_{\mathtt{C}_\times}$, the element-wise sum of the two ideals, coincides with the ideal generated by $\mathcal{L}_{\mathtt{C}_+}$ and $\mathcal{L}_{\mathtt{C}_\times}$.

**Lemma 2.8.** $\mathbb{Z}_2[\mathcal{L}]_{\mathtt{C}_{+\times}}$ *is a boolean ring.*

Proof. Because $x^2 = x \wedge y^2 = y$ implies $(x+y)^2 = x+y \wedge (xy)^2 = xy$ for all $x, y \in \mathbb{Z}_2[\mathcal{L}]$. $\square$

Since every polynomial ring over a commutative ring is a $*$-algebra for each element $f$ with

$$f^*(x) = f(-x)$$

$\mathbb{Z}_2[\mathcal{L}]$ becomes a trivial $*$-algebra over $\mathbb{Z}_2$ because of the nilpotency so that $-x = x$.

**Definition 2.9 (The $*$-ring $\mathcal{A}$ of partial isometries $p$ and $q$ (cf. [3, 8])).** The $*$-ring $\mathcal{A}$ is generated by $\{p, q\}$, which satisfies the following equations with $u$ and $v$ any elements of $\mathcal{A}$:

$$\begin{array}{ccc} 0^* = 0 & 1^* = 1 & p^*p = q^*q = 1 \\ (u+v)^* = u^* + v^* & (uv)^* = v^*u^* & p^*q = q^*p = 0 \end{array}$$

We want to augment a $\mathbb{Z}_2[\mathcal{L}]$-algebra structure on the $*$-ring $\mathcal{A}$. This is accomplished by the construction of an *extension of scalars* (also called *change of coefficient ring*), yielding a $\mathbb{Z}_2[\mathcal{L}]$-algebra structure for $\mathcal{A}$ (from that of a $\mathbb{Z}$-algebra).

**Definition 2.10 (The scalar extension $\mathcal{A}_{(\mathbb{Z}_2[\mathcal{L}])}$ in $\mathcal{A}$ to $\mathbb{Z}_2[\mathcal{L}]$ (cf. Ch.2 §5.1 of [2])).** The *extension of scalars* in $\mathcal{A}$ to $\mathbb{Z}_2[\mathcal{L}]$ is the $\mathbb{Z}_2[\mathcal{L}]$-algebra defined by the tensor product over $\mathbb{Z}$:

$$\mathcal{A}_{(\mathbb{Z}_2[\mathcal{L}])} := \mathbb{Z}_2[\mathcal{L}] \otimes_{\mathbb{Z}} \mathcal{A} \tag{1}$$

in which each element is written (not uniquely) by $\sum_i v_i \otimes x_i$ with $v_i \in \mathbb{Z}_2[\mathcal{L}]$ and $x_i \in \mathcal{A}$.

The tensor product (1) becomes a $\mathbb{Z}_2[\mathcal{L}]$-module by the following scalar action, in which $w \in \mathbb{Z}_2[\mathcal{L}]$ and $x \in \mathcal{A}_{(\mathbb{Z}_2[\mathcal{L}])}$:

$$\text{(scalar)} \qquad wx = (w \otimes 1)x$$

To be explicit, $w(v \otimes x) = (wv) \otimes x$ on generating elements.
Moreover, the tensor product becomes a $\mathbb{Z}_2[\mathcal{L}]$-algebra with the multiplication on generating elements:

$$\text{(multiplication)} \quad (v \otimes x)(v' \otimes x') = vv' \otimes xx'$$

Finally the tensor product (1) becomes a $*$-algebra (inheriting $*$ of $\mathcal{A}$ with trivial structure on $\mathbb{Z}_2[\mathcal{L}]$):

$$\text{($*$-involution)} \qquad (\textstyle\sum_i v_i \otimes x_i)^* = \sum_i v_i \otimes (x_i)^*$$

**Definition 2.11 (The reduction of $\mathcal{A}_{(\mathbb{Z}_2[\mathcal{L}])}$ modulo an ideal $\mathfrak{a}$ of the scalars $\mathbb{Z}_2[\mathcal{L}]$).** Let the ring $R := \mathbb{Z}_2[\mathcal{L}]$ and the ($*$-) $R$-algebra $\mathcal{B} := \mathcal{A}_{(R)}$. For an ideal $\mathfrak{a}$ of $R$, the *reduction of $\mathcal{B}$ modulo* $\mathfrak{a}$ is a ($*$-) $R/\mathfrak{a}$-algebra, constructed by

$$R/\mathfrak{a} \otimes_R \mathcal{B} \cong \mathcal{B}/\mathfrak{a}\mathcal{B} \tag{2}$$

in which $\mathfrak{a}\mathcal{B}$ is the sub-algebra generated by elements $\lambda x$ ($\lambda \in \mathfrak{a}$, $x \in \mathcal{B}$). The canonical isomorphism for (2) is for $R/\mathfrak{a}$-module because the ideal $\mathfrak{a}$ annihilates the $R$-module $\mathcal{B}/\mathfrak{a}\mathcal{B}$. (2) is identified more directly with the scalar extension $\mathcal{A}_{(R/\mathfrak{a})}$ in $\mathcal{A}$ as follows (because of associativity and absorbing of $\otimes$): $R/\mathfrak{a} \otimes_R (R \otimes_{\mathbb{Z}} \mathcal{A}) \cong (R/\mathfrak{a} \otimes_R R) \otimes_{\mathbb{Z}} \mathcal{A} \cong R/\mathfrak{a} \otimes_{\mathbb{Z}} \mathcal{A}$

**Remark 2.12.** We write the algebra (2) with $R := \mathbb{Z}_2[\mathcal{L}]$ and $\mathcal{B} := \mathcal{A}_{(R)}$, simply by

$$\mathcal{A}_{(\mathbb{Z}_2[\mathcal{L}])} \mod \mathfrak{a}$$

**Example 2.13.** In $\mathcal{A}_{(\mathbb{Z}_2[\mathcal{L}])}$, $(apq^* + bqp^*)^2 \equiv 0 \mod \langle ab \rangle$ and $(apq^* + bqp^*)^3 \equiv abpq^* + abqp^* \mod \langle a^2 - a \rangle \oplus \langle b^2 - b \rangle$. The latter in particular implies $(pq^* + qp^*)^3 = pq^* + qp^*$.

In Section 3, after interpreting MALL proofs by matrices of elements from $\mathcal{A}_{(\mathbb{Z}_2[\mathcal{L}])}$, the $*$-algebra considered is $\mathcal{A}_{(\mathbb{Z}_2[\mathcal{L}]_{\mathtt{C}_{+\times}})}$, that is $\mathcal{A}_{(\mathbb{Z}_2[\mathcal{L}])}$ mod the ideal generated by both $\mathcal{L}_{\mathtt{C}_+}$ and $\mathcal{L}_{\mathtt{C}_\times}$. On the other hand, in Section 4, the $*$-algebra considered is a *finer grained reduction* of $\mathcal{A}_{(\mathbb{Z}_2[\mathcal{L}])}$ modulo an ideal generated by $\mathcal{L}_{\mathtt{C}_+}$ and by a *sub-ideal* $\mathfrak{a}$ of $\mathcal{L}_{\mathtt{C}_\times}$, which we denote

$$\mathcal{A}_{(\mathbb{Z}_2[\mathcal{L}]_{\mathtt{C}_+})} \mod \mathfrak{a}$$

# 3 MALL Geometry of Interaction in $\mathcal{M}(\mathcal{A}_{(\mathbb{Z}_2[\mathcal{L}])})$ and Termination in $\mathbb{Z}_2[\mathcal{L}]_{\mathtt{C}_{+\times}}$

## 3.1 Interpretation of MALL Proofs as Matrices of $\mathcal{A}_{(\mathbb{Z}_2[\mathcal{L}])}$

**Convention:**
All the matrices considered in this paper are square and indexed by a sequence $\Gamma$ of MALL formulas (i.e., of type $(\Gamma, \Gamma)$). The $(i, j)$ entry, of row $i$ and column $j$, of a matrix $M$ is written by $M_i^j$, where in this paper, $i$ and $j$ are occurrences of formulas. The zero matrix of type $(\Gamma, \Gamma)$ is written by $0_\Gamma$. Every pair $\Gamma$ and $\Delta$ of subsequences respectively for rows and for columns of the index, determines the block matrix, written by $M_\Gamma^\Delta$. $M_-^-$ denotes (an occurrence of) a block matrix inside a bigger matrix so that -'s are automatically determined by the position where the block is put: The upper (resp. lower) - is determined by subindices for columns (resp. rows) of the bigger matrix in which the block is put. For a fixed sequence $\Gamma$, the block matrices $M_-^\Gamma$ and $M_\Gamma^-$ are defined similarly. Given a ring (or more generally, semiring) $R$ and a sequence $\Gamma$ of formulas, $\mathcal{M}_\Gamma(R)$ denotes the set of the square matrices of elements from $R$, indexed with $\Gamma$. When $\Gamma$ is clear from context, the set is denoted by $\mathcal{M}(R)$. $\mathcal{M}_\Gamma(R)$ forms a ring (resp. semiring). dg is short for diag (diagonal of matrices).

**Definition 3.1 ($\mathcal{L}(A)$ for MALL formulas $A$).**
A set $\mathcal{L}(A)$ of literals of eigenweights is associated to MALL formula $A$: $\mathcal{L}(X) = \mathcal{L}(X^\perp) = \emptyset$, $\mathcal{L}(A \otimes B) = \mathcal{L}(A \parr B) = \mathcal{L}(A \oplus B) := \mathcal{L}(A) \uplus \mathcal{L}(B)$, $\mathcal{L}(A \&_c B) := \{c, \bar{c}\} \uplus \mathcal{L}(A) \uplus \mathcal{L}(B)$ for $c$ the associated eigenweight with the &. For a sequence $\Gamma$ of $A_1, \ldots, A_n$, $\mathcal{L}(\Gamma) := \uplus_i \mathcal{L}(A_i)$.

**Definition 3.2 (interpretation $[\![\pi]\!]$).**
Every MALL proof $\pi$ of $\vdash [\Delta], \Gamma$ is interpreted by $[\![\pi]\!]$ where $[\![\pi]\!] \in \mathcal{M}_{\Delta,\Gamma}(\mathcal{A}_{(\mathbb{Z}_2[\mathcal{L}(\Delta,\Gamma)])})$. The matrix $[\![\pi]\!]$ has size $2m + n$ with $2m$ and $n$ the number of formulas in $\Delta$ and $\Gamma$ respectively.

We define $[\![\pi]\!]$ inductively in accordance with the last rule of $\pi$. In the following, the diagonal matrix dg denotes up to exchange of indices.
(axiom) $\pi$ is $ax$:

$$[\![ax]\!] = \begin{array}{c} \\ A \\ A^\perp \end{array} \begin{array}{c} \begin{array}{cc} A & A^\perp \end{array} \\ \left(\begin{array}{cc} 0 & 1 \\ 1 & 0 \end{array}\right) \end{array}.$$ That is, $[\![ax]\!]_A^A = [\![ax]\!]_{A^\perp}^{A^\perp} = 0$ and $[\![ax]\!]_{A^\perp}^A = [\![ax]\!]_A^{A^\perp} = 1$.

(cut-rule)
$[\![\pi]\!] = \mathrm{dg}\left([\![\pi_1]\!], [\![\pi_2]\!]\right)$. I.e., putting the two matrices diagonally then reordering rows and columns correspondingly to the index.

($\parr$-rule)
Contracting (in terms of summing) the rows (resp. the two columns) of $A_1$ and $A_2$ preceded by postcomposing respectively $1 \otimes p$ and $1 \otimes q$ (resp. by precomposing $(1 \otimes p)^* = 1 \otimes p^*$ and $(1 \otimes q)^* = 1 \otimes q^*$). See Remark 3.4 below for the contraction. See Fig 2.

($\otimes$-rule)
$[\![\pi]\!]$ is obtained from $\mathrm{dg}\left([\![\pi_1]\!], [\![\pi_2]\!]\right)$ the same as the $\parr$-rule by contracting rows and columns of $A_1$ and $A_2$ to obtain those of $A_1 \otimes A_2$. See Fig 2.

($\oplus_1$-rule) (same for $\oplus_2$-rule employing instead $q$ and $q^*$):
Let $[\![\pi_1]\!]^\bullet$ denote $\mathrm{dg}\left([\![\pi_1]\!], 0_{A_1}\right)$. Then $[\![\pi]\!]$ is obtained from $[\![\pi_1]\!]^\bullet$ by contracting $A_1$ and $A_2$ to obtain the row and column of $A_1 \oplus A_2$. See Fig 2.
(&-rule)
Let $[\![\pi_i]\!]^\bullet$ denote the matrix $\mathrm{dg}\left([\![\pi_i]\!], 0_{\Delta_{3-i}}\right)$. Then

$$[\![\pi]\!] := S^{\Delta_1,\Delta_2,\Gamma}_{(A_1,A_2,a)}[[\![\pi_1]\!]^\bullet, [\![\pi_2]\!]^\bullet]$$

where $S^{\Delta_1,\Delta_2,\Gamma}_{(A_1,A_2,a)}$ is the superposition introduced below Definition 3.3. See Fig 2.

**Definition 3.3 (superposition $\mathrm{S}^{\Gamma}_{(A_1,A_2,a)}[f,g]$).** Given a fresh (atomic) eigenweight $a \in \mathcal{L}$ and two square matrices $f$ and $g \in \mathcal{M}(\mathcal{A}_{(\mathbb{Z}_2[\mathcal{L}])})$ of the same size with respective indices $\Gamma, A_1$ and $\Gamma, A_2$, the matrix $\mathrm{S}^{\Gamma}_{(A_1,A_2,a)}[f,g]$ of the same size with index $\Gamma, A_1 \&_a A_2$, called *superposition of $f$ and $g$ over $\Gamma$ w.r.t $(A_1, A_2, a)$*, is defined as follows, in which the eigenweight $a$ is associated to the newly introduced & (denoted $\&_a$ explicitly):

$$\mathrm{S}^{\Gamma}_{(A_1,A_2,a)}[f,g] := \begin{array}{c} \\ \Gamma \\ \\ A_1 \&_a A_2 \end{array} \overset{\begin{array}{cc} \Gamma \qquad\qquad & \qquad A_1\&_a A_2 \end{array}}{\begin{pmatrix} af^{\Gamma}_{\Gamma} + \bar{a}g^{\Gamma}_{\Gamma} & af^{A_1}_{\Gamma}(1\otimes p^*) + \bar{a}g^{A_2}_{\Gamma}(1\otimes q^*) \\ \\ a(1\otimes p)f^{\Gamma}_{A_1} + \bar{a}(1\otimes q)g^{\Gamma}_{A_2} & a(1\otimes p)f^{A_1}_{A_1}(1\otimes p^*) + \bar{a}(1\otimes q)g^{A_2}_{A_2}(1\otimes q^*) \end{pmatrix}}$$

That is, the $(\Gamma, \Gamma)$ component is a linear combination of the corresponding components of $f$ and $g$ so that each component is scalar multiplied either by $a$ or by $\bar{a}$ accordingly to their respective origins in $f$ or in $g$. The row and column of $A_1 \& A_2$ are obtained by contracting (in terms of $+$) those of $A_1$ and $A_2$, followed by the scalar multiplication. The contraction of the rows (resp. columns) of $A_1$ and $A_2$ is obtained by respectively postcomposing $1 \otimes p$ and $1 \otimes q$ (resp. precomposing $(1 \otimes p)^* = 1 \otimes p^*$ and $(1 \otimes q)^* = 1 \otimes q^*$) and the contracted elements are multiples either of scalars $a$ or $\bar{a}$ accordingly to their respective origins in $f$ or in $g$.

**Remark 3.4 (The contraction for rows and columns).** In the construction of Def 3.2, the contraction $\mathcal{M}_{n+1}(\mathcal{A}_{(\mathbb{Z}_2[\mathcal{L}])}) \longrightarrow \mathcal{M}_n(\mathcal{A}_{(\mathbb{Z}_2[\mathcal{L}])})$ is used to shrink the size of matrices. This is induced by the retraction structure[1]:
$(1 \otimes p) + (1 \otimes q) : \quad \mathcal{A}_{(\mathbb{Z}_2[\mathcal{L}])} \oplus \mathcal{A}_{(\mathbb{Z}_2[\mathcal{L}])} \lhd \mathcal{A}_{(\mathbb{Z}_2[\mathcal{L}])} \quad : (1 \otimes p^*) \oplus (1 \otimes q^*)$. The retraction inherits from the original Girard's partial isometries [8] and Haghverdi-Scott's categorical reformulation $p + q : \quad \mathcal{A} \oplus \mathcal{A} \lhd \mathcal{A} \quad : p^* \oplus q^*$ of GoI situation [13]. The left morphism $p+q$ internalises the direct sum $(p+q)(x \oplus y) := px + qy$ and the right morphism is its left inverse $(p^* \oplus q^*)(x) := p^*x \oplus q^*x$.

**Remark 3.5.** Intuitively the matrix $[\![\pi]\!]$ describes an I/O (input/output) box à la Haghverdi-Scott [14], which input (resp. output) is into the indices for the column (resp. out of the indices for the row). See Fig 4 in Appendix B.

The following is a property of the superposition of Definition 3.3 used later in Section 3.3.

**Lemma 3.6 (commutativity of superposition and diagonal).** *For an arbitrary matrix $h$ with indices $\Delta$, the following holds in $\mathcal{A}_{(\mathbb{Z}_2[\mathcal{L}]_{\mathtt{C}_+})}$*

$$\mathrm{S}^{\Delta,\Gamma}_{(A_1,A_2,a)}[\mathrm{dg}(h,f), \mathrm{dg}(h,g)] = \mathrm{dg}\left(h, \mathrm{S}^{\Gamma}_{(A_1,A_2,a)}[f,g]\right)$$

[1] $f : C \lhd D : g$ means $g \circ f = \mathrm{Id}_C$

Each of the following denotes the respective $\llbracket \pi \rrbracket$ in Def 3.2:

$\parr$-rule

$$
\begin{array}{c|cc}
 & \Delta\ \Gamma & A_1 \parr A_2 \\
\hline
\Delta\ \Gamma & \llbracket \pi' \rrbracket^{-}_{-} & \llbracket \pi' \rrbracket^{A_1}_{-}(1 \otimes p^*) + \llbracket \pi' \rrbracket^{A_2}_{-}(1 \otimes q^*) \\
A_1 \parr A_2 & \begin{array}{c}(1 \otimes p)\llbracket \pi' \rrbracket^{-}_{A_1} \\ +(1 \otimes q)\llbracket \pi' \rrbracket^{-}_{A_2}\end{array} & \begin{array}{c}(1 \otimes p)\llbracket \pi' \rrbracket^{A_1}_{A_1}(1 \otimes p^*) + (1 \otimes p)\llbracket \pi' \rrbracket^{A_2}_{A_1}(1 \otimes q^*) \\ +(1 \otimes q)\llbracket \pi' \rrbracket^{A_1}_{A_2}(1 \otimes p^*) + (1 \otimes q)\llbracket \pi' \rrbracket^{A_2}_{A_2}(1 \otimes q^*)\end{array}
\end{array}
$$

$\otimes$-rule

$$
\begin{array}{c|ccccc}
 & \Delta_1 & \Delta_2 & \Gamma_1 & \Gamma_2 & A_1 \otimes A_2 \\
\hline
\Delta_1 & \llbracket \pi_1 \rrbracket^{-}_{-} & 0 & \llbracket \pi_1 \rrbracket^{-}_{-} & 0 & \llbracket \pi_1 \rrbracket^{A_1}_{-}(1 \otimes p^*) \\
\Delta_2 & 0 & \llbracket \pi_2 \rrbracket^{-}_{-} & 0 & \llbracket \pi_2 \rrbracket^{-}_{-} & \llbracket \pi_2 \rrbracket^{A_2}_{-}(1 \otimes q^*) \\
\Gamma_1 & \llbracket \pi_1 \rrbracket^{-}_{-} & 0 & \llbracket \pi_1 \rrbracket^{-}_{-} & 0 & \llbracket \pi_1 \rrbracket^{A_1}_{-}(1 \otimes p^*) \\
\Gamma_2 & 0 & \llbracket \pi_2 \rrbracket^{-}_{-} & 0 & \llbracket \pi_2 \rrbracket^{-}_{-} & \llbracket \pi_2 \rrbracket^{A_2}_{-}(1 \otimes q^*) \\
A_1 \otimes A_2 & (1 \otimes p)\llbracket \pi_1 \rrbracket^{-}_{A_1} & (1 \otimes q)\llbracket \pi_2 \rrbracket^{-}_{A_2} & (1 \otimes p)\llbracket \pi_1 \rrbracket^{-}_{A_1} & (1 \otimes q)\llbracket \pi_2 \rrbracket^{-}_{A_2} & \begin{array}{c}(1 \otimes p)\llbracket \pi_1 \rrbracket^{A_1}_{A_1}(1 \otimes p^*) \\ + \\ (1 \otimes q)\llbracket \pi_2 \rrbracket^{A_2}_{A_2}(1 \otimes q^*)\end{array}
\end{array}
$$

$\oplus_1$-rule

$$
\begin{array}{c|cc}
 & \Delta\ \Gamma & A_1 \oplus A_2 \\
\hline
\Delta\ \Gamma & \llbracket \pi_1 \rrbracket^{-}_{-} & \llbracket \pi_1 \rrbracket^{A_1}_{-}(1 \otimes p^*) \\
A_1 \oplus A_2 & (1 \otimes p)\llbracket \pi_1 \rrbracket^{-}_{A_1} & (1 \otimes p)\llbracket \pi_1 \rrbracket^{A_1}_{A_1}(1 \otimes p^*)
\end{array}
$$

&-rule

$$
\begin{array}{c|cccc}
 & \Delta_1 & \Delta_2 & \Gamma, \Sigma & A_1 \& A_2 \\
\hline
\Delta_1 & a\llbracket \pi_1 \rrbracket^{-}_{-} & 0 & a\llbracket \pi_1 \rrbracket^{-}_{-} & a\llbracket \pi_1 \rrbracket^{A_1}_{-}(1 \otimes p^*) \\
\Delta_2 & 0 & \bar{a}\llbracket \pi_2 \rrbracket^{-}_{-} & \bar{a}\llbracket \pi_2 \rrbracket^{-}_{-} & \bar{a}\llbracket \pi_2 \rrbracket^{A_2}_{-}(1 \otimes q^*) \\
\Gamma\ \Sigma & a\llbracket \pi_1 \rrbracket^{-}_{-} & \bar{a}\llbracket \pi_2 \rrbracket^{-}_{-} & a\llbracket \pi_1 \rrbracket^{-}_{-} + \bar{a}\llbracket \pi_2 \rrbracket^{-}_{-} & \begin{array}{c}a\llbracket \pi_1 \rrbracket^{A_1}_{-}(1 \otimes p^*) \\ +\bar{a}\llbracket \pi_2 \rrbracket^{A_2}_{-}(1 \otimes q^*)\end{array} \\
A_1 \& A_2 & a(1 \otimes p)\llbracket \pi_1 \rrbracket^{-}_{A_1} & \bar{a}(1 \otimes q)\llbracket \pi_2 \rrbracket^{-}_{A_2} & \begin{array}{c}a(1 \otimes p)\llbracket \pi_1 \rrbracket^{-}_{A_1} \\ + \\ \bar{a}(1 \otimes q)\llbracket \pi_2 \rrbracket^{-}_{A_2}\end{array} & \begin{array}{c}a(1 \otimes p)\llbracket \pi_1 \rrbracket^{A_1}_{A_1}(1 \otimes p^*) \\ + \\ \bar{a}(1 \otimes q)\llbracket \pi_2 \rrbracket^{A_2}_{A_2}(1 \otimes q^*)\end{array}
\end{array}
$$

Fig. 2. Interpretation $\llbracket \pi \rrbracket$ of Def 3.2

*Proof.* Because the $(\Delta, \Delta)$-component of LHS is $ah + \bar{a}h$, which is equal to $h \mod \langle a + \bar{a} - 1 \rangle$. The other components are automatically identical. □

**Example 3.7 (of Def 3.2 ).** Let $\pi_1$ and $\pi_2$ denote the following respective left and right proofs applying &-rules to the same premises, but choosing different superposition of the cuts. The cut formulas are superposed in $\pi_1$, hence appear contracted in the cut list of the conclusion, while in $\pi_2$, they are not superposed, hence appear separately in the cut list of the conclusion. All the premises are $cut(ax, ax)$ with $X$ and $Y$ the same formulas. The axioms links for the left (resp. right) premises of the &-rule are written upward (resp. downward).

$$\frac{\vdash Y \quad [Y^\perp * X] \quad X^\perp \qquad \vdash Y \quad [Y^\perp * X] \quad X^\perp}{\vdash Y \quad [Y^\perp * X] \quad X^\perp \,\&\, X^\perp}\,\& \qquad\qquad \frac{\vdash Y \quad [Y^\perp * X] \quad X^\perp \qquad \vdash Y \quad [Y^\perp * X] \quad X^\perp}{\vdash Y \quad [Y^\perp * X,\; Y^\perp * X] \quad X^\perp \,\&\, X^\perp}\,\&$$

$[\![ \pi_1 ]\!]$ puts the two matrices of the equation (4) (see Example 3.10 in the subsection below) anti-diagonally (indexed by $X, Y^\perp, X, X^\perp \,\&\, X^\perp$). $[\![ \pi_2 ]\!]$ puts the two matrices of the equation (5) anti-diagonally (indexed by $X, Y^\perp, X, Y^\perp, X, X^\perp \,\&\, X^\perp$).

### 3.2 Quasi-Execution Formula and Execution Formula

This subsection first investigates the quasi-execution formula $\mathsf{qEx}\,(\sigma_\Delta, [\![ \pi ]\!])$ for the interpretation $[\![ \pi ]\!]$ with a partial symmetry $\sigma_\Delta$. The partial symmetry interprets the permutations among pair-wise dual cut formulas. This $\mathsf{qEx}\,(\sigma_\Delta, [\![ \pi ]\!])$ is called *quasi*, because there remain the surplus scalars $\mathcal{L}(\Delta)$ from the cut-formulas $\Delta$, which formulas, on the syntactic side, disappear during the cut-elimination; i.e., the elements of $\mathsf{qEx}\,(\sigma_\Delta, [\![ \pi ]\!])$ are not in general from $\mathcal{A}_{(\mathbb{Z}_2[\mathcal{L}(\Gamma)])}$, but from $\mathcal{A}_{(\mathbb{Z}_2[\mathcal{L}(\Delta,\Gamma)])}$. Second, in order to eliminate the surplus scalars, the legitimate execution $\mathsf{Ex}\,(d, \sigma_\Delta, [\![ \pi ]\!])$ is formulated via a ring homomorphism $d$. The surplus scalars from the cuts become definable by the scalars from the conclusion; i.e., $d$ maps each entries of $\mathsf{qEx}\,(\sigma_\Delta, [\![ \pi ]\!])$ to $\mathcal{A}_{(\mathbb{Z}_2[\mathcal{L}(\Gamma)])}$

**Definition 3.8 (quasi-execution formula** $\mathsf{qEx}\,(\sigma_\Delta, U)$ **).** For a matrix $U \in \mathcal{M}_{\Delta,\Gamma}\left(\mathcal{A}_{(\mathbb{Z}_2[\mathcal{L}])}\right)$ with a sequence $\Delta$ of pair-wise dual formulas, the *quasi-execution formula* is defined:

$$\mathsf{qEx}\,(\sigma_\Delta, U) := U_\Gamma^\Gamma + \sum_{n \geq 0} U_\Gamma^\Delta (\sigma_\Delta U_\Delta^\Delta)^n (\sigma_\Delta U_\Delta^\Gamma) \tag{3}$$

where $\sigma_\Delta$ is a partial symmetry of pair-wise dual formulas $\Delta$:

$$\sigma_\Delta = \mathrm{dg}(\cdots, \begin{matrix} \\ C \\ C^\perp \end{matrix} \begin{matrix} C & C^\perp \\ \left(\begin{matrix} 0 \\ 1 \end{matrix}\right. & \left.\begin{matrix} 1 \\ 0 \end{matrix}\right) \end{matrix}, \cdots)$$

so that a pair $C * C^\perp$, for dg, ranges over $\Delta$, hence $\sigma_\Delta$ is indexed by $\Delta$.

If the sum (3) exists algebraically so that there exists $n_0$ and for all $m \geq n_0$, the $m$-th factor equals 0, then $\mathsf{qEx}\,(\sigma_\Delta, U)$ belongs to $\mathcal{M}_\Gamma(\mathcal{A}_{(\mathbb{Z}_2[\mathcal{L}])})$, more precisely to $\mathcal{M}_\Gamma(\mathcal{A}_{\mathbb{Z}_2[\mathcal{L}(\Delta,\Gamma)]})$.

*Why quasi ?* The $\mathsf{qEx}\,(\sigma_\Delta, [\![ \pi ]\!])$ is a matrix of $\mathcal{M}_\Gamma(\mathcal{A}_{(\mathbb{Z}_2[\mathcal{L}])})$ indexed by the conclusion $\Gamma$ (the cut-formulas list $\Delta$ of $\pi$ disappeared). Although $\Delta$ is erased from the index, the scalars for the matrix entries in general retain the occurrences of literals $\mathcal{L}(\Delta)$ of eigenweights from the cut formulas: I.e., the matrix entries range over $\mathcal{A}_{(\mathbb{Z}_2[\mathcal{L}(\Delta,\Gamma)])}$ (not over $\mathcal{A}_{(\mathbb{Z}_2[\mathcal{L}(\Gamma)])}$). This is where the terminology *quasi* comes from in the sense that the surplus scalars from $\mathcal{L}(\Delta)$ are not erased.

**Remark 3.9.** Under the intuitive convention of operator $U$ as I/O box, $\mathsf{qEx}\,(\sigma_\Delta, U)$ stipulates calculating I/O involved in the feedback $\sigma_\Delta$ on $\Delta$, see Fig 5 in Appendix B. E.g., when $U$ is an interpretation of a proof $\vdash [A{*}A^\perp],\ \Gamma, \Delta$ derived from $\vdash A, \Gamma$ and $\vdash A^\perp, \Delta$ by cut, then the I/O box results from the communication $\sigma_{A,A^\perp}$ between the matched pair $A * A^\perp$ of the cut. The communication is plugging the output of $A$ and of $A^\perp$ respectively into the input of $A^\perp$ and of $A$, hence reciprocally annihilates $A$ and $A^\perp$ from I/O. See Figure 6 in Appendix B.

**Example 3.10 (of Def 3.8 ).** For $\pi_1$ and $\pi_2$ of Example 3.7, the following (4) and (5) depict how to calculate $\mathsf{qEx}\left(\sigma_{X,Y^\perp}, [\![\, \pi_1 \,]\!]\right)$ and $\mathsf{qEx}\left(\sigma_{X,Y^\perp}, X, Y^\perp, [\![\, \pi_2 \,]\!]\right)$ respectively in terms of I/O involved in the feedback. Note that both quasi-execution formulas become indexed by $X, X^\perp \,\&\, X^\perp$:

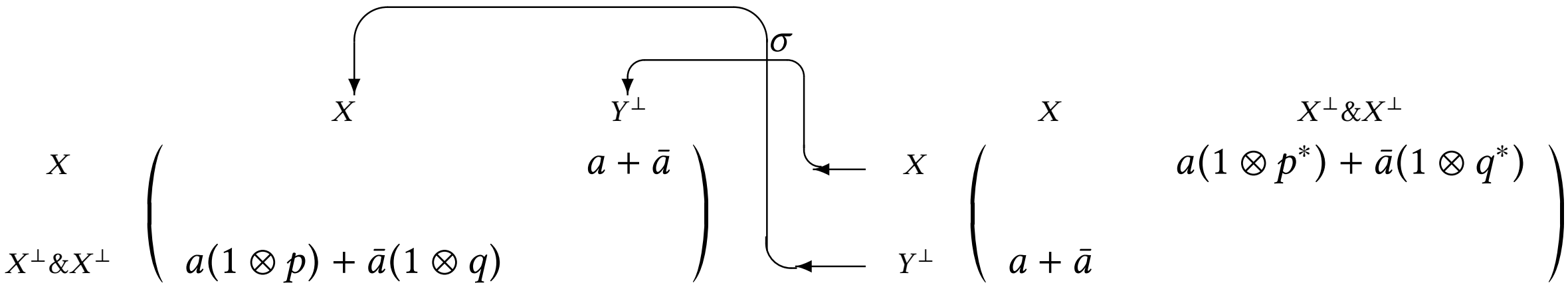

(4)

The element $\left(\mathsf{qEx}\left(\sigma_{X,Y^\perp}, [\![\, \pi_1 \,]\!]\right)\right)^{X}_{X^\perp \& X^\perp}$ (of the column $X$ and of the row $X^\perp \,\&\, X^\perp$) is obtained by chasing the paths in (4) whose input are $X$ and whose output are $X^\perp \,\&\, X^\perp$, in accordance with the matrix calculation: I.e., (4) has only one such path; entering the right $X$ vertically down to 1, exiting $Y^\perp$ horizontally, then via $\sigma$ entering the left $X$ vertically down to $a(1 \otimes p) + \bar{a}(1 \otimes q)$, exiting $X^\perp \,\&\, X^\perp$ horizontally. Multiplying the matrix elements on this path yields the sought element $a(1 \otimes p) + \bar{a}(1 \otimes q)$.

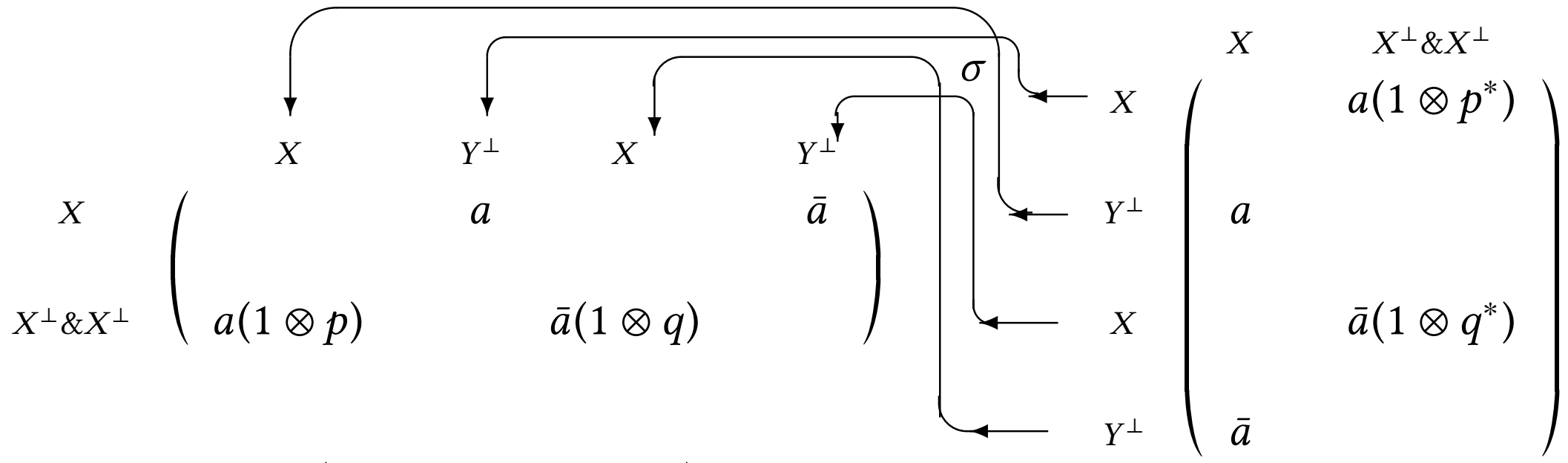

(5)

The element $\left(\mathsf{qEx}\left(\sigma_{X,Y^\perp}, X, Y^\perp, [\![\, \pi_2 \,]\!]\right)\right)^{X}_{X^\perp \& X^\perp}$ is obtained similarly by chasing the paths in (5), which on the other hand has two such paths; Entering the right $X$ vertically down to $a$ (resp. to $\bar{a}$), exiting $Y^\perp$ horizontally, then via $\sigma$ entering the left $X$ vertically down to $a(1 \otimes p)$ (resp. to $\bar{a}(1 \otimes p)$), exiting $X^\perp \,\&\, X^\perp$ horizontally. Summing the multiplications of the matrix elements on the respective paths yields the sought element $a.a(1 \otimes p) + \bar{a}.\bar{a}(1 \otimes q)$.

Similarly for the other elements, it is checked that both quasi-execution formulas equalise to the matrix (8) of Example 3.19 below.

Transitivity of cut is performed in quasi-execution formula in the same way as [8, 13]:

**Proposition 3.11 (associativity of quasi-execution formula).** *For arbitrary $U$, indexed by $\Delta_1, \Delta_2, \Gamma$, and partial symmetries $\sigma_i := \sigma_{\Delta_i}$ with a list $\Delta_i$ of pair-wise dual formulas, if* $\mathsf{qEx}\,(\sigma_1, U)$ *exists and belongs to $\mathcal{M}_{\Delta_2,\Gamma}(\mathcal{A}_{(\mathbb{Z}_2[\mathcal{L}])})$, then*

$$\mathsf{qEx}\,(\sigma_2 \otimes \sigma_1, U) = \mathsf{qEx}\,(\sigma_2, \mathsf{qEx}\,(\sigma_1, U))$$

*LHS exists and belongs to* $\mathcal{M}_\Gamma(\mathcal{A}_{(\mathbb{Z}_2[\mathcal{L}])})$ *iff RHS exists and belongs to* $\mathcal{M}_\Gamma(\mathcal{A}_{(\mathbb{Z}_2[\mathcal{L}])})$.

Proof. Direct algebraic calculation analogously to Girard's Lemma 5 of [8], called "associativity of cut". □

For defining execution formulas, we recall the following universal properties of polynomial rings. A general reference for commutative ring theory is [22].

**Lemma 3.12.** *1) The natural inclusion* $R \longrightarrow R[X]$*, which just sends an element* $r \in R$ *to the constant polynomial, is a ring homomorphism.*

*2) The polynomial ring* $R[X_1, \ldots, X_{n+1}]$ *with* $n+1$ *indeterminates is constructed inductively by* $(R[X_1, \ldots, X_n])[X_{n+1}]$*. The ring is isomorphic, for any i, to* $(R[X_1, \ldots, X_i])[X_{i+1}, \ldots, X_{n+1}]$*, which is abbreviated by* $R[X_1, \ldots, X_i][X_{i+1}, \ldots, X_{n+1}]$*. Indeterminates are considered up to permutations. In this paper,* $\mathbb{Z}_2[\mathcal{L}(\Gamma_1, \Gamma_2)]$ *is identified with* $\mathbb{Z}_2[\mathcal{L}(\Gamma_1)][\mathcal{L}(\Gamma_2)]$ *and* $\mathbb{Z}_2[\mathcal{L}(\Gamma_2)][\mathcal{L}(\Gamma_1)]$*.*

*3) For any ring homomorphism* $\phi : R \longrightarrow S$ *and any element* $s \in S$*, there exists a unique homomorphism* $\phi[s/X]$ *mapping X to the s, which makes the right diagram commute. Iteratively* $\phi[s_1/X_1, \ldots, s_n/X_n]$ *is defined.*

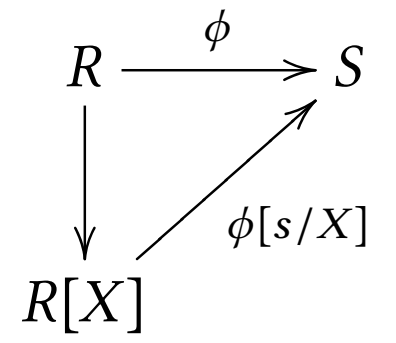

*4) For any ring homomorphisms* $\phi_1$ *and* $\phi_2$ *to S respectively from* $R[X_1]$ *and* $R[X_2]$ *such that* $\phi_1$ *and* $\phi_2$ *meet on R, there is a unique homomorphism* $\phi_1 \uplus \phi_2 : R[X_1, X_2] \longrightarrow S$ *which makes the right diagram commute:*

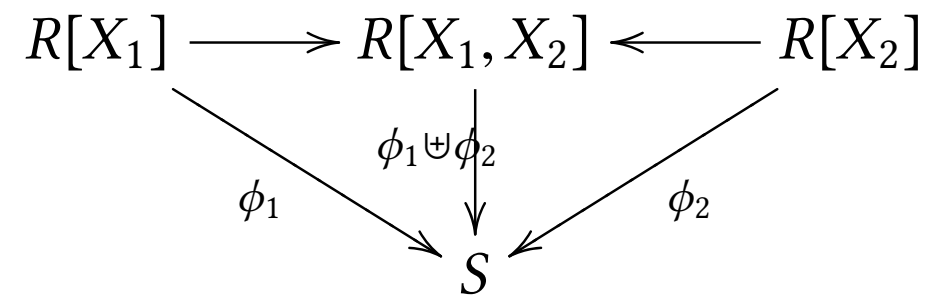

*In this paper for* $d : \mathbb{Z}_2[\mathcal{L}(\Delta)] \longrightarrow \mathbb{Z}_2[\mathcal{L}(\Gamma)]$*,* $d^\Gamma$ *is short for* $d[\mathcal{L}(\Gamma)/\mathcal{L}(\Gamma)] : \mathbb{Z}_2[\mathcal{L}(\Delta, \Gamma)] \longrightarrow \mathbb{Z}_2[\mathcal{L}(\Gamma)]$*: That is,* $d^\Gamma$ *restricted on* $\mathbb{Z}_2[\mathcal{L}(\Delta)]$ *(resp. on* $\mathbb{Z}_2[\mathcal{L}(\Gamma)]$*) is d (resp. the identity).*

Proof. Proof of (4): The unique homomorphisms coincide by universality of $\phi_1[\phi_2(X_2)/X_2]$ guaranteed by (3) for the left triangle together with $\phi_2(X_2)$ and of $\phi_2[\phi_1(X_1)/X_1]$ for the right triangle together with $\phi_1(X_1)$, which is $\phi_1 \uplus \phi_2$. □

**Applying ring homomorphism to matrix:** Every ring homomorphism $d : \mathbb{Z}_2[\mathcal{L}(\Theta)] \longrightarrow \mathbb{Z}_2[\mathcal{L}(\Xi)]$ induces the $*$-ring homomorphism $\mathcal{A}_{(d)} := d \otimes \mathrm{Id}_{\mathcal{A}} : \mathcal{A}_{(\mathbb{Z}_2[\mathcal{L}(\Theta)])} \longrightarrow \mathcal{A}_{(\mathbb{Z}_2[\mathcal{L}(\Xi)])}$. For a matrix $U \in \mathcal{M}_\Theta(\mathcal{A}_{\mathbb{Z}_2[\mathcal{L}(\Theta)]})$, $d(U) \in \mathcal{M}_\Xi(\mathcal{A}_{\mathbb{Z}_2[\mathcal{L}(\Xi)]})$ denotes the matrix which is the image of $U$ by the element-wise induced map $\mathcal{M}_\Theta(\mathcal{A}_{\mathbb{Z}_2[\mathcal{L}(\Theta)]}) \longrightarrow \mathcal{M}_\Xi(\mathcal{A}_{\mathbb{Z}_2[\mathcal{L}(\Xi)]})$.

**Definition 3.13 (Execution Formula** $\mathsf{Ex}\,(d_\Delta, \sigma_\Delta, U)$**).** Using the same $\sigma_\Delta$ and $U$ of Definition 3.8, but together with a ring homomorphism $d_\Delta : \mathbb{Z}_2[\mathcal{L}(\Delta)] \longrightarrow \mathbb{Z}_2[\mathcal{L}(\Gamma)]$,

$$\mathsf{Ex}\,(d_\Delta, \sigma_\Delta, U) = d_\Delta^\Gamma\,(\mathsf{qEx}\,(\sigma_\Delta, U))$$

Transitivity of cut is now represented, on top of Proposition 3.11, in terms of reciprocal $\mathfrak{Res}$ and $\mathfrak{Ind}$ for ring homomorphisms:

**Proposition 3.14 (associativity of execution formula).** $\sigma_{\Delta_i}$ *and U are the same as Proposition 3.11, and given a ring homomorphism* $d_{\Delta_1} : \mathbb{Z}_2[\mathcal{L}(\Delta_1)] \longrightarrow \mathbb{Z}_2[\mathcal{L}(\Gamma)]$*, the following holds:*
*(i) For a ring homomorphism* $d_{\Delta_2} : \mathbb{Z}_2[\mathcal{L}(\Delta_2)] \longrightarrow \mathbb{Z}_2[\mathcal{L}(\Gamma)]$

$$\mathsf{Ex}\,(d_{\Delta_1} \uplus d_{\Delta_2}, \sigma_{\Delta_1} \otimes \sigma_{\Delta_2}, U) = \mathsf{Ex}\,(d_{\Delta_1}, \sigma_{\Delta_1}, \mathsf{Ex}\,(\mathfrak{Res}(d_{\Delta_2}), \sigma_{\Delta_2}, U))$$

*(ii) For a ring homomorphism* $d_{\Delta_2} : \mathbb{Z}_2[\mathcal{L}(\Delta_2)] \longrightarrow \mathbb{Z}_2[\mathcal{L}(\Delta_1, \Gamma)]$*,*

$$\mathsf{Ex}\,(d_{\Delta_1} \uplus \mathfrak{Ind}(d_{\Delta_2}), \sigma_{\Delta_1} \otimes \sigma_{\Delta_2}, U) = \mathsf{Ex}\,(d_{\Delta_1}, \sigma_{\Delta_1}, \mathsf{Ex}\,(d_{\Delta_2}, \sigma_{\Delta_2}, U))$$

*where* $\mathfrak{Res}(-)$ *and* $\mathfrak{Ind}(-)$ *are postcomposition respectively by the inclusion* $R[Y] \longrightarrow (R[Y])[X]$ *and by* $d_1[Y/Y] : (R[X])[Y] \longrightarrow R[Y]$ *with* $R = \mathbb{Z}_2, X = \mathcal{L}(\Delta_1)$ *and* $Y = \mathcal{L}(\Delta_2)$.

### 3.3 Slice-wise Nilpotency of qEx and Invariance of Ex

This subsection proves invariance of the quasi-execution formula during cut-elimination up to ring homomorphisms erasing scalars from cut-formulas. We start with nilpotency of the quasi-execution formula, which property in this section is obtained in the coarse reduction $\mathcal{A}_{(\mathbb{Z}_2[\mathcal{L}]_{\mathtt{C}_{+\times}})}$.

**Proposition 3.15 (slice-wise nilpotency of $[\![\pi]\!]$).** For every MALL proof $\pi$ of $\vdash [\Delta],\ \Gamma$, the product $\sigma_\Delta [\![\pi]\!]$ is *nilpotent* in $\mathbb{Z}_2[\mathcal{L}]_{\mathtt{C}_{+\times}}$. That is, the term $(\sigma_\Delta [\![\pi]\!]^\Delta_\Delta)^n$ of (3) in $\mathsf{qEx}\,(\sigma_\Delta, [\![\pi]\!])$ becomes $0_\Delta$ for some $n$.

Proof. Simultaneously done in proving Theorem 3.17. Also a direct corollary of Theorem 4.13 in the next section. □

**Remark 3.16 (slice-wise nilpotency is (P2)).** Proposition 3.15 is slice-wise nilpotency because $\mathbb{Z}_2[\mathcal{L}]_{\mathtt{C}_{+\times}}$ imposes the condition $a.\bar{a} = 0$ uniformly for all $a \in \mathcal{L}$. This uniform imposition makes our GoI interpretation reduce to the slice-wise collection of the original Girard's GoI [8] for MLL. Thus Proposition 3.15 is our GoI counterpart of (P2) for the slice-wise DR. This subsumes the by now well known correspondence (implicit or explicit) in multiplicative GoI between the termination (in particular, nilpotency) of the execution formula and the DR acyclicity [4] of the proof-nets. Such a correspondence is directly observable in our construction $\mathsf{qEx}\,(\sigma, [\![\pi]\!])$ as follows: Chasing the paths on the I/O box of $[\![\pi]\!]$, in accordance with the matrix calculation of $\mathsf{qEx}\,(\sigma, [\![\pi]\!])$, amounts to tracing the subformula-trees with linkings on the sliced proof-net for $\pi$.

(**scalars**:) In what follows in this subsection, quasi-execution formulas are run in the scalars $\mathbb{Z}_2[\mathcal{L}]_{\mathtt{C}_{+\times}}$. That is $\mathcal{A}_{(\mathbb{Z}_2[\mathcal{L}])}$ mod the ideal generated by $\mathcal{L}_{\mathtt{C}_+}$ and $\mathcal{L}_{\mathtt{C}_\times}$.

**Theorem 3.17 (invariance of $\mathsf{qEx}\,(\sigma, [\![\pi]\!])$ under external ring homomorphism $d$ for scalars).** *For every* MALL *proof* $\pi$ *of* $\vdash [\Delta], \Gamma$,

- $\mathsf{qEx}\,(\sigma_\Delta, [\![\pi]\!])$ *belongs to* $\mathcal{M}_\Gamma(\mathcal{A}_{\mathbb{Z}_2[\mathcal{L}(\Delta,\Gamma)]})$.

*Suppose moreover* $\pi$ *reduces to* $\pi'$ *of* $\vdash [\Delta'], \Gamma$ *by any sequence of cut-eliminations. Then:*

(i) *There exists a ring homomorphism* $d : \mathbb{Z}_2[\mathcal{L}(\Delta)] \longrightarrow \mathbb{Z}_2[\mathcal{L}(\Delta', \Gamma)]$ *so that*

$$d^\Gamma\,(\mathsf{qEx}\,(\sigma_\Delta, [\![\pi]\!])) = \mathsf{qEx}\left(\sigma_{\Delta'}, [\![\pi']\!]\right)$$

(ii) *In particular, if* $\pi'$ *is cut free, hence* $\Delta'$ *being empty,*

$$\mathsf{Ex}\,(d_\Delta, \sigma_\Delta, [\![\pi]\!]) = [\![\pi']\!]$$

*for some ring homomorphism* $d_\Delta : \mathbb{Z}_2[\mathcal{L}(\Delta)] \longrightarrow \mathbb{Z}_2[\mathcal{L}(\Gamma)]$.

(iii) *There exist ring homomorphisms* $d_\Delta$ *and* $d_{\Delta'}$ *respectively from* $\mathbb{Z}_2[\mathcal{L}(\Delta)]$ *and from* $\mathbb{Z}_2[\mathcal{L}(\Delta')]$ *both to* $\mathbb{Z}_2[\mathcal{L}(\Gamma)]$ *so that*

$$\mathsf{Ex}\,(d_\Delta, \sigma_\Delta, [\![\pi]\!]) = \mathsf{Ex}\left(d_{\Delta'}, \sigma_{\Delta'}, [\![\pi']\!]\right)$$

**Note:** In (i), each literal $\ell \in \mathcal{L}(\Delta)$, in the cut-list $\Delta$ of $\pi$ is mapped to $d(\ell) \in \mathbb{Z}_2[\mathcal{L}(\Delta', \Gamma)]$, hence the homomorphism $d$ makes the literals $\ell$ of $\Delta$ disappear during the cut-elimination, definable (as $d(\ell)$) in terms of the literals in $\Delta', \Gamma$ of the reduced proof $\pi'$. In particular in (ii), for a cut-free $\pi'$, the hom $d$ makes each literal in the cut-list definable from those of the conclusion. Same for $d$ and $d'$ in (iii). Note that values of literals $a$ and $\bar{a}$ by $d$ are *unrelated* in general since in the polynomial ring $\mathbb{Z}_2[\mathcal{L}]$, the two are independent indeterminates over $\mathbb{Z}_2$.

Proof. (iii) is directly implied from (i) and (ii): Take the homomorphism $d$ in (i) and the homomorphism $d_{\Delta'}$ with a cut-free $\pi''$ in (ii), then take $d_\Delta := (d_{\Delta'})^\Gamma \circ d$ (cf. Lemma 3.12 (3) for $(-)^\Gamma$). Then $(d_\Delta)^\Gamma := ((d_{\Delta'})^\Gamma \circ d)^\Gamma = (d_{\Delta'})^\Gamma \circ d^\Gamma$, hence the equality for (iii) is derived to denote $\llbracket \pi'' \rrbracket$.
Since (ii) is a direct consequence of (i), we shall prove (i) when the reduced proof $\pi'$ (called $rpf$) is a one step reduction of the given proof $\pi$ (called $gpf$).
(Crucial Case 1)

$$\dfrac{\dfrac{}{\vdash B, B^\perp} \quad \dfrac{\vdots\ \pi'}{\vdash [\Delta], A, \Gamma}}{\vdash [\Delta, B^\perp * A], B, \Gamma}\ cut$$

(with $A$ and $B$ are the different occurrences of the same formula) is reduced to

$$\begin{array}{c}\vdots\ \pi' \\ \vdash [\Delta], B, \Gamma\end{array}$$

(with the identification of the occurrence of $A$ to $B$)

$$d : \mathbb{Z}_2[\mathcal{L}(\Delta, B^\perp, A)] \longrightarrow \mathbb{Z}_2[\mathcal{L}(\Delta, B, \Gamma)]$$

is defined so that $d$ on $\mathcal{L}(\Delta)$, on $\mathcal{L}(A)$, and on $\mathcal{L}(B^\perp)$ are respectively id, the isomorphism to $\mathcal{L}(B)$, and zero.
Since *gpf* is interpreted by dg $\left(\llbracket \pi' \rrbracket, \llbracket ax \rrbracket\right)$, in which $\llbracket ax \rrbracket$ is the anti-diagonal of non-zero components $1^B_{B^\perp}$ and $1^{B^\perp}_B$, we have ($n = 0$ in the expansion of (3))

$$\mathsf{qEx}\left(\sigma_{A,B^\perp}, \llbracket gpf \rrbracket\right) = \begin{array}{c} \\ B \\ \Delta\,\Gamma \end{array} \begin{array}{c} \begin{array}{cc} B & \Delta\,\Gamma \end{array} \\ \begin{pmatrix} 0 & 0 \\ 0 & \llbracket \pi' \rrbracket^{\Delta\Gamma}_{\Delta\Gamma} \end{pmatrix} \end{array} + \begin{array}{c} \\ B \\ \Delta\,\Gamma \end{array} \begin{array}{c} \begin{array}{cc} B & \Delta\,\Gamma \end{array} \\ \begin{pmatrix} \llbracket \pi' \rrbracket^A_A\, \sigma_{B^\perp,A}\, 1^B_{B^\perp} & 1^{B^\perp}_B\, \sigma_{B^\perp,A}\, \llbracket \pi' \rrbracket^{\Delta\Gamma}_A \\ \llbracket \pi' \rrbracket^A_{\Delta\Gamma}\, \sigma_{B^\perp,A}\, 1^B_{B^\perp} & 0 \end{pmatrix} \end{array}$$

This is $\llbracket \pi' \rrbracket$ with the canonical replacement of the indeterminates $\mathcal{L}(A)$ with those $\mathcal{L}(B)$ since $\mathcal{L}(A)$ and $\mathcal{L}(B)$ are isomorphic sets. The replacement, corresponding to $\alpha$-conversion, is realized algebraically by the homomorphism $d$ defined above.
(Crucial Case 2)

$$\dfrac{\dfrac{\left\{\begin{array}{c}\vdots\ \pi_i \\ \vdash [\Omega, \Delta_i], \Gamma, A_i\end{array}\right\}^{i=1,2}}{\vdash [\Omega, \Delta_1, \Delta_2], \Gamma, A_1 \,\&\, A_2}\ \& \qquad \dfrac{\begin{array}{c}\vdots\ \pi_3 \\ \vdash [\Delta_3], \Xi, A_1^\perp\end{array}}{\vdash [\Delta_3], \Xi, A_2^\perp \oplus A_1^\perp}\ \oplus_2}{\vdash [\Omega, \Delta_1, \Delta_2, \Delta_3, (A_1 \,\&\, A_2) * (A_2^\perp \oplus A_1^\perp)], \Gamma, \Xi}\ cut$$

is reduced to

$$\dfrac{\begin{array}{c}\vdots\ \pi_1 \\ \vdash [\Omega, \Delta_1], \Gamma, A_1\end{array} \quad \begin{array}{c}\vdots\ \pi_3 \\ \vdash [\Delta_3], \Xi, A_1^\perp\end{array}}{\vdash [\Omega, \Delta_1, \Delta_3, A_1 * A_1^\perp], \Gamma, \Xi}\ cut$$

In the following $A = A_1$ and $B = A_2$. $a$ is the eigenweight of the & introduced at the left premise of $gpf$.
$d : \mathbb{Z}_2[\mathcal{L}(\Omega, \Delta_1, \Delta_2, \Delta_3, A \,\&\, B, B^\perp \oplus A^\perp)] \longrightarrow \mathbb{Z}_2[\mathcal{L}(\Omega, \Delta_1, \Delta_3, A, A^\perp, \Gamma, \Xi)]$
is defined, whose restriction to $\mathcal{L}(\Omega, \Delta_1, \Delta_3, A, A^\perp)$ is the identity, $d(a) := 1$, $d(\bar{a}) := 0$, and $d(\mathcal{L}(\Delta_2)) := d(\mathcal{L}(B)) := d(\mathcal{L}(B^\perp)) := 0$.

We show in $\mathcal{A}_{(\mathbb{Z}_2[\mathcal{L}]_{\mathsf{C}_+})}$ mod $\langle a\bar{a} \rangle$;

$$d\left(\mathsf{qEx}\left(\sigma_{A\&B, B^\perp \oplus A^\perp}, \llbracket gpf \rrbracket\right)\right) \hookleftarrow^0 \mathsf{qEx}\left(\sigma_{A,A^\perp}, \llbracket rpf \rrbracket\right)$$

where $M \hookleftarrow^0 N$ means $M$ is dg $(N, 0)$ for a 0 matrix (of some size) modulo permutations of indices.

This assertion is how syntactical (*erasing*) is performed in a quasi-execution formula.

Let $\overline{\pi}_3$ denote the right premise of $gpf$ (i.e., $\pi_3$ applied by the $\oplus_2$).

$$[\![ gpf ]\!] = \mathrm{dg}\left(\mathsf{S}^{\Omega,\Delta_1,\Delta_2,\Gamma}_{(A,B,a)}\left[[\![\pi_1]\!]^\bullet, [\![\pi_2]\!]^\bullet\right], [\![\overline{\pi}_3]\!]\right) \overset{Lem3.6}{=} \mathsf{S}^{\Omega,\Delta_1,\Delta_2,\Gamma,\Delta_3,\Xi}_{(A,B,a)}\left[\mathrm{dg}\left([\![\pi_1]\!]^\bullet, [\![\overline{\pi}_3]\!]\right), \mathrm{dg}\left([\![\pi_2]\!]^\bullet, [\![\overline{\pi}_3]\!]\right)\right]$$

Since $[\![ gpf ]\!]^{A\&B, B^\perp\oplus A^\perp}_{A\&B, B^\perp\oplus A^\perp} = \mathrm{dg}\left([\![ gpf ]\!]^{A\&B}_{A\&B}, [\![ gpf ]\!]^{B^\perp\oplus A^\perp}_{B^\perp\oplus A^\perp}\right)$ with $[\![ gpf ]\!]^{A\&B}_{A\&B} = a(1\otimes p)[\![\pi_1]\!]^A_A(1\otimes p^*) + \bar{a}(1\otimes q)[\![\pi_2]\!]^B_B(1\otimes q^*)$ while $[\![ gpf ]\!]^{B^\perp\oplus A^\perp}_{B^\perp\oplus A^\perp} = (1\otimes p)[\![\pi_3]\!]^{A^\perp}_{A^\perp}(1\otimes p^*)$, it holds that

$$[\![ gpf ]\!]^{A\&B, B^\perp\oplus A^\perp}_{A\&B, B^\perp\oplus A^\perp}\sigma[\![ gpf ]\!]^{A\&B, B^\perp\oplus A^\perp}_{A\&B, B^\perp\oplus A^\perp} = \mathrm{dg}\left([\![ gpf ]\!]^{B^\perp\oplus A^\perp}_{B^\perp\oplus A^\perp}\sigma[\![ gpf ]\!]^{A\&B}_{A\&B}, [\![ gpf ]\!]^{A\&B}_{A\&B}\sigma[\![ gpf ]\!]^{B^\perp\oplus A^\perp}_{B^\perp\oplus A^\perp}\right)$$
$$= a\,\mathrm{dg}\left((1\otimes p)[\![\pi_3]\!]^{A^\perp}_{A^\perp}\sigma[\![\pi_1]\!]^A_A(1\otimes p^*), (1\otimes p)[\![\pi_1]\!]^A_A\sigma[\![\pi_3]\!]^{A^\perp}_{A^\perp}(1\otimes p^*)\right) \quad \text{by } p^*q = q^*p = 0.$$

Since any component $[\![ gpf ]\!]^{A\&B}_{-}$ (resp. $[\![ gpf ]\!]^{-}_{A\&B}$) is either a precomposition (resp. composition) by $1\otimes x^*$ (resp. by $1\otimes x$) with $x \in \{p, q\}$ or their sum, the following holds in $\mathbb{Z}_2[\mathcal{L}]$ for any $n \geq 0$

$$[\![ gpf ]\!]^{A\&B, B^\perp\oplus A^\perp}_\star\left(\sigma[\![ gpf ]\!]^{A\&B, B^\perp\oplus A^\perp}_{A\&B, B^\perp\oplus A^\perp}\right)^n\left(\sigma[\![ gpf ]\!]^{-}_{A\&B, B^\perp\oplus A^\perp}\right)$$
$$= a^n[\![ rpf ]\!]^{A,A^\perp}_\star\left(\sigma[\![ rpf ]\!]^{A,A^\perp}_{A,A^\perp}\right)^n\left(\sigma[\![ rpf ]\!]^{-}_{A,A^\perp}\right)$$

This means

$$\mathsf{qEx}\left(\sigma_{A\&B, B^\perp\oplus A^\perp}, [\![ gpf ]\!]\right) = a\,\mathsf{qEx}\left(\sigma_{A,A^\perp}, \mathrm{dg}\left([\![\pi_1]\!]^\bullet, [\![\pi_3]\!]\right)\right) + \bar{a}\,\mathsf{qEx}\left(\sigma_{B,B^\perp}, \mathrm{dg}\left([\![\pi_2]\!]^\bullet, 0_{\Delta_3,\Xi,B^\perp}\right)\right)$$
$$= a\,\mathsf{qEx}\left(\sigma_{A,A^\perp}, \mathrm{dg}\left([\![\pi_1]\!]^\bullet, [\![\pi_3]\!]\right)\right) + \bar{a}\,\mathrm{dg}\left([\![\pi_2]\!]^{\Omega,\Delta_2,\Gamma}_{\Omega,\Delta_2,\Gamma}, 0_{\Delta_3,\Xi}\right) \tag{6}$$

Applying $d$ to (6) is the assertion since the second term of (6) disappears by $d(\bar{a}) = 0$, and (the first term of (6) instantiated by $a := 1$) $\hookleftarrow^0$ $\mathsf{qEx}\left(\sigma_{A,A^\perp}, \mathrm{dg}\left([\![\pi_1]\!], [\![\pi_3]\!]\right)\right) = \mathsf{qEx}\left(\sigma_{A,A^\perp}, [\![ rpf ]\!]\right)$.

(Crucial Case 3.1)

$$\dfrac{\dfrac{}{\vdash [\Xi], \Delta, A}\,\vdots\,\rho \qquad \dfrac{\left\{\begin{array}{c}\vdots\,\pi_i\\ \vdash [\Omega, \Sigma_i], A^\perp, \Gamma, B_i\end{array}\right\}^{i=1,2}}{\vdash [\Omega, \Sigma_1, \Sigma_2], A^\perp, \Gamma, B_1 \,\&\, B_2}\,\&}{\vdash [\Xi, \Omega, \Sigma_1, \Sigma_2, A * A^\perp], \Delta, \Gamma, B_1 \,\&\, B_2}\,cut$$

is reduced to

$$\dfrac{\left\{\dfrac{\begin{array}{c}\vdots\,\rho\\ \vdash [\Xi], \Delta, A\end{array} \quad \begin{array}{c}\vdots\,\pi_i\\ \vdash [\Omega, \Sigma_i], A^\perp, \Gamma, B_i\end{array}}{\vdash [\Xi, \Omega, \Sigma_i, A * A^\perp], \Delta, \Gamma, B_i}\,cut\right\}^{i=1,2}}{\vdash [\Xi, \Omega, A * A^\perp, \Sigma_1, \Sigma_2], \Delta, \Gamma, B_1 \,\&\, B_2}\,\&$$

Since the cut-lists of the two proofs are identical, $d$ is the identity.

The interpretations of $[\![\pi]\!]$ and $[\![\pi']\!]$ are the same in $\mathbb{Z}_2[\mathcal{L}]_{\mathsf{C}_+}$ by Lemma 3.6: The given proof is interpreted by $\mathrm{dg}\left([\![\rho]\!], \mathsf{S}_{(B_1,B_2,a)}\left[[\![\pi_1]\!]^\bullet, [\![\pi_2]\!]^\bullet\right]\right)$ where $[\![\pi_i]\!]^\bullet = \mathrm{dg}\left([\![\pi_i]\!], 0_{\Sigma_{3-i}}\right)$. While the reduced proof is by $\mathsf{S}_{(B_1,B_2,a)}\left[\mathrm{dg}\left([\![\rho]\!], [\![\pi_1]\!], 0_{\Sigma_2}\right), \mathrm{dg}\left([\![\rho]\!], [\![\pi_2]\!], 0_{\Sigma_1}\right)\right]$ but $\mathrm{dg}\left([\![\rho]\!], [\![\pi_i]\!], 0_{\Sigma_{3-i}}\right) = \mathrm{dg}\left([\![\rho]\!], [\![\pi_i]\!]^\bullet\right)$ by the associativity of dg.

(Crucial Case 3.2)

The same reduction as Case 3.1 but without the superposition on the common cut-lists $\Xi, \Omega, A*A^\perp$

for the last rule &. WLOG, we consider the list of cut formula for the conclusion of $rpf$ is

$$[\Omega, \Xi, \Xi, \Sigma_1, \Sigma_2, A * A^{\perp}, A * A^{\perp}]$$

so that the superposition via the last rule & is only for $\Omega$.
In this case, syntactical *co-contraction* (i.e., duplication) arises from $\mathcal{L}(\Xi, A, A^{\perp})$ to $\mathcal{L}(\Xi^{\delta}, A^{\delta}, (A^{\perp})^{\delta})$, where the notation $\delta$ denotes (*co-contraction* of formulas):
For a sequence $\Gamma$ of formulas, $\Gamma^{\delta}$ denotes the duplicated sequence $\Gamma, \Gamma$. The left (resp. right) occurrence of $\Gamma$ in $\Gamma^{\delta}$ is denoted by $\Gamma^{\delta_1}$ (resp. by $\Gamma^{\delta_2}$). Note that $\mathcal{L}(\Gamma^{\delta_1})$ and $\mathcal{L}(\Gamma^{\delta_2})$ are isomorphic sets of indeterminates.

The co-contraction is interpreted by a homomorphism $d$ induced by a superposition and an identity:

$$d := \left(a\, J_{\Xi^{\delta_1}, (A,A^{\perp})^{\delta_1}} + \bar{a}\, J_{\Xi^{\delta_2}, (A,A^{\perp})^{\delta_2}}\right) \uplus \mathrm{Id}_{\Omega, \Sigma_1, \Sigma_2} :$$

$$\mathbb{Z}_2[\mathcal{L}(\Omega, \Xi, \Sigma_1, \Sigma_2, A, A^{\perp})] \longrightarrow \mathbb{Z}_2[\mathcal{L}(\Omega, \Xi^{\delta}, \Sigma_1, \Sigma_2, (A, A^{\perp})^{\delta})]$$

where $J_{\Xi^{\delta_i}, (A,A^{\perp})^{\delta_i}}$ is the following ring inclusion, complementary for $i = 1, 2$:

$$\begin{aligned} \mathbb{Z}_2[\mathcal{L}(\Xi^{\delta_i}, (A, A^{\perp})^{\delta_i})] &\hookrightarrow \mathbb{Z}_2[\mathcal{L}(\Xi^{\delta_1}, (A, A^{\perp})^{\delta_1}), \mathcal{L}(\Xi^{\delta_2}, (A, A^{\perp})^{\delta_2})] \\ &\cong \mathbb{Z}_2[\mathcal{L}(\Xi^{\delta}, (A, A^{\perp})^{\delta})] \end{aligned} \tag{7}$$

Thus, the left of the $\uplus$ is a *superposition* of the two complementary inclusions:
(*superposition*)

$$a\, J_{\Xi^{\delta_1}, (A,A^{\perp})^{\delta_1}} + \bar{a}\, J_{\Xi^{\delta_2}, (A,A^{\perp})^{\delta_2}}$$

mapping $X \in \mathcal{L}(\Xi, A, A^{\perp})$ to $aX^{\delta_1} + \bar{a}X^{\delta_2} \in \mathbb{Z}_2[\mathcal{L}((\Xi, A, A^{\perp})^{\delta})]$.

We show in $\mathcal{A}_{(\mathbb{Z}_2[\mathcal{L}]_{\mathtt{C}_+})} \mod \langle a\bar{a} \rangle$;

$$\mathsf{qEx}\left(\sigma_{\Xi, A, A^{\perp}}, [\![ gpf ]\!]\right) = \mathsf{qEx}\left(\sigma_{(\Xi, A, A^{\perp})^{\delta}}, [\![ rpf ]\!]\right)$$

This is by the following Lemma 3.18 because:
The given proof is interpreted by

$$\mathsf{S}^{\Omega, \Xi, A, A^{\perp}, \Sigma, \Delta, \Gamma}_{(B_1, B_2, a)} \left[\mathsf{dg}\left([\![ \rho ]\!], [\![ \pi_1 ]\!], 0_{\Sigma_2}\right), \mathsf{dg}\left([\![ \rho ]\!], [\![ \pi_2 ]\!], 0_{\Sigma_1}\right)\right]$$

While, the reduced proof is interpreted by

$$\mathsf{S}^{\Omega, (\Xi, A, A^{\perp})^{\delta}, \Sigma_1, \Sigma_2, \Delta, \Gamma}_{(B_1, B_2, a)} [\star_1, \star_2] \quad \text{with} \quad \star_i = \mathsf{dg}\left([\![ \rho ]\!], [\![ \pi_i ]\!], 0_{\Sigma_{3-i}}, 0_{(\Xi, A, A^{\perp})^{\delta_{3-i}}}\right)$$

□

**Lemma 3.18 (eliminating $\delta$ via qEx and superposing $d$).**
*Let* $d := (a\, J_{\Delta^{\delta_1}} + \bar{a}\, J_{\Delta^{\delta_2}}) \uplus \mathrm{Id}_{\mathbb{Z}_2[\mathcal{L}(\Gamma, B_1 \& B_2)]} : \mathbb{Z}_2[\mathcal{L}(\Delta, \Gamma, B_1 \& B_2)] \longrightarrow \mathbb{Z}_2[\mathcal{L}(\Delta^{\delta}, \Gamma, B_1 \& B_2)]$, *where* $J_{\Delta^{\delta_i}}$ *are complementary ring inclusions defined same as (7). Then the following holds in* $\mathcal{A}_{(\mathbb{Z}_2[\mathcal{L}]_{\mathtt{C}_+})}$ *under* $\mod \langle a\bar{a} \rangle$*:*

$$d(\mathsf{qEx}\left(\sigma_{\Delta}, \mathsf{S}^{\Delta, \Gamma}_{(B_1, B_2, a)} [f, g]\right)) = \mathsf{qEx}\left(\sigma_{\Delta^{\delta}}, \mathsf{S}^{\Delta^{\delta}, \Gamma}_{(B_1, B_2, a)} \left[\mathsf{dg}\left(0_{\Delta^{\delta_2}}, f\right), \mathsf{dg}\left(0_{\Delta^{\delta_1}}, g\right)\right]\right)$$

Proof. See Sec C of Appendix. □

**Example 3.19 (of Theorem 3.17).**
($i$) Both $\pi_1$ and $\pi_2$ of Example 3.10 are normalized into the cut-free proof, say $\pi_3$, of $\vdash Y,\ X^\perp \& X^\perp$, whose proof is translated into the proof-net $\vdash Y,\ X^\perp \& X^\perp$ so that $[\![\, \pi_3 \,]\!]$ is the following matrix

$$\begin{array}{c} X \\ X^\perp\&X^\perp \end{array} \begin{pmatrix} & a(1\otimes p^*) + \bar{a}(1\otimes q^*) \\ a(1\otimes p) + \bar{a}(1\otimes q) & \end{pmatrix} \quad (8)$$

(columns labelled $X$ and $X^\perp\&X^\perp$)

As depicted respectively in (4) and (5) of Example 3.10, both qEx $\left(\sigma_{X,Y^\perp}, [\![\, \pi_1 \,]\!]\right)$ and qEx $\left(\sigma_{X,Y^\perp,X,Y^\perp}, [\![\, \pi_2 \,]\!]\right)$ equate to (8).

($ii$) Let $\eta$ and $\tau$ denote respectively the two proofs, both to the sequent $\vdash X \oplus X,\ X^\perp \& X^\perp$; the $\eta$-expansion of the axioms and that whose $\oplus_i$ rules are replaced by $\oplus_{3-i}$. The following left (resp. right) is the proof-net translating $\eta$ (resp. $\tau$): $\vdash X \oplus X,\ X^\perp \& X^\perp$ $\quad$ $\vdash X \oplus X,\ X^\perp \& X^\perp$.

$[\![\, \eta \,]\!]$ (resp. $[\![\, \tau \,]\!]$) is the anti-diagonal matrix of the element $a(1\otimes p)(1\otimes p^*) + \bar{a}(1\otimes q)(1\otimes q^*)$ (resp. $a(1\otimes q)(1\otimes p^*) + \bar{a}(1\otimes p)(1\otimes q^*)$). qEx $\left(\sigma_{X\oplus X, X^\perp\& X^\perp}, \left[\!\!\left[\, cut(\tau,\tau) \,\right]\!\!\right]\right)$ is the anti-diagonal of the element $\prod_{i=1,2}(a_i(1\otimes q)(1\otimes p^*) + \bar{a}_i(1\otimes p)(1\otimes q^*))$, which element equates to $a_1\bar{a}_2(1\otimes q)(1\otimes q^*) + \bar{a}_1 a_2(1\otimes p)(1\otimes p^*)$, where $a_1$ (resp. $a_2$) is the eigenweight for $X^\perp \& X^\perp$ inside (resp. outside) the cut-list. When the ring iso $d$ is defined to map $a_2 \mapsto \bar{a}_1$ and $\bar{a}_2 \mapsto a_1$ between $\mathbb{Z}_2[\mathcal{L}(a_i, \bar{a}_i)]$'s ($i = 1, 2$), Ex $\left(d, \sigma_{X\oplus X, X^\perp\& X^\perp}, \left[\!\!\left[\, cut(\tau,\tau) \,\right]\!\!\right]\right)$ equates to $[\![\, \eta \,]\!]$.

**Remark 3.20 (Refinement of Theorem 3.17 in Section 5 in terms of reduction free ring homs).** Theorem 3.17 has one blind side that the choice of ring homomorphisms depends on the cut-elimination itself for $\pi$. This is why Theorem 3.17 is unsatisfactory as GoI semantics, which semantics is supposed to capture dynamism of cut-elimination autonomously, hence freely from any cut-elimination procedure. In the last section, the theorem is improved into Theorem 5.12 (autonomous invariance) to overcome the defect so that ring homomorphisms arise autonomously while running quasi-execution formula for $\pi$, without any reference to the cut elimination for $\pi$.

**Remark 3.21 (Relationship to Girard's GoI III [9]).**
The MALL GoI of this section as well as that improved in Section 5 can provide a general construction for Girard's GoI III [9] when his implicit use of monomial weights (from [10]) is relaxed by our method to accommodate polynomial weights. In particular, Girard's equivalence relation of variant, in terms of his algebra of resolution, is replaced by our ring homomorphisms for the scalars $\mathbb{Z}_2[\mathcal{L}]$. A precise formulation needs to be examined, in terms of our polynomial ring, of his syntactical $*$-algebra consisting of clauses for predicates.

## 4 HvG Nilpotency of Quasi-Execution in $\mathcal{A}_{(\mathbb{Z}_2[\mathcal{L}])}$ under mod $\mathfrak{m}_\pi$ for Scalars

The aim of this section is to present a finer quasi-execution formula, intuitively satisfying the following:

(*finer grained execution*) Running quasi-execution formulas recognizes, in terms of termination (i.e., nilpotency), not only DR acyclicity of (P2), but also the legal cycles for (P3).

This concerns a construction $\mathfrak{m}_\pi$ of GoI interpretation to satisfy roughly the following schema:

$$\text{(P2) versus (P3)} \quad = \quad [\![\, \pi \,]\!] \text{ versus } \mathfrak{m}_\pi$$

The combination of $[\![\pi]\!]$ of Section 3 and $\mathfrak{m}_\pi$ of this section is shown to realise the finer grained execution. It is here in this section the schema arises naturally from the following question: How is the toggling condition (P3) accommodated into GoI semantics, given that Section 3 is considered as a GoI counterpart of (P2) (cf. Remark 3.16) ? This section gives an answer by constructing a faithful interpretation of (P3) using GoI semantics in the same spirit of the previous sections. Our concern in this section is not to give a rigorous correspondence, but to show how (P3) shed a new light to GoI semantics. Thus, the GoI construction presented in this section may provide a self-contained understanding of (P3) from the GoI viewpoint.

The shallow algebraic structure $\mathcal{A}_{(\mathbb{Z}_2[\mathcal{L}]_{\mathtt{C}_{+\times}})}$ imposing the quotient does not suffice for the aim, which structure was used in Subsection 3.3 when running qEx. This is because the shallow structure, uniformly imposing $a.\bar{a} = 0$ over $a$'s in $\mathcal{L}$, collapses the toggling (P3), so that legal cycles cannot be discriminated from other illegal ones. To overcome this, we construct a certain measure $\mathfrak{m}_\pi$ providing us to recognise legal cycles in a proof $\pi$.

We start this subsection by associating a formal language $m$ consisting of finite words over $\mathcal{L}$ to each occurrence of an element $x \in \mathcal{A}_{(\mathbb{Z}_2[\mathcal{L}])}$. The association is expressed by $x : m$, and is done for each element of the matrix $[\![\pi]\!] \in \mathcal{M}(\mathcal{A}_{(\mathbb{Z}_2[\mathcal{L}])})$ to yield the matrix $\mathfrak{m}_\pi$ of the same size, indexed by the same formulas. Since the set of all the formal languages forms a semiring, the quasi-execution formula $\mathfrak{qex}\ (\sigma_\Delta, \mathfrak{m}_\pi)$ is run for $\mathfrak{m}_\pi$ w.r.t the partial symmetry $\sigma_\Delta$. The $\mathfrak{qex}\ (\sigma_\Delta, \mathfrak{m}_\pi)$ provides a measure allowing to determine a finer grained termination of qEx $(\sigma_\Delta, [\![\pi]\!])$ in $\mathbb{Z}_2[\mathcal{L}]_{\mathtt{C}_+}$. Nilpotency corresponds to HvG (P3).

Recall that a *semiring* is a set with two binary operations (addition +) and (multiplication $\cdot$) with respective constants 0 and 1 satisfying the ring axiom *except that there may not be additive inverses (i.e., without subtraction).*

**Definition 4.1 (The semiring $2^{W(\mathcal{L})}$ of formal languages over $\mathcal{L}$ (cf. [6])).** Let $W(\mathcal{L})$ denote the free monoid generated by $\mathcal{L}$ (i.e., all the finite words over $\mathcal{L}$). Each subset of $W(\mathcal{L})$ is called a *formal language* over $\mathcal{L}$. The power set $2^{W(\mathcal{L})}$, consisting of the formal languages over $\mathcal{L}$, forms a semiring under set-theoretical *union* as addition and element-wise *concatenation* of words as multiplication. The zero of the semiring $2^{W(\mathcal{L})}$ is the empty-set (empty language) $\emptyset$ and the unit is the singleton $\{\epsilon\}$ of the empty string $\epsilon$. $(2^{W(\mathcal{L})}, \cup, \cdot, \emptyset, \{\epsilon\})$ is called the *semiring of formal languages over $\mathcal{L}$* .

**Definition 4.2 (interpretation $\mathfrak{m}_\pi$ for MALL proof $\pi$).** For a MALL proof $\pi$, the interpretation $\mathfrak{m}_\pi \in \mathcal{M}_{\Gamma,\Delta}(2^{W(\mathcal{L}(\Gamma,\Delta))})$ is defined. The matrix $\mathfrak{m}_\pi$ is indexed with the same formulas as $[\![\pi]\!]$ of Definition 3.2, hence of the same size, but of elements from the semiring $2^{W(\mathcal{L})}$.

In the definition to follow except for the &-rule, the interpretation $\mathfrak{m}_\pi$ is the same instance of Definition 3.2, but simpler since it is needless to employ $p^{(*)}$s and $q^{(*)}$s for the contraction of row and column. For the construction, set-theoretical union suffices in the definition. However the interpretation of & differs owing to the difference between Definition 3.3 above and Definition 4.3 below.

$$\text{(axiom) } \mathfrak{m}_{ax} = \begin{array}{c} \\ A \\ A^\perp \end{array} \begin{array}{c} \begin{array}{cc} A & A^\perp \end{array} \\ \left( \begin{array}{cc} \emptyset & \{\epsilon\} \\ \{\epsilon\} & \emptyset \end{array} \right) \end{array}$$

(cut-rule) $\mathfrak{m}_\pi = \mathrm{dg}\ (\mathfrak{m}_{\pi_1}, \mathfrak{m}_{\pi_2})$

($\invamp$-rule)

Contracting the rows (resp. the columns) of $A_1$ and $A_2$ in terms of set-theoretical union to obtain those of $A_1 \invamp A_2$. See Figure 3.

($\otimes$-rule)
$\mathfrak{m}_\pi$ is obtained from dg $(\mathfrak{m}_{\pi_1}, \mathfrak{m}_{\pi_2})$ the same as the $\parr$-rule by contracting rows and columns of $A_1$ and $A_2$ to obtain those of $A_1 \otimes A_2$. See Fig 3.

($\oplus_1$-rule) (same for $\oplus_2$-rule): $\mathfrak{m}_\pi$ is obtained from dg $(\mathfrak{m}_{\pi_1}, \emptyset_{A_2})$ by contracting rows and columns of $A_1$ and $A_2$ to obtain those of $A_1 \oplus A_2$. See Fig 3.

(&-rule) Let $\mathfrak{m}^\bullet_{\pi_i}$ denote dg $(\mathfrak{m}_{\pi_i}, \emptyset_{A_{3-i}})$.

$$\mathfrak{m}_\pi := \mathfrak{s}^{\Delta_1,\Delta_2,\Gamma}_{(A_1,A_2,a)} \left[\mathfrak{m}^\bullet_{\pi_1}, \mathfrak{m}^\bullet_{\pi_2}\right]$$

where $\mathfrak{s}^{\Delta_1,\Delta_2,\Gamma}_{(A_1,A_2,a)}$ is the superposition introduced below Definition 4.3. See Fig 3.
Note that due to Definition 4.3 no words belonging to the elements on the row and the column of $A_1 \& A_2$ have occurrences of $a$ nor $\bar{a}$. While for the other rows and columns, the words belonging to the element there (except $\emptyset$) are scalar multiplied either by $a$ or $\bar{a}$ inheriting its respective origin either in $\mathfrak{m}_1$ or $\mathfrak{m}_2$.

**Definition 4.3 (superposition $\mathfrak{s}^{\Gamma}_{(A_1,A_2,a)}[h,\ell]$).** $\mathfrak{s}^{\Gamma}_{(A_1,A_2,a)}(h,\ell)$ of the indices $\Gamma, A_1 \& A_2$ is defined for $h, \ell \in \mathcal{M}(2^{W(L)})$ of the respective indices $\Gamma, A_1$ and $\Gamma, A_2$.

$$\mathfrak{s}^{\Gamma}_{(A_1,A_2,a)}(h,\ell) \quad := \quad \begin{array}{r} \\ \Gamma \\ \\ A_1\&_a A_2 \end{array} \begin{array}{c} \begin{array}{cc} \Gamma & A_1\&_a A_2 \end{array} \\ \begin{pmatrix} \{a\}h^\Gamma_\Gamma \cup \{\bar{a}\}\ell^\Gamma_\Gamma & h^{A_1}_\Gamma \cup \ell^{A_2}_\Gamma \\ & \\ h^\Gamma_{A_1} \cup \ell^\Gamma_{A_2} & h^{A_1}_{A_1} \cup \ell^{A_2}_{A_2} \end{pmatrix} \end{array}$$

This is neither the same nor parallel instance of Definition 3.3 to $\mathcal{M}(2^{W(\mathcal{L})})$. Although the same for the $(\Gamma, \Gamma)$-block, the row and column of $A_1 \& A_2$ are contracted from those of $A_1$ and $A_2$ without employing the multiplication by $a$ nor by $\bar{a}$. The definition is the key to reflect the HvG's toggling condition.

Each of the following denotes the respective $\mathfrak{m}_\pi$ in Def 4.2, in which $\mathfrak{m}_{\pi_i} := \mathfrak{m}_i$ and $\mathfrak{m}_{\pi'} := \mathfrak{m}'$.

$$\begin{array}{c|cc} & \Delta\ \Gamma & A_1 \parr A_2 \\ \hline \begin{array}{c}\Delta\\ \Gamma\end{array} & \mathfrak{m}'^{-}_{-} & \bigcup_i \mathfrak{m}'^{A_i}_{-} \\ A_1 \parr A_2 & \bigcup_i \mathfrak{m}'^{-}_{A_i} & \bigcup_{i,j} \mathfrak{m}'^{A_i}_{A_j} \end{array}$$

$\parr$-rule

$$\begin{array}{c|ccccc} & \Delta_1 & \Delta_2 & \Gamma_1 & \Gamma_2 & A_1 \otimes A_2 \\ \hline \Delta_1 & \mathfrak{m}_1{}^{-}_{-} & \emptyset & \mathfrak{m}_1{}^{-}_{-} & \emptyset & \mathfrak{m}_1{}^{A_1}_{-} \\ \Delta_2 & \emptyset & \mathfrak{m}_2{}^{-}_{-} & \emptyset & \mathfrak{m}_2{}^{-}_{-} & \mathfrak{m}_2{}^{A_2}_{-} \\ \Gamma_1 & \mathfrak{m}_1{}^{-}_{-} & \emptyset & \mathfrak{m}_1{}^{-}_{-} & \emptyset & \mathfrak{m}_1{}^{A_1}_{-} \\ \Gamma_2 & \emptyset & \mathfrak{m}_2{}^{-}_{-} & \emptyset & \mathfrak{m}_2{}^{-}_{-} & \mathfrak{m}_2{}^{A_2}_{-} \\ A_1 \otimes A_2 & \mathfrak{m}_1{}^{-}_{A_1} & \mathfrak{m}_2{}^{-}_{A_2} & \mathfrak{m}_1{}^{-}_{A_1} & \mathfrak{m}_2{}^{-}_{A_2} & \mathfrak{m}_1{}^{A_1}_{A_1} \cup \mathfrak{m}_2{}^{A_2}_{A_2} \end{array}$$

$\otimes$-rule

$$\begin{array}{c|cc} & \Delta\ \Gamma & A_1 \oplus A_2 \\ \hline \begin{array}{c}\Delta\\ \Gamma\end{array} & \mathfrak{m}_1{}^{-}_{-} & \mathfrak{m}_1{}^{A_1}_{-} \\ A_1 \oplus A_2 & \mathfrak{m}_1{}^{-}_{A_1} & \mathfrak{m}_1{}^{A_1}_{A_1} \end{array}$$

$\oplus_1$-rule

$$\begin{array}{c|cccc} & \Delta_1 & \Delta_2 & \Gamma & A_1 \& A_2 \\ \hline \Delta_1 & \{a\}\mathfrak{m}_1{}^{-}_{-} & \emptyset & \{a\}\mathfrak{m}_1{}^{-}_{-} & \mathfrak{m}_1{}^{A_1}_{-} \\ \Delta_2 & \emptyset & \{\bar{a}\}\mathfrak{m}_2{}^{-}_{-} & \{\bar{a}\}\mathfrak{m}_2{}^{-}_{-} & \mathfrak{m}_2{}^{A_2}_{-} \\ \Gamma & \{a\}\mathfrak{m}_1{}^{-}_{-} & \{\bar{a}\}\mathfrak{m}_2{}^{-}_{-} & \{a\}\mathfrak{m}_1{}^{-}_{-} \cup \{\bar{a}\}\mathfrak{m}_2{}^{-}_{-} & \bigcup_i \mathfrak{m}_i{}^{A_i}_{-} \\ A_1 \& A_2 & \mathfrak{m}_1{}^{-}_{A_1} & \mathfrak{m}_2{}^{-}_{A_2} & \bigcup_i \mathfrak{m}_i{}^{-}_{A_i} & \bigcup_i \mathfrak{m}_i{}^{A_i}_{A_i} \end{array}$$

&-rule

Fig. 3. Interpretation $\mathfrak{m}_\pi$ of Def 4.2

**Remark 4.4.** Let us describe intuitively how arose the interpretation of & in Definition 4.3. The definition will later shown in Remark 4.14 to be reflected by HvG (P3):

In terms of matrix as I/O box, any flow on $\mathfrak{s}^{\Gamma}_{(A_1, A_2, a)}[h, \ell]$ passing through $A_1 \&_a A_2$ (i.e., either with input or output of $A_1 \& A_2$) does not memorise that it toggles $\&_a$; e.g., see the right picture for the respective flows I and O, whose respective elements $*$ do not memorise the toggling of $\&_a$: This reflects that the row and the column of $A_1 \& A_2$ are free from $a$ and $\bar{a}$.

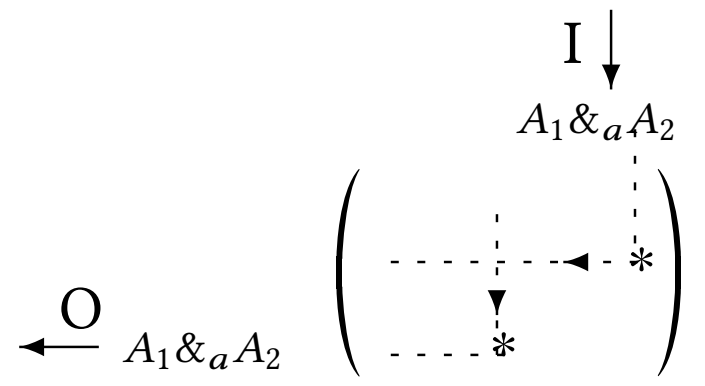

While other flows memorise whether it toggles $\&_a$ or not, in terms of the multipliers of the scalars $a$ or $\bar{a}$ in the semiring $2^{W(\mathcal{L})}$: Consequently, each element of the other row and column than $A_1 \& A_2$ designates its origin in terms of the left $a$, the right $\bar{a}$ and the (right and left) superposition $\cup$.

**Example 4.5 (of Def 4.2).** For $\pi_1$ and $\pi_2$ of Example 3.7, $\mathfrak{m}_{\pi_1}$ (resp. $\mathfrak{m}_{\pi_2}$) puts the following two matrices of (9) (resp. (10)) anti-diagonally:

$$\begin{array}{c} \\ X \\ X^{\perp}\&X^{\perp} \end{array} \begin{array}{c} \begin{array}{cc} X & Y^{\perp} \end{array} \\ \begin{pmatrix} & \{a, \bar{a}\} \\ \{\epsilon\} & \end{pmatrix} \end{array} \quad \begin{array}{c} \\ X \\ Y^{\perp} \end{array} \begin{array}{c} \begin{array}{cc} X & X^{\perp}\&X^{\perp} \end{array} \\ \begin{pmatrix} & \{\epsilon\} \\ \{a, \bar{a}\} & \end{pmatrix} \end{array} \tag{9}$$

$$\begin{array}{c} \\ X \\ X^{\perp}\&X^{\perp} \end{array} \begin{array}{c} \begin{array}{cccc} X & Y^{\perp} & X & Y^{\perp} \end{array} \\ \begin{pmatrix} & \{a\} & & \{\bar{a}\} \\ \{\epsilon\} & & \{\epsilon\} & \end{pmatrix} \end{array} \quad \begin{array}{c} \\ X \\ Y^{\perp} \\ X \\ Y^{\perp} \end{array} \begin{array}{c} \begin{array}{cc} X & X^{\perp}\&X^{\perp} \end{array} \\ \begin{pmatrix} & \{\epsilon\} \\ \{a\} & \\ & \{\epsilon\} \\ \{\bar{a}\} & \end{pmatrix} \end{array} \tag{10}$$

The interpretation $\mathfrak{m}_{\pi_1}$ is observed to be a different instance of $[\![ \pi_1 ]\!]$ of Definition 3.2: The element $\{a, \bar{a}\}$ of the row $X$ (resp. $Y^{\perp}$) and the column $Y^{\perp}$ (resp. $X$) in (9) memorises that it toggle the $\&_a$, while the corresponding element $a + \bar{a}$ in (4), becoming 1 by the cancellation in $\mathbb{Z}_2[\mathcal{L}]_{\mathsf{C}_+}$, cannot tell the toggling consequently.

Parallel to Definition 3.8 for $[\![ \pi ]\!]$, we define the quasi-execution formula for $\mathfrak{m}_{\pi}$. Note that the quasi-execution formula, as a formal algebraic expression, is applicable generally to matrices of entries from any semiring, in particular the semiring $2^{W(\mathcal{L})}$.

**Definition 4.6 (quasi-execution formula $\mathfrak{qex}\,(\sigma_{\Delta}, \mathfrak{m})$).** For a matrix $\mathfrak{m} \in \mathcal{M}_{\Delta,\Gamma}\left(2^{W(\mathcal{L})}\right)$ with a sequence $\Delta$ of pair-wise dual formulas and $\sigma_{\Delta}$ of Definition 3.8,

$$\mathfrak{qex}\,(\sigma_{\Delta}, \mathfrak{m}) := \mathfrak{m}^{\Gamma}_{\Gamma} \cup \bigcup_{n \geq 0} \mathfrak{m}^{\Delta}_{\Gamma}(\sigma_{\Delta}\mathfrak{m}^{\Delta}_{\Delta})^n(\sigma_{\Delta}\mathfrak{m}^{\Gamma}_{\Delta}) \tag{11}$$

The equation (11) always makes sense and belongs to $\mathcal{M}_{\Delta,\Gamma}(2^{W(\mathcal{L}(\Delta,\Gamma))})$ (see the contrast to the formal sum (instead of the set-union) in (3) of Definition 3.8 ).

To emphasize the parallel definition of $\mathsf{qEx}\,(\sigma_{\Delta}, -)$ for $[\![ \pi ]\!]$, we can define quasi-execution simultaneously to a pair $[\![ \pi ]\!] : \mathfrak{m}_{\pi}$ to yield the pair $\mathsf{qEx}\,(\sigma_{\Delta}, [\![ \pi ]\!]) : \mathfrak{qex}\,(\sigma_{\Delta}, \mathfrak{m}_{\pi})$. The simultaneous definition makes sense since the expressions $(\alpha : \mathfrak{m})$s form a semiring under addition and multiplication defined element-wise; $(\alpha : m) + (\beta : n) := (\alpha + \beta : m \cup n)$ and $(\alpha : m)(\beta : n) := (\alpha\beta : mn)$ for $\alpha, \beta \in \mathcal{A}_{(\mathbb{Z}_2[\mathcal{L}])}$ and $m, n \in 2^{W(\mathcal{L})}$.

**Example 4.7 (of Def 4.6).** The same depiction as (4) and (5) in Example 3.10 respectively for (9) and (10), $(\mathfrak{qex}\ (\sigma_{X,Y^\perp}, \mathfrak{m}_{\pi_1}))^X_{X^\perp \& X^\perp} = \{\epsilon\}\{a, \bar{a}\}$ and $(\mathfrak{qex}\ (\sigma_{X,Y^\perp,X,Y^\perp}, \mathfrak{m}_{\pi_2}))^X_{X^\perp \& X^\perp} = \{\epsilon\}\{a\} \cup \{\epsilon\}\{\bar{a}\}$.

**Definition 4.8.** The mapping $\| \quad \| : 2^{W(\mathcal{L})} \longrightarrow 2^{\mathcal{L}}$ is defined by sending $v \in 2^{W(\mathcal{L})}$ to

$$\|v\| := \left\{ a, \bar{a} \;\middle|\; \begin{array}{l} \exists\, w_1 \in v\ \exists\, w_2 \in v \quad \text{such that } a \text{ (resp. } \bar{a}) \\ \text{appears in the word } w_1 \text{ (resp. } w_2) \end{array} \right\}$$

That is, the mapping $\| \quad \|$ takes the largest set (as the indicator function) belonging to $\{\chi \in 2^{\mathcal{L}} \mid \chi(a) = \chi(\bar{a})\ \forall a \in \mathcal{L}\}$, contained in the value of the forgetful map $2^{W(\mathcal{L})} \longrightarrow 2^{\mathcal{L}}$, where the forgetful map sends $v$ to $\bigcup_{w \in v}\{\,\ell \mid \text{literal } \ell \text{ appears in the word } w\}$.

The mapping $\| \quad \|$ is lifted for matrices $\mathcal{M}(2^{W(\mathcal{L})}) \longrightarrow \mathcal{M}(2^{\mathcal{L}})$ element-wise so that for a matrix $V = (V_i^j)$ in the domain, $\|V\|$ denotes the matrix whose $(i, j)$-element is $\|V_i^j\|$. For matrices $V$ and $W$ of the same size, we denote $\|V\| \supseteq \|W\|$ when $\|V\|_i^j \supseteq \|W\|_i^j$ for every $(i, j)$-element.

**Example 4.9.** *Let* $\mathfrak{m}$ *be* $\mathfrak{m}_\pi$ *of &-rule in Definition 4.2 with the eigenweight a for the & introduced. Then no element of* $\|\mathfrak{m}\|_{-}^{A_1 \& A_2}$ *and of* $\|\mathfrak{m}\|^{-}_{A_1 \& A_2}$ *contains a nor* $\bar{a}$*, while every (non-empty) element of the other components* $\|\mathfrak{m}\|_{-}^{*}$*, ∗, - ∈* $\{\Delta_1, \Delta_2, \Gamma\}$ *contains either a or* $\bar{a}$*. E.g.,* $\|\mathfrak{m}\|^{A_1 \& A_2}_{A_1 \& A_2} = \bigcup_i \|\mathfrak{m}\|^{A_i}_{A_i}$*, while* $\|\mathfrak{m}\|^{\Gamma}_{\Gamma} = \|\{a\}\mathfrak{m}_1\|^{\Gamma}_{\Gamma} \cup \|\{\bar{a}\}\mathfrak{m}_2\|^{\Gamma}_{\Gamma} \subseteq \{a, \bar{a}\} \cup \bigcup_i \|\mathfrak{m}_i\|^{\Gamma}_{\Gamma}$.

**Notation:**( $(\sigma, \mathfrak{m})$ and $(\{\sigma_1, \ldots, \sigma_k\}, \mathfrak{m})$)
The expression $(\sigma, \mathfrak{m})$ (resp. more generally $(\{\sigma_1, \ldots, \sigma_k\}, \mathfrak{m})$) is an abbreviation of $\mathfrak{qex}\ (\sigma, \mathfrak{m})$ (resp. $\mathfrak{qex}\ (\sigma_1 \otimes \cdots \otimes \sigma_k, \mathfrak{m})$). The expression $(\sigma_1, (\sigma_2, \cdots, (\sigma_k, \mathfrak{m})))$ abbreviating $\mathfrak{qex}\ (\sigma_1, \mathfrak{qex}\ (\sigma_2, \cdots, \mathfrak{qex}\ (\sigma_k, \mathfrak{m})))$, is equal to $(\{\sigma_{\Delta_1}, \ldots, \sigma_{\Delta_k}\}, \mathfrak{m})$ by associativity of $\mathfrak{qex}$.

**Lemma 4.10 (Monotonic increase of $\mathfrak{qex}\ (\sigma, -)$ w.r.t $\| \quad \|$).** *For* $\mathfrak{m} \in \mathcal{M}_{\Delta_1, \ldots, \Delta_k, \Gamma}(2^{W(\mathcal{L})})$,

$$\|\mathfrak{m}\|^{\Gamma}_{\Gamma} \subseteq \|(\sigma_{\Delta_i}, \mathfrak{m})\|^{\Gamma}_{\Gamma} \subseteq \|(\{\sigma_{\Delta_1}, \ldots, \sigma_{\Delta_k}\}, \mathfrak{m})\|$$

Proof. Direct from (11) □

**Definition 4.11 (mod $\mathfrak{m}$ in $\mathcal{M}(\mathcal{A}_{(\mathbb{Z}_2[\mathcal{L}])})$ ).**
(1) For $m \in 2^{W(\mathcal{L})}$, mod $m$ denotes

$$\mathcal{A}_{(\mathbb{Z}_2[\mathcal{L}]_{\mathtt{C}_+})} \quad \text{mod the ideal } \bigoplus_{a, \bar{a} \in \|m\|} \langle a\bar{a} \rangle.$$

(cf. Remark 2.12.) Note, by the definition of $\|m\|$ of Definition 4.8, that $a$ and $\bar{a}$ range over (pairs of) literals, each appears in some word (not necessarily the same) contained in the language $m$.
(2) For a matrix $\mathfrak{m} \in \mathcal{M}(2^{W(\mathcal{L})})$, mod $\mathfrak{m}$ denotes the entry-wise mod $\mathfrak{m}_i^j$ of a matrix $\mathfrak{m}$ to designate mod to be taken in each entry of the matrix $\mathfrak{m}$.

**Definition 4.12 (nilpotency for a pair $U : \mathfrak{m}$).**
(1) For $x \in \mathcal{A}_{(\mathbb{Z}_2[\mathcal{L}])}$ and $m \in 2^{W(\mathcal{L})}$, we say an expression $x : m$ *is zero* when $x = 0 \mod m$. For matrices $U \in \mathcal{M}(\mathcal{A}_{(\mathbb{Z}_2[\mathcal{L}])})$ and $\mathfrak{m} \in \mathcal{M}(2^{W(\mathcal{L})})$ of the same size, we say an expression $U : \mathfrak{m}$ *is zero* when $U_i^j : \mathfrak{m}_i^j$ is zero element-wise for every $(i, j)$-element.
(2) For matrices $U \in \mathcal{M}_{\Delta,\Gamma}(\mathcal{A}_{\mathbb{Z}_2[\mathcal{L}(\Delta,\Gamma)]})$ and $\mathfrak{m} \in \mathcal{M}_{\Delta,\Gamma}(2^{W(\mathcal{L}(\Delta,\Gamma))})$ with the partial symmetry $\sigma_\Delta$, we say $(\sigma_\Delta,\ U : \mathfrak{m})$ is *nilpotent* when there exists a natural number $n$ so that $(\sigma_\Delta U^\Delta_\Delta)^n : (\sigma_\Delta \mathfrak{m}^\Delta_\Delta)^n$ is zero. Note that each $n$-power in the pair appears respectively in the summation symbol of $\mathrm{qEx}\,(\sigma_\Delta, [\![\, \pi \,]\!])$ and in the union symbol of $\mathfrak{qex}\ (\sigma_\Delta, \mathfrak{m}_\pi)$. Hence the condition is equivalent to say $\left(\sigma_\Delta U^\Delta_\Delta : \sigma_\Delta \mathfrak{m}^\Delta_\Delta\right)^n$ is zero (given that the pairs $U : \mathfrak{m}$ form the semiring).

(**scalars**): In what follows in this subsection, quasi-execution formulas are run in the scalars $\mathbb{Z}_2[\mathcal{L}]_{\mathbb{C}_+}$ with certain quotients explicitly mentioned in terms of $\mathfrak{m}_{(-)}$s.

Now that every MALL proof $\pi$ of $\vdash [\Delta], \Gamma$ is interpreted by the pair $[\![\pi]\!] : \mathfrak{m}_\pi$ by associating $\mathfrak{m}_\pi \in \mathcal{M}_{\Delta,\Gamma}(2^{W(\mathcal{L}(\Delta,\Gamma))})$ to $[\![\pi]\!] \in \mathcal{M}_{\Delta,\Gamma}(\mathcal{A}_{\mathbb{Z}_2[\mathcal{L}(\Delta,\Gamma)]})$, Proposition 3.15 is extended to the following Theorem.

**Theorem 4.13.** $(\sigma_\Delta, [\![\pi]\!] : \mathfrak{m}_\pi)$ *is* nilpotent *for any* MALL *proof* $\pi$ *of* $\vdash [\Delta], \Gamma$.

Proof. We shall prove the theorem in accordance with the cases of the proof of Theorem 3.17:
(Case 1)
As seen in the proof of Theorem 3.17, $\sigma_{B^\perp,A} [\![\pi]\!]^{B^\perp,A}_{B^\perp,A} : \emptyset$ is zero. This means nilpotency of $(\sigma_{B^\perp,A}, [\![\pi]\!] : \emptyset)$, hence automatically (by $\|\emptyset\| \subseteq \|\mathfrak{m}_\pi\|$) that of $(\sigma_{B^\perp,A}, [\![\pi]\!] : \mathfrak{m}_\pi)$. Because of qEx $(\sigma_{B^\perp,A}, [\![\pi]\!])$ is $[\![\pi']\!]$ up to the isomorphism $\mathcal{L}(A) \cong \mathcal{L}(B)$ of indeterminates shown in the proof, whose iso also induces the set iso, for each $(i, j)$ entry, $(\mathfrak{qex}\ (\sigma_{B^\perp,A}, \mathfrak{m}_\pi))^j_i \cong (\mathfrak{m}_{\pi'})^j_i$, the following holds:
If $(\sigma_\Delta, [\![\pi']\!] : \mathfrak{m}_{\pi'})$ is nilpotent, so is $(\sigma_\Delta \otimes \sigma_{B^\perp,A}, [\![\pi]\!] : \mathfrak{m}_\pi)$.
(Case 2)
Let $\vartheta$ denote qEx $\left(\sigma_{A\&B, B^\perp\oplus A^\perp}, [\![gpf]\!]\right)$. we write (6) of Theorem 3.17 by the sum $\vartheta = a\,\widetilde{\text{qEx}} + \bar{a}\,\widetilde{\text{dg}}$ where $\widetilde{\text{qEx}}$ and $\widetilde{\text{dg}}$ are the indicated occurrences in (6). Then qEx $(\sigma_\Omega, \vartheta)$ distributes over the sum (the last equation of the following) under mod $(\{\sigma_{A\&B}, \sigma_\Omega\}, gpf)^{\Gamma,\Xi}_{\Gamma,\Xi}$, while qEx $(\sigma_{\Delta_i}, \vartheta)$, $i = 1, 2, 3$, is absorbed into one component of the sum (second equation).

$$
\begin{aligned}
&\text{qEx}\,(\sigma_\Omega \otimes \sigma_{\Delta_1} \otimes \sigma_{\Delta_2} \otimes \sigma_{\Delta_3}, \vartheta)\\
&= \text{qEx}\left(\sigma_\Omega \otimes \sigma_{\Delta_1} \otimes \sigma_{\Delta_2} \otimes \sigma_{\Delta_3}, a\,\widetilde{\text{qEx}} + \bar{a}\,\widetilde{\text{dg}}\right)\\
&= \text{qEx}\left(\sigma_\Omega, \text{qEx}\left(\sigma_{\Delta_1} \otimes \sigma_{\Delta_3}, a\,\widetilde{\text{qEx}}\right) + \text{qEx}\left(\sigma_{\Delta_2}, \bar{a}\,\widetilde{\text{dg}}\right)\right)\\
&= \text{qEx}\left(\sigma_\Omega, \text{qEx}\left(\sigma_{\Delta_1} \otimes \sigma_{\Delta_3}, a\,\widetilde{\text{qEx}}\right)\right) + \text{qEx}\left(\sigma_\Omega, \text{qEx}\left(\sigma_{\Delta_2}, \bar{a}\,\widetilde{\text{dg}}\right)\right) \quad \text{mod } (\{\sigma_{A\&B}, \sigma_\Omega\}, gpf)^{\Gamma,\Xi}_{\Gamma,\Xi}
\end{aligned}
$$

The second equation holds because $\Delta_1$, $\Delta_2$ and $\Delta_3$ are not superposed in the indicated &-rule introducing $A\&B$. The last equation (the distribution) holds under mod $\langle a\bar{a}\rangle$ in general, hence under the mod described at the last equation owing that $\Omega$ is superposed in the &-rule:
This is because, for an arbitrary $n \geq 0$, if the $(*, \text{-})$-element of $\vartheta^\Omega_*(\sigma\vartheta^\Omega_\Omega)^n\sigma\vartheta^-_\Omega$ has an occurrence of scalar $a\bar{a}$, then $\{a, \bar{a}\}$ is a subset of the associated element $\|\mathfrak{m}^\Omega_*(\sigma\mathfrak{m}^\Omega_\Omega)^n\sigma\mathfrak{m}^-_\Omega\|$ where $\mathfrak{m} = \mathfrak{qex}\left(\sigma_{A\&B,B^\perp\oplus A^\perp}, \mathfrak{m}_{gpf}\right)$ (cf. Example 4.9).
Note that the mod in the last equation is finer grained than the mod $(\{\sigma_{A\&B}, \sigma_\Omega\} \cup \{\sigma_{\Delta_1}, \sigma_{\Delta_2}, \sigma_{\Delta_3}\}, gpf)$ by Lemma 4.10, hence the last equation holds automatically under the coarser grained mod. Thus from I.Hs for dg $\left([\![\pi_1]\!]^\bullet, [\![\pi_3]\!]\right)$ and for $[\![\pi_2]\!]$, the assertion is derived.
(Case 3.1) Nothing to prove in this case.
(Case 3.2)
Lemma 3.18 is redone to assert:
If $\left(\sigma_{\Delta^\delta},\ \mathsf{S}^{\Delta^\delta,\Gamma}_{(A_1,A_2,a)}\ [\text{dg}\ (0_{\Delta^{\delta_2}}, f)\,, \text{dg}\ (0_{\Delta^{\delta_1}}, g)] : \mathfrak{s}^{\Delta^\delta,\Gamma}_{(A_1,A_2,a)}\ [\text{dg}\ (0_{\Delta^{\delta_2}}, \mathfrak{f})\,, \text{dg}\ (0_{\Delta^{\delta_1}}, \mathfrak{g})]\right)$ is nilpotent, so is $\left(\sigma_\Delta,\ \mathsf{S}^{\Delta,\Gamma}_{(A_1,A_2,a)}\ [f, g] : \mathfrak{s}^{\Delta,\Gamma}_{(A_1,A_2,a)}\ [\mathfrak{f}, \mathfrak{g}]\right)$.
The assertion follows from the following, in which $\mathsf{S}^\delta$, $\mathfrak{s}^\delta$, $\mathsf{S}$ and $\mathfrak{s}$ are the indicated four occurrences (of superposition) in the assertion.

- For any eigenweight $b \in \mathcal{L}(\Delta)$, let $b_j = b^{\delta_j}$ and $\bar{b}_j = \bar{b}^{\delta_j}$. In $(\mathsf{S}^\delta)^{\Delta^\delta}_\star(\sigma(\mathsf{S}^\delta)^{\Delta^\delta}_{\Delta^\delta})^n\sigma(\mathsf{S}^\delta)^-_{\Delta^\delta}$ for any $n \geq 0$, there appear neither scalars $b_i b_{3-i}$ nor $\bar{b}_i b_{3-i}$ owing to the indicated $0_{\Delta^{\delta_1}}$ and $0_{\Delta^{\delta_2}}$ in the construction.

- The matrix $\mathfrak{qex}\left(\sigma_{\Delta^\delta}, \mathfrak{s}^\delta\right)$ is sent to $\mathfrak{qex}\,(\sigma_\Delta, \mathfrak{s})$ (of the same size) by the induced map on $2^{W(\mathcal{L})}$ from the *contraction* map $\{b_1, b_2\} \longrightarrow \{b\}$. Hence in particular the $(*, \text{-})$ component $(\mathfrak{s}^\delta)_*^{\Delta^\delta}(\sigma(\mathfrak{s}^\delta)_{\Delta^\delta}^{\Delta^\delta})^n\sigma(\mathfrak{s}^\delta)_{\Delta^\delta}^{-}$ is sent to the $(*, \text{-})$ component $\mathfrak{s}_*^\Delta(\sigma(\mathfrak{s})_\Delta^\Delta)^n\sigma(\mathfrak{s})_\Delta^{-}$.

□

**Remark 4.14 (Nilpotency of $[\![\pi]\!] : \mathfrak{m}_\pi$ is (P3)).**
The nilpotency of Theorem 4.13 is a property captured by simultaneously running both quasi-executions $\mathsf{qEx}\,(\sigma_\Delta, [\![\pi]\!])$ and $\mathfrak{qex}\,(\sigma_\Delta, \mathfrak{m}_\pi)$. Theorem 4.13 for $(\sigma_\Delta, [\![\pi]\!] : \mathfrak{m}_\pi)$ is a GoI counterpart of HvG (P3) in the following sense:

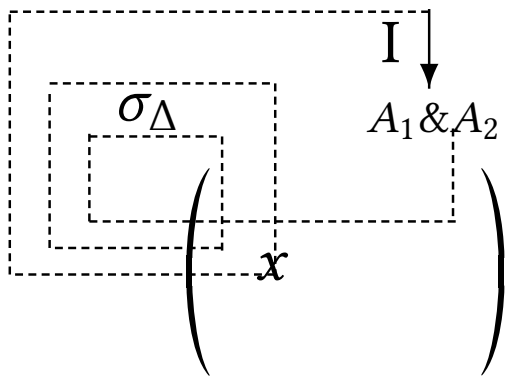

Every legal cycle passing through $A_1 \&_a A_2$, arisen accordingly to the expansion of the execution formula $\mathsf{qEx}\left(\sigma_\Delta, [\![\pi_1 \&_a \pi_2]\!]\right)$, must toggle another $\&_b$ ($b$ differs from $a$) by virtue of (P3): E.g., the right picture describes a cycle with the input $A_1 \& A_2$. Then the cycle is legal iff there exists a matrix element $x$ lying on the cycle, which memorises the toggling $\&_b$, i.e., $\|x\|$ contains both $b$ and $\bar{b}$. In particular, any flow passing through $A_1 \&_a A_2$ does not have to memorise $a$ nor $\bar{a}$ in order to judge the legality of a loop arisen in expanding the qEx.

In the following example, it is shown how $\mathfrak{m}_\pi$ on top of $[\![\pi]\!]$ recognises the legal cycle of Fig.1.

**Example 4.15 (Interpretation $[\![\pi]\!] : \mathfrak{m}_\pi$ of the proof-net of Fig 1).** Given that the two proofs for $a = 1$ and $a = 0$ are interpreted respectively by (12) and (13), The interpretation $[\![\pi]\!] : \mathfrak{m}_\pi$ is (14) for the proof-net in Fig 1 of Section 1:

$$
\begin{array}{c|ccccccc|}
 & X & Y^\perp & X^\perp & Y & X^\perp & X\otimes Y^\perp & Y_a \\
X & & & 1:\{\epsilon\} & & & & \\
Y^\perp & & & & & & & 1:\{\epsilon\} \\
X^\perp & 1:\{\epsilon\} & & & & & & \\
Y & & & & & & 1\otimes q^*:\{\epsilon\} & \\
X^\perp & & & & & & 1\otimes p^*:\{\epsilon\} & \\
X\otimes Y^\perp & & & & 1\otimes q:\{\epsilon\} & 1\otimes p:\{\epsilon\} & & \\
Y_a & & 1:\{\epsilon\} & & & & &
\end{array} \tag{12}
$$

$$
\begin{array}{c|ccccccc|}
 & X & Y^\perp & X^\perp & Y & X^\perp & X\otimes Y^\perp & Y_{\bar{a}} \\
X & & & & & 1:\{\epsilon\} & & \\
Y^\perp & & & & 1:\{\epsilon\} & & & \\
X^\perp & & & & & & 1\otimes p^*:\{\epsilon\} & \\
Y & & 1:\{\epsilon\} & & & & & \\
X^\perp & 1:\{\epsilon\} & & & & & & \\
X\otimes Y^\perp & & & 1\otimes p:\{\epsilon\} & & & & 1\otimes q:\{\epsilon\} \\
Y_{\bar{a}} & & & & & & 1\otimes q^*:\{\epsilon\} &
\end{array} \tag{13}
$$

$\sigma_{X, Y^\perp, X^\perp, Y}$

$$
\begin{array}{c|ccccccc|}
 & X & Y^\perp & X^\perp & Y & X^\perp & X\otimes Y^\perp & Y\&_a Y \\
X & & & a:\{a\} & & \bar{a}:\{\bar{a}\} & & \\
Y^\perp & & & & \bar{a}:\{\bar{a}\} & & & a\otimes p^*:\{\epsilon\} \\
X^\perp & a:\{a\} & & & & & \bar{a}\otimes p^*:\{\bar{a}\} & \\
Y & & \bar{a}:\{\bar{a}\} & & & & a\otimes q^*:\{a\} & \\
X^\perp & \bar{a}:\{\bar{a}\} & & & & & a\otimes p^*:\{a\} & \\
X\otimes Y^\perp & & & \bar{a}\otimes p:\{\bar{a}\} & a\otimes q:\{a\} & a\otimes p:\{a\} & & \bar{a}\otimes qq^*:\{\epsilon\} \\
Y\&_a Y & & a\otimes p:\{\epsilon\} & & & & \bar{a}\otimes qq^*:\{\epsilon\} &
\end{array} \tag{14}
$$

$[\![\pi]\!]^{\Delta}_{\Delta} : \mathfrak{m}^{\Delta}_{\Delta}$ is the left upper component of (14), denoted by $U^{X,Y^{\perp},X^{\perp},Y}_{X,Y^{\perp},X^{\perp},Y}$, and $(\sigma_{\Delta}, [\![\pi]\!] : \mathfrak{m}_{\pi})$ is nilpotent because:

$$(U^{X,Y^{\perp},X^{\perp},Y}_{X,Y^{\perp},X^{\perp},Y})\sigma_{\Delta}(U^{X,Y^{\perp},X^{\perp},Y}_{X,Y^{\perp},X^{\perp},Y}) = \begin{matrix} \\ X \\ Y^{\perp} \\ X^{\perp} \\ Y \end{matrix} \begin{matrix} X & Y^{\perp} & X^{\perp} & Y \\ \left( \begin{matrix} & a\bar{a} : \{a,\bar{a}\} & & \\ \bar{a}a : \{\bar{a},a\} & & & \\ & & & a\bar{a} : \{a,\bar{a}\} \\ & & \bar{a}a : \{\bar{a},a\} & \end{matrix} \right) \end{matrix} \tag{15}$$

, which is zero since

$$a\bar{a} = \bar{a}a = 0 \mod \{a, \bar{a}\}$$

In the submatrix $U^{X,Y^{\perp},X^{\perp},Y}_{X,Y^{\perp},X^{\perp},Y}$ of (14), the path is augmented to represent the permutation of cuts (left upper paths outside the matrix) together with dotted lines inside the matrix. The legal cycle emerges in (14) by tracing the path as follows: Entering $X^{\perp}$ vertically down to $a : \{a\}$, then horizontally exiting $Y^{\perp}$ followed by $\sigma$, entering $Y^{\perp}$ vertically down to $\bar{a} : \{\bar{a}\}$, then horizontally exiting $Y$, finally returning via $\sigma$, the starting entrance $X^{\perp}$. Chasing the path (starting with the $X^{\perp}$) yields the third column (non-zero) element $\bar{a}a : \{\bar{a}, a\}$ of (15).

## 5 Refining the Status of Ring Homs for the Invariance Theorem 3.17 [2]

This final section refines the definition of the ring homomorphism $d$ from an external one (employed in Theorem 3.17 of Section 3) to an internal one (Theorem 5.12 below). This is a remaining problem from Section 3.3 (cf. Remark 3.20) that the ring homomorphism $d_{\Delta}$ for obtaining $\mathsf{Ex}\,(d_{\Delta}, \sigma_{\Delta}, [\![\pi]\!])$ from $\mathsf{qEx}\,(\sigma_{\Delta}, [\![\pi]\!])$ for a MALL proof $\pi_{[\Delta],\Gamma}$ was externally defined so that the choice $d_{\Delta}$ depends plainly on the cut-elimination procedure. This problem is resolved in this section by presenting an equational system $\mathrm{eq}(\sigma_{\Delta}, [\![\pi]\!])$ in $\mathbb{Z}_2[\mathcal{L}]$, arising autonomously, parallel to the quasi-execution formula $\mathsf{qEx}\,(\sigma_{\Delta}, [\![\pi]\!])$. Each equation appears between monomials in $\mathbb{Z}_2[\mathcal{L}(\Delta, \Gamma)]$, while calculating the quasi-execution formula. The system $\mathrm{eq}(\sigma_{\Delta}, [\![\pi]\!])$ is shown to be satisfiable, so that the literals from $\mathcal{L}(\Delta)$ in the equations are definable (in other words, solvable) in terms of polynomials from the rest of $\mathcal{L}(\Gamma)$ in $\mathbb{Z}_2[\mathcal{L}(\Gamma)]$, where $\Delta$ (resp. $\Gamma$) is a cut-list (resp. a conclusion) of $\pi$ (Proposition 5.9). The satisfiability is seen as a kind of consistency of the equational system, which is inherited from the correctness structure of MALL proof $\pi$ among more relaxed proof-structures. Each solution defines a ring homomorphism $d$ from $\mathbb{Z}_2[\mathcal{L}(\Delta, \Gamma)]$ to $\mathbb{Z}_2[\mathcal{L}(\Gamma)]$. Note that the choice of a solution $d$ is not deterministic, as $d$ arises freely from proof-reduction. The goal of this section is to show that the satisfiability of $\mathrm{eq}(\sigma_{\Delta}, [\![\pi]\!])$ is compatible with cut-elimination for the proof $\pi$. The compatibility is an invariant of the quasi-execution formula modulo *reduction-free* ring homomorphisms (Theorem 5.12), which is the refinement of the invariance theorem.

### 5.1 Autonomous Equational System $\mathrm{eq}(\sigma_{\Delta}, [\![\pi]\!])$ and Satisfiability

This section starts with observing there is a $\mathbb{Z}$-module structure on the $*$-ring $\mathcal{A}$. The ring $\mathcal{A}$ is observed to be a *free-module* over $\mathbb{Z}$, and we fix once and for all its basis consisting of *paths*, which are a certain class of *monomial* elements of $\mathcal{A}$. Every *monomial* except 0 and 1 of $\mathcal{A}$ is expressed by a finite path written from left to right, accordingly by the configuration of the following directed graph:

[2] This section does not require Section 4, but directly succeeds Section 3.

E.g., $p^2(p^*)^3(q^*)^5$ and $q^2pqp^3q^*(p^*)^2$. The diagram lacks directed arrows from $q^*$ (resp. $p^*$) to $p$ and to $q$ (resp. to $q$ and to $p$), which lacking these two arrows respectively annihilates a path by $q^*p = 0$ (resp. by $p^*q = 0$) and cuts down a path by $p^*p = 1$ (resp. by $q^*q = 1$), thus both are dispensable.

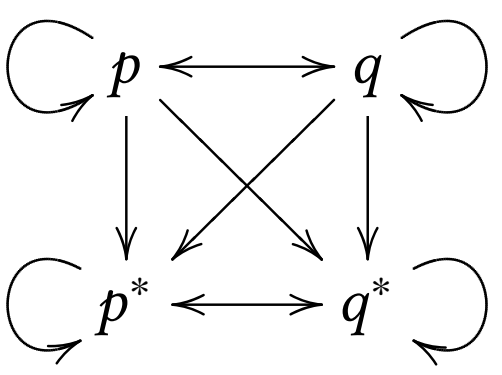

(16)

**Proposition 5.1 (independence of the paths of $\mathcal{A}$ over $\mathbb{Z}$).** *Any finite set $S$ of different paths of $\mathcal{A}$ configured by the graph (16) is independent over $\mathbb{Z}$.*

Proof. By double induction on (#$S$, #occurrences of $p, q, p^*, q^*$ in the paths of $S$).
For the base case, $S$ is a singleton, the assertion is direct; that is , any non-empty path is *torsion free* over $\mathbb{Z}$ (i.e., the empty path is the only path annihilated by a non zero element of $\mathbb{Z}$). The following proves the induction cases:
(Case 1) $S$ contains a path starting with $p$.
(Case 1.1) All the paths of $S$ start with $p$.
Each path is written in the form $p\,\boldsymbol{s}_i$. For the assertion, we assume $\sum \alpha_i\, p\, \boldsymbol{s}_i = 0$ for the linear combination. Multiplying $p^*$ from the left, by $p^*p = 1$, we have $\sum \alpha_i\, \boldsymbol{s}_i = 0$. By I.H for the second argument of the double induction, $\alpha_i = 0$ for all $i$.
(Case 1.2) Otherwise. $S$ contains a path $\boldsymbol{t}_j$ which does not start with $p$. For the assertion, we assume $\sum \alpha_i\, p\, \boldsymbol{s}_i + \sum \beta_i\, \boldsymbol{t}_j = 0$. Multiplying this equation by $q^*$ from the left, $\sum \beta_i\, \boldsymbol{t}'_j = 0$, where $\boldsymbol{t}'_j$ is $\boldsymbol{t}_j$ rid of the first element when $\boldsymbol{t}_j$ starts with $q$ and otherwise $\boldsymbol{t}'_j$ is $q^*\boldsymbol{t}_j$. By I.H on the first argument of the double induction, $\beta_j = 0$ for all $j$. Then the first equation $\sum \alpha_i\, p\, \boldsymbol{s}_i = 0$ instantiated by the zero to $\beta_j$'s yields, by I.H on the first argument, $\alpha_i = 0$ for all $i$.
(Case 2) $S$ contains a path starting with $q$.
Because of the symmetry between $q$ and $p$, same as Case 1.
(Case 3) Every path of $S$ starts either with $p^*$ or $q^*$.
From the configuration graph (16), in this case any path of $S$ ends either with $p^*$ or $q^*$. Then the symmetric argument (with reading the path *inverse* (from right to left) ) as Case 1 and 2 is done. Here the path *starting* with $p$ (resp. $q$) is read instead by that *ending* $p^*$ (resp. $q^*$) and multiplying $p^*$ (resp. $q^*$) from the *left* is read instead by multiplying $p$ (resp. $q$) from the *right*. □

Proposition 5.1 directly implies the following corollary because the short exact sequence of $\mathbb{Z}$-modules $0 \longrightarrow \mathcal{P} \longrightarrow \mathcal{A} \longrightarrow \mathbb{Z}1 \longrightarrow 0$ *splits*, where $\mathcal{P}$ denotes the $\mathbb{Z}$ module spanned by the basis of all the different paths of Proposition 5.1:

**Corollary 5.2 (monomial basis of $\mathcal{A}$ over $\mathbb{Z}$).** *All the different paths configured by the graph (16) together with* 1 *form a basis of the $\mathbb{Z}$-module $\mathcal{A}$. The basis is denoted by $\{\boldsymbol{s}_i\}_{i\in I}$.*

**Definition 5.3 (monomial basis of $\mathcal{A}_{(\mathbb{Z}_2[\mathcal{L}])}$ over $\mathbb{Z}_2$).** Let $\{m_j\}_{j\in J}$ be all the monomials in the polynomial ring $\mathbb{Z}_2[\mathcal{L}]$, which forms a basis $\mathbb{Z}_2[\mathcal{L}]$ over $\mathbb{Z}_2$. Together with the basis $\{\boldsymbol{s}_i\}_{i\in I}$ of $\mathcal{A}$ of Corollary 5.2, $\{m_j \otimes \boldsymbol{s}_i\}_{(j,i)\in J\times I}$ forms a basis of $\mathcal{A}_{(\mathbb{Z}_2[\mathcal{L}])}$ over $\mathbb{Z}_2$. Every element $x \in \mathcal{A}_{(\mathbb{Z}_2[\mathcal{L}])}$ is uniquely expressed by a linear combination $x = \sum_{i,j} m_j \otimes \boldsymbol{s}_i$. Indeed the combination is simply the sum of monomials in $\mathcal{A}_{(\mathbb{Z}_2[\mathcal{L}])}$, each $m_j \otimes \boldsymbol{s}_i$ is called *monomial component* of $x$.

**Definition 5.4 (system $\mathrm{eq}(\sigma, U)$ of equations in $\mathbb{Z}_2[\mathcal{L}]$ ).**

(1) (collapsing map $w$:) The ring homomorphism $w$ is defined to be induced from the collapsing ring homomorphism $\mathcal{A} \longrightarrow \mathbb{Z}_2$, which maps both generators $p$ and $q$ to 1.

$$w : \mathcal{A}_{(\mathbb{Z}_2[\mathcal{L}])} = \mathbb{Z}_2[\mathcal{L}] \otimes \mathcal{A} \longrightarrow \mathbb{Z}_2[\mathcal{L}] \otimes \mathbb{Z}_2 \cong \mathbb{Z}_2[\mathcal{L}]$$

(2) For any pair of elements $\alpha = \Sigma_i \alpha_i$ and $\beta = \Sigma_j \beta_j$ expressed by the unique linear combinations of monomials of $\mathcal{A}_{(\mathbb{Z}_2[\mathcal{L}])}$ in Definition 5.3, a set consisting of equations between $\mathbb{Z}_2[\mathcal{L}]$ monomial elements is defined;

$$\mathsf{E}\,[\alpha, \beta] = \mathsf{E}\left[\textstyle\sum_j \alpha_i, \sum_i \beta_j\right] := \{w(\alpha_i) = w(\beta_j) \mid \beta_i \cdot \alpha_i \neq 0\} \tag{17}$$

Note that $\alpha_i$ and $\beta_i$ are monomials in $\mathcal{A}_{(\mathbb{Z}_2[\mathcal{L}])}$, thus so are $w(\alpha_i)$ and $w(\beta_j)$ in $\mathbb{Z}_2[\mathcal{L}]$.

(3) For $\Gamma$ of formula occurrences and $\Delta$ of pair-wise dual formula occurrences, let $\gamma_j, \delta_n, \ldots, \delta_1, \gamma_i$ be a path of formulas starting with a formula $\gamma_i$ in $\Gamma$, succeeded by formulas $\delta_1, \ldots, \delta_n$ in $\Delta$, and ending with a formula $\gamma_j$ in $\Gamma$. Each path determines a system $\mathsf{eq}_{\gamma_j,\delta_n,\ldots,\delta_1,\gamma_i}$ of equations as follows:

$$\mathsf{eq}_{\gamma_j,\delta_n,\ldots,\delta_1,\gamma_i} := \mathsf{E}\left[U_{\gamma_j}^{\sigma(\delta_n)}, U_{\delta_n}^{\sigma(\delta_{n-1})}\right] \cup \bigcup_{i=1}^{n-2} \mathsf{E}\left[U_{\delta_{i+2}}^{\sigma(\delta_{i+1})}, U_{\delta_{i+1}}^{\sigma(\delta_i)}\right] \cup \mathsf{E}\left[U_{\delta_2}^{\sigma(\delta_1)}, U_{\delta_1}^{\gamma_i}\right] \tag{18}$$

Note that the path determines a product of $U$'s elements

$$U_{\gamma_j}^{\sigma(\delta_n)} U_{\delta_n}^{\sigma(\delta_{n-1})} \cdots U_{\delta_2}^{\sigma(\delta_1)} U_{\delta_1}^{\gamma_i} \tag{19}$$

so that the $(\gamma_j, \gamma_i)$ element of $(\mathsf{qEx}\,(\sigma_\Delta, U) - U_\Gamma^\Gamma)$ equates with the sum of (19) over different paths $\delta_n, \ldots, \delta_1$. See the sigma term of the equation (3) whose term is the sum of (19). Note that the occurrences inside the E's of the RHS of (18) are those from (19).

(Notation for 3): For a formula $\nu$ in a cut list $\Delta$, the formula $\sigma(\nu)$ denotes its counterpart of the cut in the list. E.g., $\sigma(A)$ and $\sigma(A^\perp)$ are respectively $A^\perp$ and $A$. $U_\nu^{\nu'} \in \mathcal{A}_{(\mathbb{Z}_2[\mathcal{L}])}$ denotes the $(\nu, \nu')$ element of the matrix $U \in \mathcal{M}_{\Gamma,\Delta}(\mathcal{A}_{(\mathbb{Z}_2[\mathcal{L}])})$.

(4) Finally by ranging over the paths,

$$\mathsf{eq}(\sigma, U) = \bigcup \mathsf{eq}_{\gamma_j,\delta_n,\ldots,\delta_1,\gamma_i} \quad \text{where } \gamma_i, \gamma_j \in \Gamma \text{ and } \delta_1, \ldots, \delta_n \in \Delta.$$

The equations in the system consist of those in $\mathbb{Z}_2[\mathcal{L}(\Delta, \Gamma)]$.

**Example 5.5 (of Def 5.4).** Let $\tau$ be from (ii) in Example 3.19. In the following, two $\tau$'s are discriminated by $\tau_1$ and $\tau_2$ with the respective eigenweights $a_1$ and $a_2$ for $X^\perp \mathbin{\&} X^\perp$. Then

$$\mathsf{eq}\left(\sigma_{X\oplus X, X^\perp \&_{a_1} X^\perp}, \llbracket cut(\tau_1, \tau_2) \rrbracket\right) = \mathsf{eq}_{X^\perp \&_{a_2} X^\perp, X^\perp \&_{a_1} X^\perp, X\oplus X} \tag{20}$$

The equation holds because the other paths (than the subscription of the right hand side) do not contribute to yield any equation. Then

$$\begin{aligned}
\text{RHS of (20)} &= \mathsf{E}\left[\llbracket \tau_1 \rrbracket_{X^\perp \&_{a_2} X^\perp}^{\sigma(X^\perp \&_{a_1} X^\perp)}, \ \llbracket \tau_2 \rrbracket_{X^\perp \&_{a_1} X^\perp}^{X\oplus X}\right] \\
&= \mathsf{E}\,[\, a_1(1 \otimes q)(1 \otimes p^*) + \bar{a}_1(1 \otimes p)(1 \otimes q^*),\ a_2(1 \otimes q)(1 \otimes p^*) + \bar{a}_2(1 \otimes p)(1 \otimes q^*)\,] \\
&= \{w(a_1(1 \otimes q)(1 \otimes p^*)) = w(\bar{a}_2(1 \otimes p)(1 \otimes q^*)),\ w(\bar{a}_1(1 \otimes p)(1 \otimes q^*)) = w(a_2(1 \otimes q)(1 \otimes p^*))\} \\
&= \{a_1 = \bar{a}_2,\ \bar{a}_1 = a_2\}
\end{aligned}$$

**Definition 5.6 (Satisfiability of the equational system $\mathsf{eq}(\sigma_\Delta, \llbracket \pi \rrbracket)$).**
Let $\pi$ be a proof of $\vdash [\Delta], \Gamma$.

**Applying ring homomorphism:** For a set $\mathsf{E}[\alpha, \beta]$ of equations of (17) in $\mathsf{eq}(\sigma_\Delta, \llbracket \pi \rrbracket)$, and a ring homomorphism $r : \mathbb{Z}_2[\mathcal{L}(\Delta, \Gamma)] \longrightarrow \mathbb{Z}_2[\mathcal{L}(\Gamma)]$, the application $\mathsf{E}[r(\alpha), r(\beta)]$ is defined:

$$\begin{aligned}
\mathsf{E}[r(\alpha), r(\beta)] &:= \mathsf{E}[\mathcal{A}_{(r)}(\alpha), \mathcal{A}_{(r)}(\beta)] \qquad \text{where } \mathcal{A}_{(r)} := r \otimes \mathrm{Id}_{\mathcal{A}} : \mathcal{A}_{(\mathbb{Z}_2[\mathcal{L}(\Delta,\Gamma)])} \longrightarrow \mathcal{A}_{(\mathbb{Z}_2[\mathcal{L}(\Gamma)])} \\
&= \{rw(\alpha_i) = rw(\beta_j) \mid \alpha_i \cdot \beta_j \neq 0\}
\end{aligned}$$

Note that while $w(\alpha_i)$ and $w(\beta_j)$ are monomials of $\mathbb{Z}_2[\mathcal{L}(\Gamma, \Delta)]$ by definition, $rw(\alpha_i)$ and $rw(\beta_j)$ are not necessarily monomials in $\mathbb{Z}_2[\mathcal{L}(\Gamma)]$.

**Satisfiable:** The equational system $\mathsf{eq}(\sigma_\Delta, [\![\pi]\!])$ is *satisfiable* if there exists a ring homomorphism, called a *solution* of the system,

$$d : \mathbb{Z}_2[\mathcal{L}(\Delta,\Gamma)] \longrightarrow \mathbb{Z}_2[\mathcal{L}(\Gamma)] \tag{21}$$

such that for every set $\mathsf{E}[\alpha,\beta]$ of equations of (17) consisting of $\mathsf{eq}(\sigma_\Delta, [\![\pi]\!])$, every equation in the application $\mathsf{E}[d(\alpha), d(\beta)]$ becomes valid in $\mathbb{Z}_2[\mathcal{L}(\Gamma)]_{\mathsf{C}_{+\times}}$. Satisfiability equivalently says that for every literal $b \in \mathcal{L}(\Delta,\Gamma)$, the simultaneous substitution $b := d(b)$ for all the $b$'s makes all the equations in $\mathsf{eq}(\sigma_\Delta, [\![\pi]\!])$ valid in $\mathbb{Z}_2[\mathcal{L}(\Gamma)]_{\mathsf{C}_{+\times}}$. Each image $d(b)$ is a polynomial defining a literal $b \in \mathcal{L}(\Delta,\Gamma)$ from $\boldsymbol{a} = (a_1,\ldots,a_n)$, with each $a_i \in \mathcal{L}(\Gamma)$, in $\mathbb{Z}_2[\mathcal{L}(\Gamma)]$. Note that $d$ restricted to the subdomain $\mathbb{Z}_2[\mathcal{L}(\Gamma)]$ is not necessarily the identity.

**Example 5.7.** (20) of Example 5.5 is satisfiable because: $d : \mathbb{Z}_2[\mathcal{L}(X \oplus X, X^\perp \&_{a_1} X^\perp, X \oplus X, X^\perp \&_{a_2} X^\perp)] \longrightarrow \mathbb{Z}_2[\mathcal{L}(X \oplus X, X^\perp \&_{a_2} X^\perp)]$ is defined by $d(a_1) := \bar{a}_2$ and $d(\bar{a}_1) = a_2$.

The following lemma is automatic.

**Lemma 5.8 (partitioning $\mathsf{eq}(\sigma_{\Delta_1,\Delta_2}, [\![\pi]\!])$).** *For every* MALL *proof of* $\vdash [\Delta_1,\Delta_2],\Gamma$,

$$\mathsf{eq}(\sigma_{\Delta_1} \otimes \sigma_{\Delta_2}, [\![\pi]\!]) = \mathsf{eq}(\sigma_{\Delta_1}, [\![\pi]\!]) \cup \mathsf{eq}(\sigma_{\Delta_2}, [\![\pi]\!])$$

Proof. Direct calculation: Every $\sigma(\delta_j)$ in the path (19) is *exclusively* $\sigma_{\Delta_1}(\delta_j)$ or $\sigma_{\Delta_2}(\delta_j)$ since $\Delta_1$ and $\Delta_2$ are disjoint. Thus each $\mathsf{E}$ constituting of $\mathsf{eq}_{\gamma_j,\delta_n,\ldots,\delta_1,\gamma_i}$ yet arises one in $\mathsf{eq}(\sigma_{\Delta_i}, [\![\pi]\!])$ with $i = 1,2$. □

Note in Lemma 5.8 even when both $\mathsf{eq}(\sigma_{\Delta_i}, [\![\pi]\!])$ are satisfiable for $i = 1,2$, it is not necessarily so for $\mathsf{eq}(\sigma_{\Delta_1} \otimes \sigma_{\Delta_2}, [\![\pi]\!])$. This is because $\mathsf{eq}(\sigma_{\Delta_i}, [\![\pi]\!])$ determines merely a subsystem of the equations and each solution is not necessarily extendable consistently to each other over the full system.

The equational system $\mathsf{eq}(\sigma, [\![\pi]\!])$ so arising autonomously still retains the property of satisfiability, as shown below.

**Proposition 5.9 (satisfiability of $\mathsf{eq}(\sigma_\Delta, [\![\pi]\!])$).**
$\mathsf{eq}(\sigma_\Delta, [\![\pi]\!])$ *is satisfiable for any* MALL *proof* $\pi$ *of* $\vdash [\Delta],\Gamma$. *Note that a solution for the satisfiability is not necessarily unique.*

Proof. By construction of a MALL proof $\pi$, we construct a solution $d$. A solution in some cases is given by a partial homomorphism, whose domain is a subring of $\mathbb{Z}_2[\mathcal{L}(\Delta,\Gamma)]$, but for which any extension over the whole domain becomes a solution. Such partial $d$ is also called as solution by abuse of notation. In the proof, $a\mathsf{E}[x,y]$ denotes $\mathsf{E}[ax,ay]$ for literal $a \in \mathcal{L}$, $x,y \in \mathcal{A}_{(\mathbb{Z}_2[\mathcal{L}])}$, and $a\,\mathsf{eq}(\sigma,U)$ is $\mathsf{eq}(\sigma,U)$ with replacing all its $\mathsf{E}[x,y]$'s with $a\mathsf{E}[x,y]$.
(Case 1) $\pi$ ends with a cut-rule.
(Case 1.1) The two premises (proofs) of the cut both end with logical rules (dual to each other) introducing the cut formulas.

(Case 1.1.1) The dual logical rules are $\otimes_1$ and $\parr$. (cf. Figure 2 for the interpretations of the two premises, say $\pi_1 \otimes \pi_2$ and $\pi'$.)

$$\mathsf{eq}(\sigma_{A_1 \otimes A_2, A_2^\perp \parr A_1^\perp}, [\![\pi]\!]) = \mathsf{eq}(\sigma_{A_1,A_1^\perp,\Delta'} \otimes \sigma_{A_2,A_2^\perp,\Delta'}, [\![\pi_1]\!] \otimes [\![\pi_2]\!] \otimes [\![\pi']\!])$$

where $\Delta$ is $\Delta', A_1 \otimes A_2, A_2^\perp \parr A_1^\perp$. By I.H on the proof $cut(\pi_1, cut(\pi_2,\pi'))$, the RHS is satisfiable.

(Case 1.1.2) The dual logical rules are $\oplus$ and &. (cf. Fig 2 for the interpretations of the two premises, say $\pi_1 \,\&\, \pi_2$ and $\oplus_1(\pi')$.) In the following $\Omega$ (resp. $\Delta_i$) denotes the cut list for $\pi'$ (resp. $\pi_i$).

$$\mathsf{eq}(\sigma_{A_1 \& A_2, A_2^\perp \oplus A_1^\perp}, [\![\pi]\!]) = a\,\mathsf{eq}(\sigma_{A_1, A_1^\perp}, [\![\pi_1]\!] \otimes [\![\pi']\!])$$

$$\text{and} \quad \mathsf{eq}(\sigma_{\Delta_1,\Delta_2,\Omega}, [\![\pi]\!]) = a\,\mathsf{eq}(\sigma_{\Delta_1,\Delta_2,\Omega}, [\![\pi_1]\!] \otimes [\![\pi']\!]) \qquad \text{modulo idempotency of } a$$

Note that the terms scalar multiplied by $\bar{a}$, occurring in the $A_1 \,\&\, A_2$ column and row of the proof $\pi_1 \,\&\, \pi_2$, do not contribute to yield any equation because they are orthogonal (because $p^*q = q^*p = 0$) to the $A_2^\perp \oplus A_1^\perp$ row and of $A_2^\perp \oplus A_1^\perp$ column of the proof $\oplus_1(\pi')$.
Then taking the union of the above two equations by Lemma 5.8:

$$\mathsf{eq}(\sigma_{\Delta_1,\Delta_2,\Omega, A_1 \& A_2, A_2^\perp \oplus A_1^\perp}, [\![\pi]\!]) = a\,\mathsf{eq}(\sigma_{\Delta_1,\Delta_2,\Omega, A_1, A_1^\perp}, [\![\pi_1]\!] \otimes [\![\pi']\!])$$

By I.H on the proof $cut(\pi_1, \pi')$, $\mathsf{eq}(\sigma_{\Delta_1,\Delta_2,\Omega,A_1,A_1^\perp}, [\![\pi_1]\!] \otimes [\![\pi']\!])$ has a solution, which also yields a solution of RHS of the above equation, independently of the value $d(a)$. Thus the assertion holds. Note that $d$ for the assertion is partial since any solution for the I.H provides one for the assertion, thus in particular, both $d(a)$ and $d(\bar{a})$ are arbitrary.

In the following cases $U_\kappa^-$ and $U_*^\kappa$ denote respectively $(\kappa, -)$ and $(*, \kappa)$ components of a matrix $U$ so that $-$ and $*$ are arbitrary.
(Case 1.2) One premise of the cut is an axiom. $\pi$ is $cut(ax, \pi')$ hence $[\![\pi]\!] = \mathsf{dg}\left([\![ax]\!], [\![\pi']\!]\right)$. $\mathsf{eq}(\sigma_{B^\perp, A}, [\![\pi]\!])$ is $\mathsf{E}[[\![ax]\!]_B^{B^\perp}, [\![\pi]\!]_A^-] \cup \mathsf{E}[[\![\pi]\!]_*^A, [\![ax]\!]_{B^\perp}^B]$. Since $[\![ax]\!]_B^{B^\perp}$ and $[\![ax]\!]_{B^\perp}^B$ are 1, the equational system is valid automatically.

(Case 1.3) Other than Cases 1.1 and 1.2.
The most crucial case is one premise (proof) of the cut ends with a &-rule not introducing the cut-formula (other logical rules are direct). This is the case where $\pi$ is the given proof in Case 3.1 (of the proof of Theorem 3.17). The sub-proofs $\rho$ to $\vdash [\Xi], \Delta, A$ and $\pi_i$ to $\vdash [\Omega, \Sigma_i], A^\perp, \Gamma, B_i$ are those referred to in the case there. Let $a$ be the eigenweight of the last &-rule. In the following $a_i$ when $i = 1$ and $i = 2$ denotes respectively $a$ and $\bar{a}$.

$\mathsf{eq}(\sigma_{\Xi,\Omega,\Sigma_1,\Sigma_2 A, A^\perp}, [\![\pi]\!])$ is, by Lemma 5.8, the union of the following $\mathsf{eq}(\sigma_\Xi, [\![\rho]\!])$, (22) and (23):
$\mathsf{eq}(\sigma_\Xi, [\![\pi]\!]) = \mathsf{eq}(\sigma_\Xi, [\![\rho]\!])$
$\mathsf{eq}(\sigma_{\Omega,\Sigma_1,\Sigma_2}, [\![\pi]\!]) = \mathsf{eq}(\sigma_{\Omega,\Sigma_1,\Sigma_2}, [\![\pi_1 \,\&\, \pi_2]\!])$ is (because $\Omega, \Sigma_1, \Sigma_2$ does not contain $B_1 \,\&\, B_2$)

$$\bigcup_{i=1,2,\,\omega\in\Omega} \mathsf{E}\left[a_i [\![\pi_i]\!]_*^{\sigma(\omega)}, a_i [\![\pi_i]\!]_\omega^-\right] \cup \bigcup_{i=1,2,\,\kappa_i \in \Sigma_i} \mathsf{E}\left[a_i [\![\pi_i]\!]_*^{\sigma(\kappa_i)}, a_i [\![\pi_i]\!]_{\kappa_i}^-\right] \tag{22}$$

$\mathsf{eq}(\sigma_{A,A^\perp}, [\![\pi]\!])$ is (because $A$ and $A^\perp$ differ from $B_1 \,\&\, B_2$)

$$\begin{aligned} &\mathsf{E}\left[[\![\pi_1 \,\&\, \pi_2]\!]_*^{\sigma(A)}, [\![\rho]\!]_A^-\right] \cup \mathsf{E}\left[[\![\rho]\!]_*^A, [\![\pi_1 \,\&\, \pi_2]\!]_{\sigma(A)}^-\right] \\ = &\bigcup_{i=1,2} \mathsf{E}\left[a_i [\![\pi_i]\!]_*^{\sigma(A)}, [\![\rho]\!]_A^-\right] \cup \bigcup_{i=1,2} \mathsf{E}\left[[\![\rho]\!]_*^A, a_i [\![\pi_i]\!]_{\sigma(A)}^-\right] \end{aligned} \tag{23}$$

On the other hand consider the proofs $cut(\rho, \pi_i)$ with $i = 1, 2$:
Note first that $[\![cut(\rho, \pi_i)]\!]$ is $\mathsf{dg}\left([\![\rho]\!], [\![\pi_i]\!]\right)$:
$\mathsf{eq}(\sigma_\Xi, \mathsf{dg}\left([\![\rho]\!], [\![\pi_i]\!]\right)) = \mathsf{eq}(\sigma_\Xi, [\![\rho]\!])$
$\mathsf{eq}(\sigma_{\Omega,\Sigma_i}, \mathsf{dg}\left([\![\rho]\!], [\![\pi_i]\!]\right)) = \mathsf{eq}(\sigma_{\Omega,\Sigma_i}, [\![\pi_i]\!])$ with $i = 1, 2$ is

$$\bigcup_{\omega\in\Omega} \mathsf{E}\left[[\![\pi_i]\!]_*^{\sigma(\omega)}, [\![\pi_i]\!]_\omega^-\right] \cup \bigcup_{\kappa_i \in \Sigma_i} \mathsf{E}\left[[\![\pi_i]\!]_*^{\sigma(\kappa_i)}, [\![\pi_i]\!]_{\kappa_i}^-\right] \tag{24}$$

$\mathrm{eq}(\sigma_{A,A^\perp}, \mathrm{dg}\left(\llbracket \rho \rrbracket, \llbracket \pi_i \rrbracket\right))$ with $i = 1, 2$ is

$$\mathsf{E}\left[\llbracket \pi_i \rrbracket_*^{\sigma(A)}, \llbracket \rho \rrbracket_A^-\right] \cup \mathsf{E}\left[\llbracket \rho \rrbracket_*^A, \llbracket \pi_i \rrbracket_{\sigma(A)}^-\right] \tag{25}$$

By I.H on the proof $cut(\rho, \pi_i)$ for each $i = 1, 2$, $\mathrm{eq}(\sigma_{\Xi,\Omega,\Sigma_i,A,A^\perp}, \left\llbracket cut(\rho, \pi_i) \right\rrbracket)$, which by Lemma 5.8 is the union $\mathrm{eq}(\sigma_\Xi, \llbracket \rho \rrbracket) \cup (24) \cup (25)$, has a solution $d_i : \mathbb{Z}_2[\mathcal{L}(\Xi, A, \Omega, \Sigma_i, A^\perp)] \longrightarrow \mathbb{Z}_2[\mathcal{L}(\Delta, \Gamma, B_i)]$.

Then the following $d$ gives a solution of $\mathrm{eq}(\sigma_{\Xi,\Omega,\Sigma_1,\Sigma_2,A,A^\perp}, \llbracket \pi \rrbracket)$:

$$d : \mathbb{Z}_2[\mathcal{L}(\Xi, A, \Omega, \Sigma_1, \Sigma_2, A^\perp, a, \bar{a})] \longrightarrow \mathbb{Z}_2[\mathcal{L}(\Delta, \Gamma, B_1 \,\&\, B_2)] \qquad \text{so that}$$

$$d(c) := \begin{cases} d_i(c) & c \in \mathcal{L}(\Sigma_i) \\ ad_1(c) + \bar{a}d_2(c) & \text{otherwise (i.e., } c \in \mathcal{L}(\Omega, A^\perp) \text{ and } c \in \mathcal{L}(\Xi, A)) \\ 1 & c \in \{a_i \mid i = 1, 2\} \end{cases}$$

This is because: (i) Satisfiability of (22) by $d$ is implied by that of (24) by $d_1$ and by $d_2$, independently of the values of $d(a)$ and of $d(\bar{a})$, and (ii) Depending on the values $d(a) = 1 = d(\bar{a})$, satisfiability of (23) by $d$ is implied from that of (25) by $d_1$ and by $d_2$.

(Case 2) Other than Case 1 so that $\pi$ ends with a logical rule.

We consider the case where $\pi$ ends with a &-rule (other logical rules are direct).

$\mathrm{eq}(\sigma_{\Delta_1,\Delta_2,\Sigma}, \llbracket \pi \rrbracket)$ is

$$\bigcup_{\delta_1 \in \Delta_1} \mathsf{E}\left[\llbracket \pi \rrbracket_*^{\sigma(\delta_1)}, \llbracket \pi \rrbracket_{\delta_1}^-\right] \;\cup\; \bigcup_{\kappa \in \Sigma} \mathsf{E}\left[\llbracket \pi \rrbracket_*^{\sigma(\kappa)}, \llbracket \pi \rrbracket_{\kappa}^-\right] \;\cup\; \bigcup_{\delta_2 \in \Delta_2} \mathsf{E}\left[\llbracket \pi \rrbracket_*^{\sigma(\delta_2)}, \llbracket \pi \rrbracket_{\delta_2}^-\right] \tag{26}$$

By the definition of $\llbracket \pi \rrbracket$ (cf. & of Fig 2), each constituent of the first and the third big unions are equal respectively to the following;

$$a\mathsf{E}\left[\llbracket \pi_1 \rrbracket_*^{\sigma(\delta_1)}, \llbracket \pi_1 \rrbracket_{\delta_1}^-\right] \qquad \text{and} \qquad \bar{a}\mathsf{E}\left[\llbracket \pi_2 \rrbracket_*^{\sigma(\delta_2)}, \llbracket \pi_2 \rrbracket_{\delta_2}^-\right]$$

On the other hand, each constituent of the middle big union is equal to one of the following three, in which $p_i$ denotes $\llbracket \pi_i \rrbracket$ for $i = 1, 2$;

$$\begin{cases} \mathsf{E}\left[ap_1|_*^{\sigma(\kappa)} + \bar{a}p_2|_*^{\sigma(\kappa)},\, ap_1|_\kappa^- + \bar{a}p_2|_\kappa^-\right] = a\mathsf{E}\left[p_1|_*^{\sigma(\kappa)},\, p_1|_\kappa^-\right] \cup \bar{a}\mathsf{E}\left[p_2|_*^{\sigma(\kappa)},\, p_2|_\kappa^-\right] \\ \text{or} \\ \mathsf{E}\left[ap_1|_*^{\sigma(\kappa)} + \bar{a}p_2|_*^{\sigma(\kappa)},\, ap_1|_\kappa^-\right] = a\mathsf{E}\left[p_1|_*^{\sigma(\kappa)},\, p_1|_\kappa^-\right] \\ \text{or} \\ \mathsf{E}\left[ap_1|_*^{\sigma(\kappa)} + \bar{a}p_2|_*^{\sigma(\kappa)},\, \bar{a}p_2|_\kappa^-\right] = \bar{a}\mathsf{E}\left[p_2|_*^{\sigma(\kappa)},\, p_2|_\kappa^-\right] \end{cases}$$

$\mathrm{eq}(\sigma_{\Delta_i,\Sigma}, \llbracket \pi_i \rrbracket)$ for $i = 1, 2$, on the other hand, is

$$\bigcup_{\delta_i \in \Delta_i} \mathsf{E}\left[\llbracket \pi_i \rrbracket_-^{\sigma(\delta_i)}, \llbracket \pi_i \rrbracket_{\delta_i}^-\right] \;\cup\; \bigcup_{\kappa \in \Sigma} \mathsf{E}\left[\llbracket \pi_i \rrbracket_-^{\sigma(\kappa)}, \llbracket \pi_i \rrbracket_{\kappa}^-\right] \tag{27}$$

By I.H, let $d_1$ (resp. $d_2$) be a solution of (27) with $i = 1$ (resp. $i = 2$). Then the following $d$ gives a solution of (26), in which the two solutions $d_i$ with $i = 1, 2$ are superposed for a literal $c$ from the superposed context $\Sigma$:

$$d(c) =^{3} \begin{cases} d_1(c) \quad (\text{resp.}\, d_2(c)) & \text{if} \quad c \in \mathcal{L}(\Delta_1) \quad (\text{resp.}\, c \in \mathcal{L}(\Delta_2)) \\ ad_1(c) + \bar{a}d_2(c) & \text{if} \quad c \in \mathcal{L}(\Sigma) \end{cases} \tag{28}$$

This is because $a_i\mathsf{E}\left[d(\alpha), d(\beta)\right] = a_i\mathsf{E}\left[d_i(\alpha), d_i(\beta)\right]$, whose LHS is the satisfiability of (26) by $d$, and RHS is implied from $\mathsf{E}\left[d_i(x), d_i(y)\right]$ (the satisfiability of (27) by $d_i$).

[3]This $d$ will be denoted by $d_1$ & $d_2$ in the proof of Theorem 5.12 below.

This has shown that if $\mathrm{eq}(\sigma_{\Delta_i,\Sigma}, [\![ \pi_i ]\!])$ with $i = 1, 2$ are satisfiable, then so is $\mathrm{eq}(\sigma_{\Delta_1,\Delta_2,\Sigma}, [\![ \pi ]\!])$. The implication is independent of the values $d(a)$ and $d(\bar{a})$ so that $d$ constructed is partial. □

**Remark 5.10 (non uniqueness of the solution).** Non-uniqueness of the algebraic solution of Prop 5.9 inherits not only from the non-locality of the additive cut-elimination (such as erasing subproofs) but also from the non-communication of & with cuts (such as case 1.3). Consequently, this makes a solution partial since certain subparts of the equational system are discarded and certain variables of literals are not stipulated by the equational system. The non-uniqueness is the origin of the modularity used in this paper for the invariance of Ex during the cut elimination. This kind of modularity was originally pointed out by Girard's GoI III as a difficulty in accommodating the additives. Although the ring hom $d$ arises autonomously, the construction $d$ is up to existence of solutions (satisfiability) rather than deterministic choice.

**Remark 5.11 (internal ring homomorphisms arise autonomously).** Proposition 5.9 guarantees that the autonomous $\mathrm{eq}(\sigma_\Delta, [\![ \pi ]\!])$ is consistent because it has a solution. Although the solution is not uniquely determined as seen in the construction of the proof, it arises autonomously consequently. Existence of such self-arising solutions, yielding ring homomorphisms, is crucial in improving Theorem 3.17 in the subsection below. In Theorem 5.12, invariance of the quasi-execution formula becomes modulo the existence of the genuine autonomous ring homomorphisms, independently of the syntactic cut elimination. For the invariance of Theorem 5.12, the solution $d_\Delta$ restricted to $\mathbb{Z}_2[\mathcal{L}(\Delta)]$, for the cut list $\Delta$, suffices because the restriction makes all the literals of $\mathcal{L}(\Delta)$ definable from $\mathbb{Z}_2[\mathcal{L}(\Gamma)]$ for the conclusion $\Gamma$.

### 5.2 Invariance of $\mathsf{qEx}(\sigma, [\![ \pi ]\!])$ Modulo Autonomous Ring Homomorphisms

Finally, this subsection refines Theorem 3.17 into Theorem 5.12 in terms of solutions of the autonomous system of $\mathrm{eq}(\sigma_\Delta, [\![ \pi ]\!])$. Invariance of the execution formula is up to existence of ring homomorphisms which emerge autonomously as solutions of $\mathrm{eq}(\sigma, [\![ \pi ]\!])$.

**Theorem 5.12 (generalization of Theorem 3.17: autonomous invariance of $\mathsf{qEx}(\sigma, [\![ \pi ]\!])$).** *Suppose* MALL *proof* $\pi$ *of* $\vdash [\Delta], \Gamma$ *reduces to* $\pi'$ *of* $\vdash [\Delta'], \Gamma$ *by any sequence of cut-eliminations. Then the following holds.*

(i) *There exist solutions* $d'$ *and* $d$ *respectively of* $\mathrm{eq}(\sigma_{\Delta'}, [\![ \pi' ]\!])$ *and of* $\mathrm{eq}(\sigma_\Delta, [\![ \pi ]\!])$ *such that*

$$d_\Delta^\Gamma\left(\mathsf{qEx}(\sigma_\Delta, [\![ \pi ]\!])\right) = d'^{\,\Gamma}_{\Delta'}\left(\mathsf{qEx}\left(\sigma_{\Delta'}, [\![ \pi' ]\!]\right)\right)$$

(ii) *In particular, when* $\pi'$ *is cut-free so that* $\Delta'$ *is empty, there exists a solution* $d$ *of* $\mathrm{eq}(\sigma_\Delta, [\![ \pi ]\!])$ *such that*

$$d_\Delta^\Gamma\left(\mathsf{qEx}(\sigma_\Delta, [\![ \pi ]\!])\right) = [\![ \pi' ]\!]$$

*(Terminology for (i) and (ii))*
*For a solution* $d$ *of (21),* $d_\Delta$ *denotes its restriction to the subdomain* $\mathbb{Z}_2[\mathcal{L}(\Delta)]$*, so that* $d_\Delta : \mathbb{Z}_2[\mathcal{L}(\Delta)] \longrightarrow \mathbb{Z}_2[\mathcal{L}(\Gamma)]$*.* $d_\Delta^\Gamma$ *is short for* $d_\Delta[\mathcal{L}(\Gamma)/\mathcal{L}(\Gamma)]$*, so that* $d_\Delta : \mathbb{Z}_2[\mathcal{L}(\Delta, \Gamma)] \longrightarrow \mathbb{Z}_2[\mathcal{L}(\Gamma)]$*. Note that for the assertion, solutions restricted to the subdomain* $\mathbb{Z}_2[\mathcal{L}(\Delta)]$ *matter, regardless of values outside the subdomain.*

Proof. Since (ii) is a direct consequence of (i), we shall prove (i) when the reduced proof (called $rpf$) is a one step reduction of the given proof (called $gpf$) according to the same cases of the proof of Theorem 3.17. In the proof, $\mathrm{Id}_\Gamma$ is short for $\mathrm{Id}_{\mathbb{Z}_2[\mathcal{L}(\Gamma)]}$, and $0_\Gamma$ is short for $\mathrm{Id}_{\mathbb{Z}_2}[\mathbf{0}/\mathcal{L}(\Gamma)]$ mapping all the indeterminates $\mathcal{L}(\Gamma)$ to 0 while identity on the constants $\mathbb{Z}_2$. All the indeterminates $\mathcal{L}(\Gamma)$ in $\mathbb{Z}_2[\mathcal{L}(\Gamma)]$ are considered up to the permutations. We use the canonical iso $\mathbb{Z}_2[\mathcal{L}(A_1 \,\&\, A_2)] \cong \mathbb{Z}_2[\mathcal{L}(A_1), a, \bar{a}, \mathcal{L}(A_2)]$, where $a$ is the eigenweight associated with the &. Note on the other hand, for the dual connective, $\mathbb{Z}_2[\mathcal{L}(A_1 \oplus A_2)] \cong \mathbb{Z}_2[\mathcal{L}(A_1), \mathcal{L}(A_2)]$. Scalar product and sum for ring homomorphisms are pointwise.

In the proof, $d_\Delta$ and $d'_{\Delta'}$ are abbreviated respectively by $f$ and by $f'$ since the cut-lists $\Delta$ and $\Delta'$ are clear in each cases.
(crucial case 1) This is compatible with case 1.2 of Proposition 5.9 so that $d$ gives a solution of the satisfiability there. Take any $f' : \mathbb{Z}_2[\mathcal{L}(\Delta)] \longrightarrow \mathbb{Z}_2[\mathcal{L}(B,\Gamma)]$, then $f$ is defined by;

$$f := f' \uplus 0_{B^\perp} \uplus I : \mathbb{Z}_2[\mathcal{L}(\Delta), \mathcal{L}(B^\perp), \mathcal{L}(A)] \longrightarrow \mathbb{Z}_2[\mathcal{L}(B,\Gamma)],$$
$$\text{where} \quad I : \mathbb{Z}_2[\mathcal{L}(A)] \simeq \mathbb{Z}_2[\mathcal{L}(B)] \quad \text{is an isomorphism.}$$

Since $f(\mathcal{L}(A)) = f(\mathcal{L}(B))$, the proof of the case of Theorem 3.17 directly implies the assertion.

In the following cases, we show existence of two ring homomorphisms $f'$ and $e$

$$f' : \mathbb{Z}_2[\mathcal{L}(\Delta')] \longrightarrow \mathbb{Z}_2[\mathcal{L}(\Gamma)] \qquad\qquad e : \mathbb{Z}_2[\mathcal{L}(\Delta)] \longrightarrow \mathbb{Z}_2[\sharp][\mathcal{L}(\Delta')]$$

where $\sharp$ is a subset [4] of $\mathcal{L}(\Gamma)$ of indeterminates, so that $f$ is $f'[\sharp/\sharp] \circ e$ as follows:

$$f : \mathbb{Z}_2[\mathcal{L}(\Delta)] \xrightarrow{\quad e \quad} \mathbb{Z}_2[\sharp][\mathcal{L}(\Delta')] \xrightarrow{f'[\sharp/\sharp]} \mathbb{Z}_2[\mathcal{L}(\Gamma)].$$

(crucial case 2) This is compatible with case 1.1.2 of Proposition 5.9 so that $f' \circ e$ (with $\sharp = \emptyset$) gives a special solution restricted on the subdomain $\mathbb{Z}_2[\mathcal{L}(\text{the cut formulas of } \pi)]$ of the satisfiability there. We define;

$$e := 0_{A_2, A_2^\perp, \Delta_2} \uplus \mathrm{Id}_{\mathbb{Z}_2}[\,0/\bar{a}, 1/a\,] \uplus \mathrm{Id}_{\Delta_1, \Delta_3, \Omega, A_1, A_1^\perp} :$$
$$\mathbb{Z}_2[\mathcal{L}(\Delta_1, \Delta_2, \Delta_3, \Omega, A_1 \,\&\, A_2, A_2^\perp \oplus A_1^\perp)] \longrightarrow \mathbb{Z}_2[\mathcal{L}(\Delta_1, \Delta_3, \Omega, A_1, A_1^\perp)]$$

Then an arbitrary $f'$ and $f := f' \circ e$ are respective solutions for the assertion. Note that the solution $f$ here is more specified than the compatible case in the proof of Proposition 5.9, as in particular $\bar{a}$ (resp. $a$) is mapped to 0 (resp. to $f'(1)$) by $f$.

(crucial case 3.1)
This is compatible with case 1.3 of Proposition 5.9 so that $f' \circ e$ gives a solution restricted on the subdomain $\mathbb{Z}_2[\mathcal{L}(\text{the cut formulas of } \pi)]$ of the satisfiability there.
For any restriction $f_i : \mathbb{Z}_2[\mathcal{L}(\Xi, \Omega, \Sigma_1, \Sigma_2, A, A^\perp)] \longrightarrow \mathbb{Z}_2[\mathcal{L}(\Delta, \Gamma, B_i)]$ of a solution of $\mathrm{eq}(\sigma_{\Xi,\Omega,\Sigma_1,\Sigma_2,A,A^\perp}, [\![ cut(\rho, \pi_i) ]\!])$, we take $f'$ to be $f_1 \,\&\, f_2$ (defined (28) in Case 2 of Proposition 5.9 as a solution of $\mathrm{eq}(\sigma_{\Xi,\Omega,\Sigma_1,\Sigma_2,A,A^\perp}, [\![ \pi' ]\!])$, where $\pi' = cut(\rho, \pi_1) \,\&\, cut(\rho, \pi_2)$) and define $e$ identity as follows:

$$f' := f_1 \,\&\, f_2 : \mathbb{Z}_2[\mathcal{L}(\Xi, \Omega, \Sigma_1, \Sigma_2, A, A^\perp)] \longrightarrow \mathbb{Z}_2[\mathcal{L}(\Delta, \Gamma, B_1 \,\&\, B_2)] \qquad e := \mathrm{Id}_{\Xi,\Omega,\Sigma_1,\Sigma_2,A,A^\perp}$$

In the corresponding case in the proof of Theorem 3.17, recall that $[\![ gpf ]\!]$ and $[\![ rpf ]\!]$ are identical except for the $(\Xi, \Delta, A,\ \Xi, \Delta, A)$-component, each whose component is the same in $\mathbb{Z}_2[\mathcal{L}]_{\mathtt{C}_+}$. Thus the assertion holds directly.

(crucial case 3.2) [5]
For the same $f_i$ of the above case 3.1, we take $f'$ to be $f_1 \& f_2$ (defined (28) in Case 2 of Proposition 5.9

[4]The set $\sharp$ of indeterminates arises typically for superposing: For two homomorphisms $f, g : \mathbb{Z}_2[\mathcal{L}(\Delta)] \longrightarrow \mathbb{Z}_2[\mathcal{L}(\Delta')]$, their superposition $af + bg$ is $\mathbb{Z}_2[\mathcal{L}(\Delta)] \longrightarrow \mathbb{Z}_2[\sharp][\mathcal{L}(\Delta')]$ with $\sharp = \{a, b\}$.

[5]This case has no counterpart in the proof of Proposition 5.9 because only the special reduction case 3.1 suffices to reduce the size of given $\pi$ for proving Prop 5.9. Recall that Prop 5.9 is a claim on proof not on proof reduction.

as a solution of $\mathrm{eq}(\sigma_{\Xi^\delta,\Omega,\Sigma_1,\Sigma_2,(A,A^\perp)^\delta}, [\![\pi']\!])$ and define $e$ (no longer an endomorphism contrary to the case 3.1 above ) as follows so that $f = f'[\sharp/\sharp]\circ e$ with $\sharp = \{a, \bar{a}\}$:

$$f' := f_1 \,\&\, f_2 : \mathbb{Z}_2[\mathcal{L}(\Omega, \Xi^\delta, \Sigma_1, \Sigma_2, (A, A^\perp)^\delta)] \longrightarrow \mathbb{Z}_2[\mathcal{L}(\Delta, \Gamma, B_1 \,\&\, B_2)]$$

$$e := \left(a\, J_{\Xi^{\delta_1},(A,A^\perp)^{\delta_1}} + \bar{a}\, J_{\Xi^{\delta_2},(A,A^\perp)^{\delta_2}}\right) \uplus \mathrm{Id}_{\Omega,\Sigma_1,\Sigma_2} \quad \text{is :}$$

$$\mathbb{Z}_2[\mathcal{L}(\Xi, A, A^\perp), \mathcal{L}(\Omega, \Sigma_1, \Sigma_2)] \longrightarrow \mathbb{Z}_2[\sharp][\mathcal{L}(\Xi^\delta, (A, A^\perp)^\delta), \mathcal{L}(\Omega, \Sigma_1, \Sigma_2)],$$

where $J_{\Xi^{\delta_i},(A,A^\perp)^{\delta_i}}$ is the ring inclusion (7), complementary for $i = 1, 2$, and the left of the $\uplus$ is a *superposition* of the two complementary inclusions:

$$a\, J_{\Xi^{\delta_1},(A,A^\perp)^{\delta_1}} + \bar{a}\, J_{\Xi^{\delta_2},(A,A^\perp)^{\delta_2}}$$

First, for any polynomial $h_i \in \mathbb{Z}_2[\mathcal{L}(\Delta, \Gamma, B_1\&B_2)]$ with $i = 1, 2$, and its copy $h_i^{\delta_i} \in \mathbb{Z}_2[\mathcal{L}(\Delta^{\delta_i}, \Gamma, B_1\& B_2)]$, the following holds where the substitutions are simultaneous for $c$ (resp. its copy $c^{\delta_i}$) ranging in $\mathcal{L}(\Delta)$ (resp. in $\mathcal{L}(\Delta^{\delta_i})$);

$$\begin{aligned} ah_1^{\delta_1}[c^{\delta_1} := f'(c^{\delta_1})] &= ah_1^{\delta_1}[c^{\delta_1} := af'(c^{\delta_1})] && \text{mod } a^2 = a \\ &= ah_1^{\delta_1}[c^{\delta_1} := af'(c^{\delta_1}) + \bar{a}f'(c^{\delta_2})] && \text{mod } a.\bar{a} = 0 \\ &= ah_1[c := f(c)] && \text{definition of } d \end{aligned}$$

Samely, $\bar{a}h_2^{\delta_2}[c^{\delta_2} := f'(c^{\delta_2})] = \bar{a}h_2[c := f(c)]$.
These mean $f'(ah_1^{\delta_1}) = af(h_1)$ and $f'(\bar{a}h_2^{\delta_2}) = \bar{a}f(h_2)$, respectively.

The instance of (30), in the proof of Lemma 3.18 (Sec C of Appendix), holds with the substitution $c := e(c)$ in RHS, thus so does also the instance (31) with the same substitution in RHS. By applying $f'[\sharp/\sharp]$ to both sides of this instance (31), the assertion is obtained because $f'[\sharp/\sharp](c^{\delta_i}) = f'[\sharp/\sharp]\circ e(c)$ by the definition of $e$. □

## Conclusion

The two main contributions of this paper are

(i) MALL GoI modelling to accommodate polynomial weights. The ingredient is the change of coefficient ring for the partial isometries to the polynomial ring of characteristic 2 in literals of eigenweights. An execution formula is formulated, invariant under cut-elimintaion, in terms of a ring homomorphism of the polynomial boolean ring. The ring homomorphism is first given dependent on proof-reduction, but finally it is improved so that it is independent of the reductions.

(ii) Constructing a finer grained quasi-execution formula, using the semiring of formal languages over eigenweights. The formula together with (i) captures the HvG correctness criteria (P2) and (P3)

We now discuss some future directions. Our GoI is inspired directly from Girard's operator theoretic interpretation of GoI I [8] and GoI III [9], thus the categorical counterpart of our construction has to be studied. For this, the question of how to accommodate the homomorphisms on the boolean polynomial ring, used in this paper, needs to be examined in accordance with the axioms of traced monoidal categories [19]. To be more concrete, how do we enrich polynomial boolean weights on top of Haghverdi-Scott's Σ-mon category [12], and obtain a unique decomposition category, a structure which is the main ingredient for categorical GoI I. Also, we believe their trace class using partial traces [13, 21] may be useful to distinguish consistently weighted traced morphisms.

Extension of our GoI to exponentials for the full LL involves checking directly whether the scalar extension of the paper is consistent with Girard's axiomatization [8] in terms of tensor product for comultiplication and dereliction for the LL modal connectives. This is work under development.

Another open problem is applying our GoI to Heijltjes's additive proof-nets [15] for graph rewriting of HvG, which must explain a GoI characterization of Joyal's *softness* [1], a nice categorical meaning of additive proof-theory for sum-product logic.

From a different perspective on the decision problem, the toggling condition (P3) is studied by De Naurois-Mogbil [23] to lower the complexity of the correctness of MALL proofs. How our GoI may relate to this perspective remains a future work.

A comparison needs to be investigated with the most recent work of Seiller's *graphings* [25] of general GoI for MALL, encompassing the standard GoI as well as von Neumann algebras. Because of the similarity between his monoid weightings to graphical edges and our algebraic scalar extension, we hope to understand how the graphings could accommodate superposition of slices, peculiar to the HvG proof-nets, hence indispensable to our additive GoI framework.

## Acknowledgment

The author would like to thank the referees for detailed and very helpful comments that have improved the presentation.

## A Definition of Tensor Product

For modules $M$ and $N$ over a commutative ring $R$, the tensor product $M \otimes_R N$ is the unique (up to iso) $R$-module together with the unique (up to iso) bilinear map $b$ satisfying the following universal property:

For any bilinear map $f$ to any $R$-module $P$, there exists a unique $R$-module homomorphism $f'$ so that the right diagram commutes:

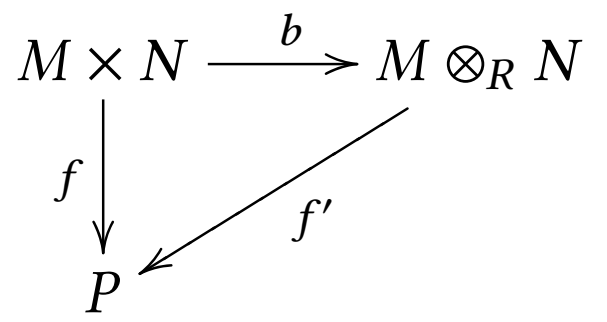

By writing $b(x, y) = x \otimes y$ for $x \in M$ and $y \in N$, the following holds for $x, x' \in R$, $y, y' \in N$ and $r \in R$:

$$(x + x') \otimes y = x \otimes y + x' \otimes y \qquad x \otimes (y + y') = x \otimes y + x \otimes y' \qquad rx \otimes y = x \otimes ry = r(x \otimes y)$$

## B I/O box for $[\![\pi]\!]$ and for qEx $(\sigma_\Delta, [\![\pi]\!])$

Figures 4, 5 and 6 below.

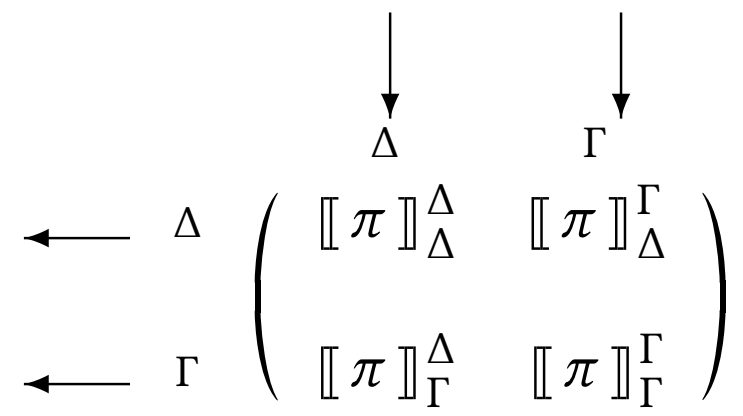

Fig. 4. I/O box for $[\![\pi]\!]$

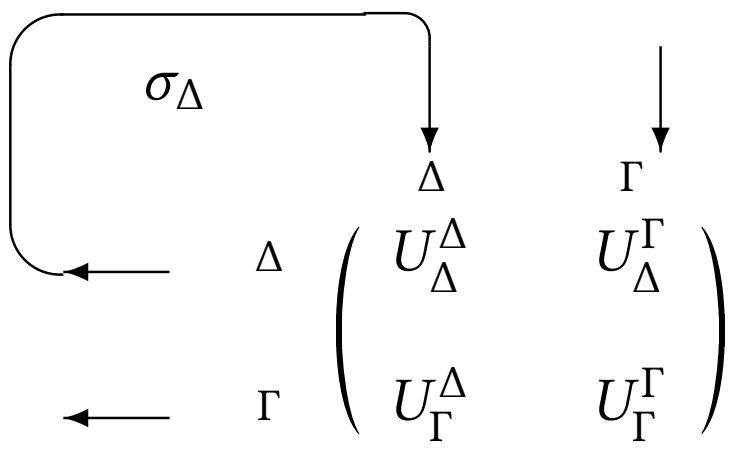

Fig. 5. I/O box for qEx $(\sigma_\Delta, U)$

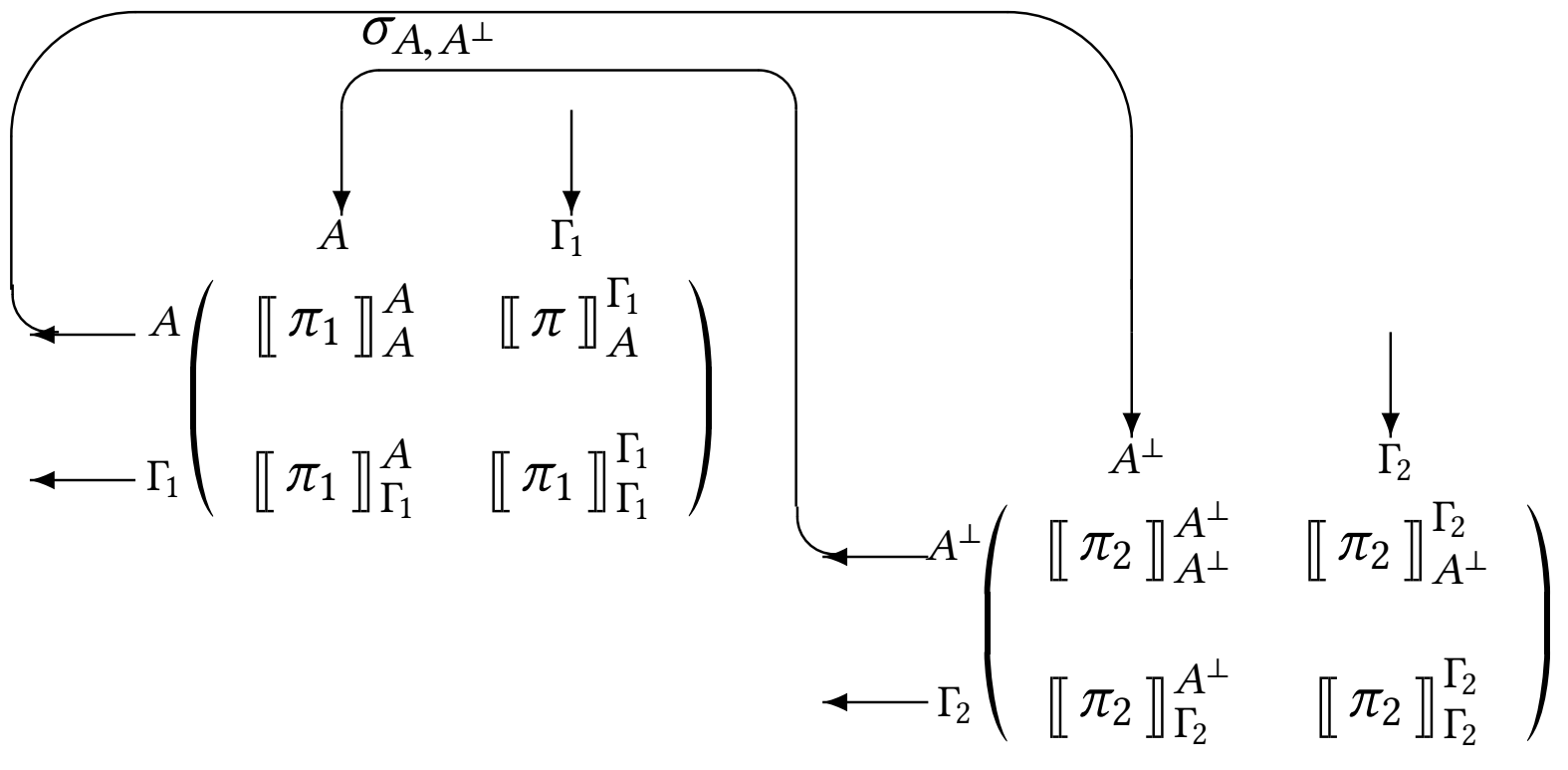

Fig. 6. figure 5 when $U = [\![\pi]\!]$ and $\pi$ is cut of $\pi_1$ and $\pi_2$

## C Proof of Lemma 3.18

Let S and $S^\delta$ denote the indicated occurrences (of the superposition) in LHS and in RHS respectively. Then S and $S^\delta$ have the identical $(\ )^{\Gamma, B_1\&B_2}_{\Gamma, B_1\&B_2}$ component. For S's other components, the component $S^\Delta_\Delta$ is $af^\Delta_\Delta + \bar{a}g^\Delta_\Delta$ and the component $S^-_\Delta$ (resp. $S^\Delta_*$) is written of the form $aM_1 + \bar{a}M_2$ (resp. $aN_1 + \bar{a}N_2$). For $S^\delta$'s other components, $(S^\delta)^{\Delta^{\delta_1},\Delta^{\delta_2}}_{\Delta^{\delta_1},\Delta^{\delta_2}} = \mathrm{dg}\left(af^\Delta_\Delta,\ \bar{a}g^\Delta_\Delta\right)$, hence $((S^\delta)^{\Delta^{\delta_1},\Delta^{\delta_2}}_{\Delta^{\delta_1},\Delta^{\delta_2}})^n = \mathrm{dg}\left(a^n(f^\Delta_\Delta)^n,\ \bar{a}^n(g^\Delta_\Delta)^n\right)$, and $(S^\delta)^-_{\Delta^{\delta_1},\Delta^{\delta_2}}$ (resp. $(S^\delta)^{\Delta^{\delta_1},\Delta^{\delta_2}}_*$) is ${}^t(1,1)\ \mathrm{dg}\,(aM_1, \bar{a}M_2)$ (resp. ${}^t(1,1)\mathrm{dg}\,(aN_1, \bar{a}N_2)$).

In the proof the following equality is used (since $d$ is a homomorphism) for any polynomial $h \in \mathbb{Z}_2[\mathcal{L}(\Delta,\Gamma)]$,

$$d(h) = h\,[\,c := d(c)\,]\,,$$

where the substitution of RHS is the simultaneous so that $c$ ranges in $\mathcal{L}(\Gamma)$. All the substitutions in the following are simultaneous where $c$ (resp. its copy $c^{\delta_i}$) ranges in $\mathcal{L}(\Gamma)$ (resp. $\mathcal{L}(\Gamma^{\delta_i})$).

First, let the pair of $h_1$ and $h_2$, occurring in S, denote either that of $f^\Delta_\Delta$ and $g^\Delta_\Delta$, of $M_1$ and $M_2$, or of $N_1$ and $N_2$, and $h_1^{\delta_1}$ and $h_2^{\delta_2}$ denote their occurrences in $S^\delta$. Then the following holds for every $c \in \mathcal{L}(\Delta)$ so that $c^{\delta_i} \in \mathcal{L}(\Delta^{\delta_i})$ with $i = 1, 2$:

$$\begin{aligned} ah_1^{\delta_1} &= ah_1^{\delta_1}[c^{\delta_1} := ac^{\delta_1}] && \text{mod } a^2 = a \\ &= ah_1[c := ac^{\delta_1} + \bar{a}c^{\delta_2}] && \text{mod } a.\bar{a} = 0 \\ &= ah_1[c := d(c)] && \text{by the definition of } d \end{aligned} \tag{29}$$

Samely, $\bar{a}h_2^{\delta_2} = \bar{a}h_2[c := d(c)]$.

Second,

$$\begin{aligned} \left(\sigma_{\Delta^\delta}(S^\delta)^{\Delta^{\delta_1},\Delta^{\delta_2}}_{\Delta^{\delta_1},\Delta^{\delta_2}}\right)^n &= (\sigma_{\Delta^{\delta_1}} af^{\Delta^{\delta_1}}_{\Delta^{\delta_1}})^n + (\sigma_{\Delta^{\delta_2}}\bar{a}g^{\Delta^{\delta_2}}_{\Delta^{\delta_2}})^n && \text{by } a\bar{a} = 0 \\ &= (\sigma_{\Delta^{\delta_1}} af^\Delta_\Delta\,[c := d(c)]\,)^n + (\sigma_{\Delta^{\delta_2}}\bar{a}g^\Delta_\Delta\,[c := d(c)]\,)^n && \text{by (29)} \\ &= (\sigma_\Delta(af^\Delta_\Delta\,[c := d(c)]\ + \bar{a}g^\Delta_\Delta\,[c := d(c)]\,))^n && \text{by } a\bar{a} = 0 \\ &= (\sigma_\Delta(af^\Delta_\Delta + \bar{a}g^\Delta_\Delta)\ [c := d(c)]\,)^n \end{aligned} \tag{30}$$

On the other hand, $(S^\delta)^-_{\Delta^{\delta_1},\Delta^{\delta_2}} =^t (1,1)\ \mathrm{dg}\,(aM_1, \bar{a}M_2) =^t (1,1)\ \mathrm{dg}\,(aM_1\,[c = d(c)]\,, \bar{a}M_2\,[c := d(c)])$, again by (29). Same calculation for $(S^\delta)^{\Delta^{\delta_1},\Delta^{\delta_2}}_*$. Thus

$$\begin{aligned} &(S^\delta)^{\Delta^{\delta_1},\Delta^{\delta_2}}_*\ \left(\sigma_{\Delta^\delta}(S^\delta)^{\Delta^{\delta_1},\Delta^{\delta_2}}_{\Delta^{\delta_1},\Delta^{\delta_2}}\right)^n\ \sigma_{\Delta^\delta}(S^\delta)^-_{\Delta^{\delta_1},\Delta^{\delta_2}} \\ &= (aN_1 + \bar{a}N_2)\,[c := d(c)]\ (30)\ (aM_1 + \bar{a}M_2)\,[c := d(c)] \\ &= \left(S^\Delta_*\ (\sigma_\Delta S^\Delta_\Delta)^n\ \sigma_\Delta S^-_\Delta\right)[c := d(c)] \end{aligned} \tag{31}$$

Thus the lemma holds.